\documentclass{iopjournal}

\usepackage{graphicx}
\usepackage{multirow}
\usepackage{amsmath,amssymb,amsfonts}
\usepackage{amsthm}
\usepackage{mathrsfs}
\usepackage[title]{appendix}
\usepackage{xcolor}
\usepackage{textcomp}
\usepackage{booktabs}
\usepackage{url}
\usepackage{xr-hyper}          % cross-document references to the standalone SI

\providecommand{\botrule}{\bottomrule}

\usepackage{braket}           % Dirac notation: \ket, \bra, \braket
\usepackage{tikz}
\usetikzlibrary{shapes,arrows.meta,positioning,fit,backgrounds,calc,decorations.pathreplacing}

\usepackage{soul}
\newif\ifhighlight
\highlightfalse
\sethlcolor{yellow}
\newcommand{\rev}[1]{\ifhighlight\hl{#1}\else#1\fi}

\newcommand{\nano}{n}
\providecommand{\micro}{\ensuremath{\mu}}
\newcommand{\milli}{m}
\newcommand{\kilo}{k}
\newcommand{\mega}{M}
\newcommand{\giga}{G}
\providecommand{\second}{s}
\newcommand{\hertz}{Hz}
\providecommand{\kelvin}{K}
\newcommand{\radian}{rad}
\newcommand{\decibel}{dB}
\newcommand{\SI}[2]{#1\,#2}
\newcommand{\si}[1]{#1}

\newcommand{\Tr}{\operatorname{Tr}}
\newcommand{\dF}{\delta F}                    % Filter function discrepancy
\newcommand{\heom}{\textrm{HEOM}}
\newcommand{\emuplat}{\textsc{EmuPlat}}
\newcommand{\vppu}{\textrm{VPPU}}
\newcommand{\cpmg}{\textrm{CPMG}}
\newcommand{\dac}{\textrm{DAC}}
\newcommand{\nco}{\textrm{NCO}}
\newcommand{\isa}{\textrm{ISA}}

\newcommand{\Hq}{H_{\mathrm{q}}}             % Qubit Hamiltonian
\newcommand{\Hd}{H_{\mathrm{d}}}             % Drive Hamiltonian
\newcommand{\Hsys}{H_{\mathrm{sys}}}         % System Hamiltonian

\newcommand{\ketbra}[2]{\ket{#1}\!\bra{#2}}

\newcommand{\Xq}{\hat{X}_{\mathrm{q}}}
\newcommand{\Yq}{\hat{Y}_{\mathrm{q}}}

\DeclareMathOperator{\sinc}{sinc}

\newcommand{\includegraphicsorplaceholder}[2]{%
  \IfFileExists{#2}{%
    \includegraphics[width=#1]{#2}%
  }{%
    \fbox{\parbox{0.9\linewidth}{\centering\vspace{1em}%
    Supplementary figure unavailable for this build.\\[0.5em]
    \texttt{\detokenize{#2}}\vspace{1em}}}%
  }%
}

\newif\ifsiseparate
\siseparatefalse
\ifsiseparate
\else
  \hypersetup{hypertexnames=false}
\fi

\begin{document}

% Single-PDF / arXiv build (\siseparatefalse): the SI is inlined as an appendix,
% so main text and SI share ONE .aux. The separate build's cross-document keys have
% no xr target here: the main text writes \ref{supp:<SI label>} and the SI writes
% \ref{main:<main-text label>}. Strip the supp:/main: prefix at the REFERENCE side
% so each key resolves to its native in-document label -- this avoids editing the
% 90+ \ref call sites. We patch \ref (the single stable entry point; \eqref calls
% \ref, so it is covered too) rather than \label, because float caption and amsmath
% environments route \label through internals that bypass a redefined \label.
% No-op in separate mode.
\ifsiseparate\else
  % Deep-copy the robust \ref. NOT \let: \ref is a robust command whose real body
  % lives in the inner \ref␣; \let copies only the \protect\ref␣ shell, and our
  % \DeclareRobustCommand below then overwrites \ref␣, so \jyorigref would recurse
  % into the NEW \ref and leak \endcsname into the key. \NewCommandCopy clones it
  % properly (inner cs included).
  \NewCommandCopy\jyorigref\ref
  \ExplSyntaxOn
  % Drop a leading "supp:" or "main:" (both 5 chars); leave any other key untouched.
  % \exp_args:Ne fully expands the stripped key (\expanded semantics) before handing
  % it to the original robust \ref. Two non-obvious requirements, both learned by
  % compiling: (1) a plain \edef is NOT enough -- \str_range relies on \expanded to
  % finish its \csname...\endcsname, so \edef leaks a bare \endcsname into the key;
  % (2) redefine with \DeclareRobustCommand (the very mechanism hyperref uses for
  % \ref), NOT \RenewDocumentCommand -- wrapping this high-frequency kernel macro in
  % xparse derails later parsing and the run never reaches \bibliography.
  \cs_new:Npn \jy_strip_crossdoc:n #1 {
    \str_if_eq:eeTF { \str_range:nnn {#1} {1} {5} } { supp: }
      { \str_range:nnn {#1} {6} {-1} }
      { \str_if_eq:eeTF { \str_range:nnn {#1} {1} {5} } { main: }
          { \str_range:nnn {#1} {6} {-1} }
          { \exp_not:n {#1} } }
  }
  \DeclareRobustCommand{\ref}[1]{\exp_args:Ne \jyorigref { \jy_strip_crossdoc:n {#1} }}
  \ExplSyntaxOff
\fi

\articletype{Paper}

\title{Non-perturbative CPMG scaling and qutrit-driven breakdown
  under compiled superconducting-qubit control:
  a single-qubit study}

\author{Jun Ye$^{1,*}$\orcid{0000-0003-1963-0865}}

\affil{$^1$CETC International Cornerstone Quantum Industry (Suzhou) Co., Ltd., Suzhou, Jiangsu Province, P.R. China}

\affil{$^*$Author to whom any correspondence should be addressed.}

\email{yjmaxpayne@hotmail.com}


%%% Abstract %%%
% abstract.tex — target <=300 words for QST

\begin{abstract}
Decoherence in superconducting qubits \rev{arises from both}
multilevel dynamics and structured environmental noise, yet
perturbative models cannot capture all resulting signatures.
Here, \emuplat{} couples instruction-set-architecture-level waveform
generation to the hierarchical equations of motion (\heom{}) under
$1/f$ non-Markovian pure dephasing.
In the resulting non-perturbative regime---where filter-function
predictions become quantitatively uninformative---\cpmg{} scaling of a
three-level superconducting transmon yields one calibration result, two physical findings,
and one structural null.
Y-\cpmg{} exhibits axis-dependent scaling-law breakdown---non-monotonic
decoherence, partial coherence revival, and pronounced X--Y population
asymmetry ($0.204$ vs ${<}\,0.01$)---driven by third-level anharmonicity amplified by bath memory;
X-\cpmg{} maintains well-behaved power-law scaling with a finite-$n$
transient excess consistent with non-Markovian bath-memory effects.
\rev{This null result is equally informative}: waveform-level
differences---Standard versus \vppu{} realizations---remain
undetectable across all coupling strengths. \rev{This shows that}
rotating-frame pure-dephasing coupling renders control-layer detail
invisible to scaling observables.
These findings define testable predictions, the most experimentally
accessible requiring only qualitative verification.
\end{abstract}

%%% Main body %%%
% introduction.tex — §1 Introduction (~800 words)

\section{Introduction}\label{sec:intro}

Superconducting transmon processors operate at the intersection of two
structured sources of error.
On the control side, gate instructions are converted into finite-resolution
waveforms by DACs, numerically controlled oscillators (NCOs), and FPGA timing
logic.
On the environment side, the same qubits interact with strongly colored baths,
most notably $1/f$ charge noise and fluctuating two-level systems.
As coherence improves, these two mechanisms can no longer be treated as
independent perturbations: control artefacts that appear negligible at the
waveform level can be selectively amplified once the bath retains long
memory~\cite{krantz2019,vandijk2019,falci2024}.

This issue is no longer only theoretical.
Burkard derived exact non-Markovian qubit dynamics under $1/f$ noise without a
Markov approximation~\cite{burkard2009}.
Bylander~et~al.\ used dynamical decoupling to measure the low-frequency noise
spectrum over multiple frequency decades~\cite{bylander2011}, and subsequent
work identified coherence revivals and other signatures of long bath memory in
superconducting-qubit settings~\cite{gulacsi2023}.
At the same time, dynamical-decoupling performance depends experimentally on
the realized pulse sequence rather than on idealized protocol labels
alone~\cite{biercuk2009,ezzell2023}.
Recent experiments further show that pulse distortions, leakage-control
envelopes, and cryogenic controller nonidealities can all become directly
visible in superconducting-qubit gate performance~\cite{hyyppa2024leakage,
underwood2023cmos,hellings2025flux}.
The physical question is therefore not whether control hardware matters in
principle, but what physics emerges---and what becomes invisible---when
compiled control interacts with a non-Markovian bath.

Existing simulation paradigms address only parts of that question.
Digital-twin, calibration, and pulse-level platforms reproduce waveform
generation and control-stack constraints faithfully but usually stop at
Lindblad noise models~\cite{wittler2021c3,teske2022qopt,mueller2025,
fruitwala2024}.
Conversely, recent studies using the hierarchical equations of motion
(\heom{}), a numerically exact method beyond the Born--Markov
approximation, treat bath memory rigorously but still start from
idealized mathematical
pulses~\cite{chen2025heom,nakamura2024single,nakamura2025two,nakamura2025cpmg}.
The filter-function formalism provides an analytic bridge between control and
noise spectroscopy~\cite{cerfontaine2021,hangleiter2021,barnes2016}, yet its
own literature identifies slow correlated baths as a boundary of perturbative
reliability.
We refer to the conditions beyond this boundary---where bath-induced
decoherence at experimental timescales is no longer a small perturbation
amenable to first-order cumulant expansion---as the
\emph{non-perturbative regime} throughout this work (quantitative
calibration in Section~\ref{sec:results:ff}).
What is still missing is a controlled path from hardware-faithful instruction
execution to beyond-Lindblad open-system dynamics.

This gap is one of observability.
Single-pulse metrics can remain almost unchanged while cumulative multi-pulse
observables, such as Carr--Purcell--Meiboom--Gill (\cpmg{}) scaling laws, can be qualitatively altered by
the interplay of compiled control, bath memory, and qutrit-level physics.
Ideal-pulse \heom{} cannot resolve that distinction because it omits the
compiled control path, whereas realistic-pulse Lindblad simulation cannot
resolve it because it omits bath memory.
What is needed is a framework in which waveform realization and bath dynamics
can be varied independently while all other ingredients are held fixed.

\emuplat{}~\cite{emuplat2026} \rev{is that enabling framework rather than an end in itself}.
It takes a pulse specification through four successive representation
stages \rev{(from gate-level definition to time-domain waveform)} and can pass
the resulting drive to either a Lindblad or \heom{} solver within the same
simulation run, \rev{so that} solver comparisons share identical input.
Two waveform-compilation routes are provided: the Standard route constructs
analytic envelopes, while the Virtual Pulse Processing Unit (\vppu{})~\cite{nvqlink2025} route reproduces QubiC-style FPGA
waveform generation, including amplitude quantization, phase accumulation,
and timing discretization~\cite{xu2021qubic,fruitwala2024}.
For the single-qubit Ramsey and \cpmg{} comparisons reported here, all
compiled-pulse calculations are performed in the qubit rotating frame so that
differences between waveform realizations can be assigned to control
realization rather than to carrier-resolved numerical stiffness.

We therefore frame the present \rev{study} as a controlled numerical
\rev{investigation} within deliberately bounded conditions, not as a general
bound for superconducting-qubit control.
\rev{These bounds are deliberate rather than incidental: holding the device,
control architecture, and frame fixed isolates the
compiled-control--bath-memory interplay, so that an observed effect can be
attributed to that interplay rather than to confounding parameter variation,
whereas a genuinely general bound would demand parameter, bath-model, and
architecture sweeps beyond the reach of exact \heom{} dynamics. Which of the
resulting signatures are expected to transfer to real devices, and which
remain specific to this model, is delimited in
Section}~\ref{sec:disc:limitations}\rev{.}
The evidentiary chain is deliberately narrower than the hardware platform
itself: a single $d = 3$ transmon with the spectator qubit decoupled
(\rev{SI Sec.}~\ref{supp:si:platform_model}), one QubiC-like compiled-control
architecture, and rotating-frame RWA dynamics.
For the strongest sequence-level claims we further restrict to the Tier-0
(literature-calibrated) window---the lowest of three bath-coupling strengths
spanning one decade in~$\eta$ (Methods~\S\ref{sec:meth:bath})---where stable
compiled-pulse fits remain available.
Section~\ref{sec:discussion} discusses which control-stack features the
rotating-frame choice preserves and which it averages away.
Within those boundaries, the comparison is physically informative in
either direction: it can
expose bath-memory phenomena invisible to simpler approaches, and
equally establish when a waveform null is structural rather than a
sensitivity artefact.

Using this controlled comparison, we report one calibration finding,
two core physical findings, and one structural null.
First-order filter-function theory becomes quantitatively uninformative
in the present $1/f$ window, predicting infidelities orders of magnitude
below exact dynamics; this breakdown is expected in strong-coupling
regimes~\cite{cerfontaine2021} but had not previously been calibrated
against exact dynamics, and its quantification establishes the
non-perturbative baseline for what follows.
Even-$n$ X-\cpmg{} analysis provides suggestive evidence of a
bath-memory-driven transient scaling excess: fitting
$\chi = -\ln(1 - \varepsilon) \propto n^\gamma$, X-\cpmg{} yields
$\gamma$ marginally exceeding the perturbative prediction
$\gamma \approx 1$ (Cywi\'{n}ski Eq.\,25, fixed-$\tau$ protocol),
with a finite-$n$ transient absent under matched Lindblad dynamics.
The more striking finding is an axis-dependent scaling-law
breakdown: $d = 3$ qutrit physics causes Y-\cpmg{} to exhibit
non-monotonic decoherence and partial coherence revival, while
X-\cpmg{} maintains well-behaved power-law scaling under identical
conditions---an asymmetry traceable to anharmonicity-driven per-pulse
population redistribution amplified by non-Markovian bath dynamics
(\rev{SI Sec.}~\ref{supp:si:cpmg_params}).
Waveform-realization differences, by contrast, remain undetectable
across all coupling strengths and metrics examined.

These results define concrete experimental targets ordered by
accessibility, from a qualitative Y-\cpmg{} non-monotonicity test to
the more demanding X-\cpmg{} scaling exponent measurement.
\rev{Together, these findings} show that non-Markovian bath memory and
qutrit-level physics produce physically distinct phenomena---most
robustly the axis-dependent breakdown driven by qutrit physics, and
more tentatively a bath-memory-driven scaling excess---that are
invisible to either idealized-pulse or Lindblad-dynamics pictures alone.

\rev{The remainder of the paper follows the model-specific scope set out
above---a single rotating-frame, single-qutrit transmon in the Tier-0
coupling window. Section}~\ref{sec:results}
\rev{presents the results in four steps: calibration of the first-order
perturbative baseline against exact \heom{} dynamics, the even-\mbox{$n$}
\cpmg{} transient structure, the axis-dependent Y-\cpmg{} scaling-law
breakdown, and the paired waveform comparison establishing the structural
null. Section}~\ref{sec:discussion}
\rev{interprets these findings, states explicitly which of the bounded
conditions each result depends on, and derives the resulting experimental
predictions; the platform model, \heom{} validation, and statistical
methodology are collected in the Supplementary Information.}

\rev{The Supplementary Information is supplied as a separate
Supplementary Data File, with sections numbered S1 onward and all
figures and tables carrying an ``S'' prefix.}
\rev{It collects, in five parts, the HEOM solver validation and
benchmarks; the analytical theory framework, covering the
filter-function method and the qutrit breakdown model; the
control-platform architecture and effective-model assumptions; the
pipeline-isolation and cross-solver validation study; and the
statistical methodology and convergence checks.}
\rev{To help the reader locate the evidence behind each claim,
Table}~\ref{tab:si_map}~\rev{maps every principal main-text result to
its supporting supplementary section.}

\begin{table}[htbp]
\centering
\caption{Map from principal main-text results to their supporting
Supplementary Information (SI) sections. Section numbers of the form
``S$n$'' refer to the standalone supplementary document.}
\label{tab:si_map}
\begin{tabular}{@{}p{0.45\textwidth}p{0.45\textwidth}@{}}
\toprule
\textbf{Main-text result} & \textbf{Supporting SI section(s)} \\
\midrule
Platform architecture and effective single-qutrit model
  (Fig.~\ref{fig:architecture}) &
  SI Sec.~\ref{supp:si:terminology}, \ref{supp:si:platform_model} \\
Calibration of the perturbative filter-function baseline &
  SI Sec.~\ref{supp:si:ff_method}, \ref{supp:si:burkard} \\
Even-$n$ CPMG transient structure &
  SI Sec.~\ref{supp:si:espira_cpmg}, \ref{supp:si:convergence} \\
Axis-dependent Y-CPMG scaling-law breakdown &
  SI Sec.~\ref{supp:si:qutrit_model}, \ref{supp:si:ycpmg_robustness} \\
Paired Standard-versus-VPPU waveform null &
  SI Sec.~\ref{supp:si:pipeline}, \ref{supp:si:vppu_phase} \\
Non-perturbative HEOM ground truth and stability &
  SI Sec.~\ref{supp:si:numerical_validation}, \ref{supp:si:heom_boundary} \\
Scaling-exponent statistical robustness &
  SI Sec.~\ref{supp:si:fitting_robustness}, \ref{supp:si:bootstrap} \\
\bottomrule
\end{tabular}
\end{table}

% results.tex — §2 Results (~3000 words, 5 main figures)

\section{Results}\label{sec:results}

% fig1_architecture.tex — Three-panel architecture diagram for EmuPlat
% Usage: \input{figures/fig1_architecture.tex}  (from main.tex)
%
% Required in preamble:
%   \usepackage{tikz}
%   \usetikzlibrary{shapes,arrows.meta,positioning,fit,backgrounds,calc}

\begin{figure*}[t]
\centering

% ── Wong colourblind-safe palette ──────────────────────────────────────
\definecolor{wongBlue}{HTML}{0072B2}
\definecolor{wongVermilion}{HTML}{D55E00}
\definecolor{wongSky}{HTML}{56B4E9}
\definecolor{wongGreen}{HTML}{009E73}
\definecolor{wongGrey}{HTML}{999999}
\definecolor{layerCore}{HTML}{0072B2}
\definecolor{layerDomain}{HTML}{009E73}
\definecolor{layerApp}{HTML}{D55E00}
\definecolor{layerInfra}{HTML}{999999}

% ── Global TikZ styles ────────────────────────────────────────────────
\tikzset{
  every node/.style={font=\sffamily\fontsize{7}{8.4}\selectfont},
  irbox/.style={
    draw=wongGrey, rounded corners=2pt, minimum width=22mm,
    minimum height=6mm, align=center, fill=#1!12, text=black,
    line width=0.4pt
  },
  irbox/.default=wongSky,
  routebox/.style={
    draw=#1, rounded corners=2pt, minimum width=20mm,
    minimum height=5.5mm, align=center, fill=#1!10, text=black,
    line width=0.4pt
  },
  sinkbox/.style={
    draw=#1, rounded corners=2pt, minimum width=20mm,
    minimum height=6mm, align=center, fill=#1!18, text=black,
    line width=0.5pt, font=\sffamily\fontsize{7}{8.4}\selectfont\bfseries
  },
  arrStyle/.style={-{Stealth[length=2pt,width=2pt]}, line width=0.4pt, wongGrey},
  panelLabel/.style={font=\sffamily\bfseries\fontsize{8}{9.6}\selectfont, anchor=north west},
}

% ======================================================================
%  Panel (a): 4-Level IR vertical flow
% ======================================================================
\begin{minipage}[t]{0.48\textwidth}
\centering
\begin{tikzpicture}[remember picture, baseline=(current bounding box.north)]
  % Unified bounding box: equalises panel height with panel (b) and
  % top-aligns both panels so their (a)/(b) labels sit on the same line.
  \useasboundingbox (-0.4,0.7) rectangle (4.4,-7.2);

  % Title
  \node[font=\sffamily\fontsize{7}{8.4}\selectfont\bfseries,
        anchor=south, text=wongBlue] at (1.3, 0.15) {4-Level IR};

  % panel label (above title, left-aligned)
  \node[panelLabel] at (-0.2, 0.45) {(a)};

  % IR levels
  \node[irbox] (circuit)  at (1.3, -0.45) {Circuit};
  \node[irbox] (gate)     at (1.3, -1.35) {Gate};
  \node[irbox] (pulse)    at (1.3, -2.25) {Pulse};
  \node[irbox] (waveform) at (1.3, -3.15) {Waveform};

  % Annotations (right side, small italic)
  \node[anchor=west, font=\sffamily\fontsize{6}{7.2}\selectfont\itshape,
        text=wongGrey] at ($(circuit.east)+(1.5mm,0)$) {circuit description};
  \node[anchor=west, font=\sffamily\fontsize{6}{7.2}\selectfont\itshape,
        text=wongGrey] at ($(gate.east)+(1.5mm,0)$) {native gate set};
  \node[anchor=west, font=\sffamily\fontsize{6}{7.2}\selectfont\itshape,
        text=wongGrey] at ($(pulse.east)+(1.5mm,0)$) {$(\tau,\,\Omega,\,\phi)$};
  \node[anchor=west, font=\sffamily\fontsize{6}{7.2}\selectfont\itshape,
        text=wongGrey] at ($(waveform.east)+(1.5mm,0)$) {IQ samples};

  % Arrows
  \draw[arrStyle] (circuit) -- (gate);
  \draw[arrStyle] (gate)    -- (pulse);
  \draw[arrStyle] (pulse)   -- (waveform);

\end{tikzpicture}
\end{minipage}%
\hfill
\begin{minipage}[t]{0.48\textwidth}
\centering
% ======================================================================
%  Panel (b): 3 Execution Routes
% ======================================================================
\begin{tikzpicture}[remember picture, baseline=(current bounding box.north)]
  % Unified bounding box matches panel (a) for equal height + top alignment.
  \useasboundingbox (-2.4,0.7) rectangle (4.7,-7.2);

  % panel label (above entry node)
  \node[panelLabel] at (-0.5, 0.45) {(b)};

  % ── Entry: OpenQASM ──
  \node[irbox=wongSky] (qasm) at (2.2, -0.45) {OpenQASM};

  % ── Shared pipeline (top) ──
  \node[irbox=wongSky] (circuit)   at (2.2, -1.25) {Circuit};
  \node[irbox=wongSky] (transpile) at (2.2, -2.05) {Transpile};
  \node[irbox=wongSky] (compile)   at (2.2, -2.85) {Compile};
  \node[irbox=wongSky] (ps)        at (2.2, -3.65) {PulseSequence};

  \draw[arrStyle] (qasm)      -- (circuit);
  \draw[arrStyle] (circuit)   -- (transpile);
  \draw[arrStyle] (transpile) -- (compile);
  \draw[arrStyle] (compile)   -- (ps);

  % ── Fan-out: WaveformGeneratorFactory ──
  \node[irbox=wongGrey, minimum width=28mm,
        font=\sffamily\fontsize{6}{7.2}\selectfont\bfseries]
        (factory) at (2.2, -4.45) {WaveformGeneratorFactory};
  \draw[arrStyle] (ps) -- (factory);

  % ── Column positions ──
  \def\colStd{1.1}
  \def\colVPPU{3.3}

  % ── Standard route (blue) ──
  \node[routebox=wongBlue] (std) at (\colStd, -5.35) {Standard};
  \node[font=\sffamily\fontsize{5.5}{6.6}\selectfont\itshape,
        text=wongBlue] at (\colStd, -5.85) {\texttt{float64}};

  \draw[arrStyle, wongBlue] (factory.south) to[out=-150, in=90] (std.north);

  % ── VPPU route (vermilion) ──
  \node[routebox=wongVermilion] (vppu) at (\colVPPU, -5.35) {VPPU};
  \node[font=\sffamily\fontsize{5.5}{6.6}\selectfont\itshape,
        text=wongVermilion] at (\colVPPU, -5.85) {ISA+HBL \texttt{int16}};

  \draw[arrStyle, wongVermilion] (factory.south) to[out=-30, in=90] (vppu.north);

  % ── Sink: both routes converge to the same Simulator ──
  \node[sinkbox=wongBlue, minimum width=30mm] (sim) at (2.2, -6.75)
       {Simulator (QuTiP / HEOM)};

  \draw[arrStyle, wongBlue]            (std.south)  to[out=-50, in=150] (sim.north west);
  \draw[arrStyle, wongVermilion]       (vppu.south) to[out=-130, in=30] (sim.north east);

  % ── Shared bracket label ──
  \node[font=\sffamily\fontsize{6}{7.2}\selectfont, text=wongGrey,
        anchor=east, align=right] at (-0.7, -2.05) {Shared\\front-end};
  \draw[wongGrey, line width=0.3pt, decorate,
        decoration={brace, amplitude=3pt, mirror}]
        (-0.5, -0.25) -- (-0.5, -3.95);

\end{tikzpicture}
\end{minipage}

\caption{%
  \textbf{Platform context for the paired waveform-realization comparison.}
  \textbf{(a)}~Four-level intermediate representation (IR) pipeline:
  a quantum circuit is progressively lowered through gate decomposition
  and pulse parametrisation to time-domain IQ waveform samples.
  \textbf{(b)}~Full compilation pipeline: an OpenQASM circuit is
  lowered through transpilation, compilation, and pulse parametrisation
  to a \texttt{PulseSequence}.
  A \texttt{WaveformGeneratorFactory} fans out to two execution routes:
  the \emph{Standard} route retains \texttt{float64} analytical envelopes;
  the \emph{VPPU} route applies ISA-level NCO mixing and DAC quantisation
  (\texttt{int16}).
  Both routes share the same front-end and solver interface,
  so the physics comparison isolates waveform realization rather than the rest
  of the software stack.
  Terminology: \emph{OpenQASM} is the textual circuit input;
  \emph{transpilation} rewrites the circuit into the native gate set;
  the \emph{ISA} (instruction-set architecture) is the low-level
  control-instruction layer the \vppu{} route targets;
  \emph{QubiC} is the FPGA-based qubit-control system whose waveform
  generation the \vppu{} route emulates
  (\emph{FPGA}: field-programmable gate array).
  A full glossary is given in
  Supplementary Information~\S\ref{supp:si:terminology}.%
}
\label{fig:architecture}
\end{figure*}
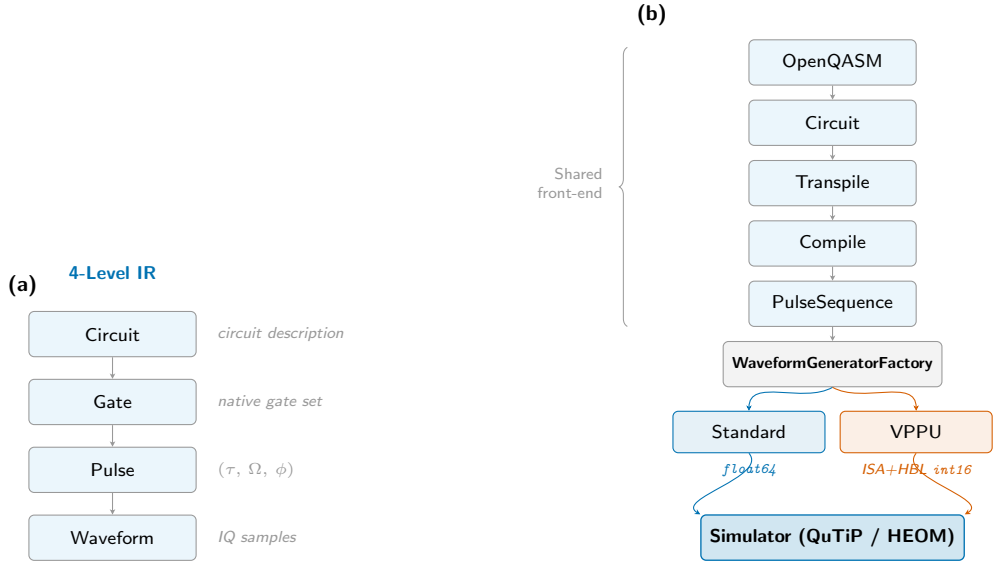

The central questions are physical: when exact long-memory bath dynamics
interact with hardware-faithful compiled control, what new phenomena emerge
in multi-pulse observables---and which of those phenomena, if any, carry a
distinguishable signature of the specific waveform realization?
Figure~\ref{fig:architecture} summarizes the controlled framework used to
address both questions.
All manuscript-level evidence below concerns single-qubit rotating-frame
calculations; within that scope, the platform keeps the circuit, pulse
parametrisation, bath model, and solver interface fixed.
Sections~\ref{sec:results:burkard}--\ref{sec:results:ycpmg} report the
non-Markovian multi-pulse phenomena that this controlled setting reveals
across one decade in coupling strength~$\eta$.
Section~\ref{sec:results:paired} then tests whether those phenomena carry a
waveform-dependent signature, restricting to the Tier-0 coupling window
(\rev{\mbox{$\eta = \SI{7.85\times10^{-7}}{\giga\hertz}$}}, the weakest of three coupling strengths
and the literature-calibrated anchor; Methods~\S\ref{sec:meth:bath}) where
decoherence is moderate enough to sustain stable power-law fits.
Within that window the platform varies only the waveform realization:
the Standard route produces analytic rectangular envelopes, the \vppu{} route
generates bit-exact waveforms matching QubiC FPGA control
electronics~\cite{xu2021qubic,fruitwala2024}, and the same
\texttt{EngineAdapter} passes either realization to Lindblad \texttt{mesolve}
or \heom{}.

\begin{figure}[t]
\centering
\includegraphics[width=\textwidth]{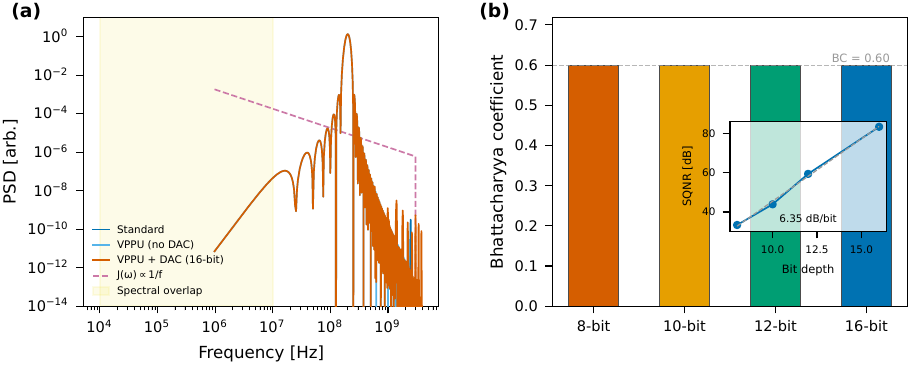}
\caption{%
  \textbf{Waveform spectral structure and $1/f$ bath overlap.}
  \textbf{a},~Power spectral density of Standard (rectangular), \vppu{}-noDac,
  and \vppu{}+\dac{} (16/12/10/8-bit) waveforms, overlaid with the $1/f$
  bath spectral density $J(\omega) = \eta/\omega$.
  The shaded region marks the spectral overlap zone
  [\SI{10}{\kilo\hertz}--\SI{10}{\mega\hertz}].
  \textbf{b},~Bhattacharyya coefficient (BC $\approx 0.60$)
  is invariant across 8--16\,bit \dac{} resolution, confirming that the
  spectral coupling channel exists independently of \dac{} bit depth.
  Inset: time-domain $\mathrm{SQNR}_t$ vs.\ bit depth
  ($\SI{6.35}{\decibel/bit}$, near theoretical $\SI{6.02}{\decibel/bit}$).%
}\label{fig:spectral}
\end{figure}

Although the \vppu{} route introduces a broadband \dac{} quantization
noise floor whose spectral profile overlaps with the $1/f$ bath
(Bhattacharyya coefficient BC~$\approx 0.60$, invariant across
8--16\,bit; Fig.~\ref{fig:spectral}), the ${\sim}10^8$ power mismatch renders this
shape overlap operationally irrelevant for the dynamics studied here
(Supplementary Information, \rev{SI Sec.}~\ref{supp:si:dac_spectral}).
We therefore turn to the dynamical validation and sequence-level results.

\paragraph{Notation guide.}
Bath coupling tiers are numbered in ascending strength:
Tier-0 (weakest, \rev{\mbox{$\eta = \SI{7.85\times10^{-7}}{\giga\hertz}$}}),
Tier-1 (intermediate, \rev{\mbox{$1.8\times10^{-6}$\,GHz}}),
Tier-2 (strongest, \rev{\mbox{$5.0\times10^{-6}$\,GHz}});
main-text compiled-pulse claims are restricted to Tier-0.
Decoherence is analysed in two complementary spaces:
$\varepsilon$-space (infidelity) and
$\chi$-space ($\chi = -\ln(1 - \varepsilon)$, the decoherence
function); the latter resolves scaling differences that
$\varepsilon$ saturation compresses.
Remaining notation ($n_\pi$, even-$n$ convention, $K$ vs.\
$N_\mathrm{d}$) is introduced at first use
(\S\ref{sec:results:cpmg} and Methods).
Table~\ref{tab:notation} collects the key symbols for reference.

\begin{table}[t]
\caption{%
  Key symbols and conventions used in the manuscript.%
}\label{tab:notation}
\footnotesize
\begin{tabular*}{\textwidth}{@{\extracolsep\fill}lll@{}}
\toprule
Symbol & Meaning & Where defined \\
\midrule
$\varepsilon$, $\chi$ & Infidelity and decoherence function
  ($\chi = -\ln(1-\varepsilon)$) & \S\ref{sec:results} \\
$\beta$, $\gamma$ & Scaling exponents in $\varepsilon$- and $\chi$-space
  ($\varepsilon \propto n^{\beta}$, $\chi \propto n^{\gamma}$) & \S\ref{sec:results:cpmg} \\
$n_\pi$, $n_\mathrm{cycle}$ & Number of $\pi$~pulses and CPMG cycles
  ($n_\mathrm{cycle} = n_\pi/2$) & \S\ref{sec:results:cpmg} \\
even-$n$ / all-$n$ & Fitting convention: even-$n$ only vs.\ all pulse counts
  & \S\ref{sec:results:cpmg} \\
Tier-0 / 1 / 2 & $\eta = 7.85/18.0/50.0\times10^{-7}$\,GHz (ascending)
  & Methods~\S\ref{sec:meth:bath} \\
$K$ & Espira-II correlation-function fit order
  & Methods~\S\ref{sec:meth:heom} \\
$D$, $N_r$ & HEOM hierarchy depth and retained bath-mode pairs
  & Methods~\S\ref{sec:meth:heom} \\
$N_\mathrm{d}$ & Number of data points in power-law fits
  & \S\ref{sec:results:cpmg} \\
Standard / \vppu{} & Waveform realization routes
  & Methods~\S\ref{sec:meth:vppu} \\
X-/Y-/XY-4 & Dynamical decoupling schemes (refocusing axis)
  & \S\ref{sec:results:cpmg} \\
\bottomrule
\end{tabular*}
\end{table}

%% ─────────────────────────────────────────────
\subsection{Burkard analytical validation and the finite-time Gaussian ratio}\label{sec:results:burkard}

\begin{figure}[t]
\centering
\includegraphics[width=\textwidth]{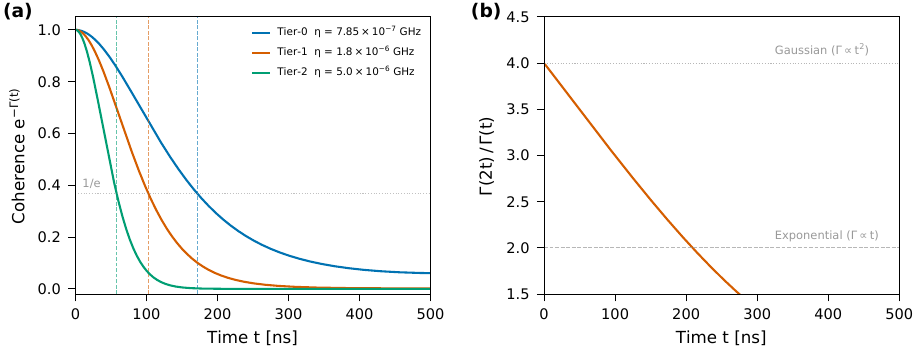}
\caption{%
  \textbf{Burkard decoherence function and finite-time Gaussian ratio for the $1/f$
  bath.}
  \textbf{a},~Free induction decay $e^{-\Gamma(t)}$ computed from Burkard's
  exact analytical formula (Eq.~\ref{eq:burkard}) for three coupling
  tiers spanning one decade in $\eta$.
  Dashed lines mark $T_2^*$ (171.7, 102.8, \SI{58.0}{\nano\second}).
  Rotating-frame \heom{} validation confirms $<1.5$\% deviation
  from these curves across all tiers.
  \textbf{b},~The finite-time ratio diagnostic
  $\mathrm{GC}_{\mathrm{ratio}}(t) = \Gamma(2t)/\Gamma(t)$
  for Tier-1 starts near $\approx$4 (Gaussian-like) at short times and
  decreases monotonically toward $\approx$2 (exponential) at
  $t \gg T_2^*$, confirming that the $1/f$ bath produces intermediate,
  non-trivially non-Gaussian decoherence at the operational timescale---a
  key reason the filter function perturbative expansion
  (Fig.~\ref{fig:ff_breakdown}) breaks down.%
}\label{fig:burkard}
\end{figure}

Burkard's exact decoherence function~\cite{burkard2009} provides the
analytical anchor for our \heom{} implementation.
For the $1/f$ bath spectral density $J(\omega) = \eta/\omega$ with cutoffs
$\omega_{\mathrm{lc}}$ and $\omega_{\mathrm{hc}}$, the decoherence integral
$\Gamma(t)$ (Eq.~\ref{eq:burkard}) yields $T_2^*$ values spanning
58--\SI{172}{\nano\second} across one decade in coupling strength $\eta$
(Fig.~\ref{fig:burkard}a).
Rotating-frame \heom{} simulation (Supplementary Information,
\rev{SI Sec.}~\ref{supp:si:burkard}) reproduces these values to within 1.5\%,
validating both the espira-II Prony bath decomposition and the \heom{}
hierarchy truncation at the parameter sets used throughout this work.

To avoid ambiguity with the short-time curvature diagnostic reported in the
Supplementary Information, we refer here to the finite-time ratio diagnostic
$\mathrm{GC}_{\mathrm{ratio}}(t) = \Gamma(2t)/\Gamma(t)$
(Fig.~\ref{fig:burkard}b).
A purely Gaussian decay gives
$\mathrm{GC}_{\mathrm{ratio}} = 4$, whereas a purely exponential decay gives
$\mathrm{GC}_{\mathrm{ratio}} = 2$.
For the $1/f$ bath, $\mathrm{GC}_{\mathrm{ratio}}(t)$ starts near $\sim$4 at
short times and decreases to $\sim$3 at $t \sim T_2^*$, showing that the
operational-time dynamics lie between exponential and Gaussian limits.
Together with the Supplementary Information curvature diagnostic, this places the present simulations
in an intermediate non-Markovian regime where the first-order filter-function
baseline of \S\ref{sec:results:ff} becomes practically uninformative.

%% ─────────────────────────────────────────────
\subsection{Calibrating the perturbative baseline: filter-function estimates versus \texorpdfstring{\heom{}}{HEOM} ground truth}\label{sec:results:ff}

\begin{figure}[t]
\centering
\includegraphics[width=\textwidth]{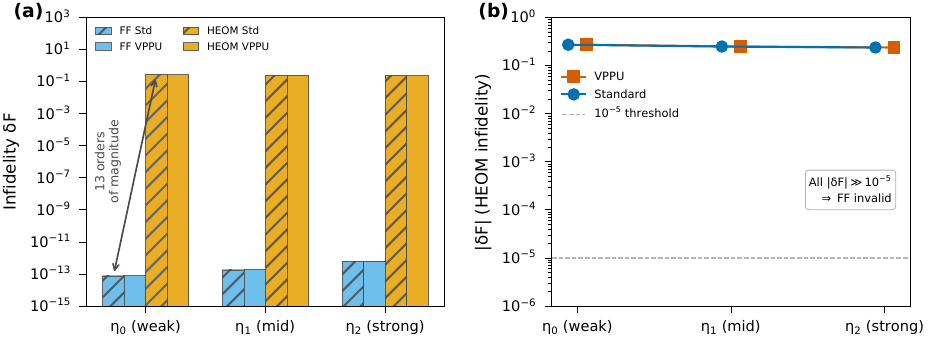}
\caption{%
  \textbf{First-order FF becomes quantitatively uninformative in the present
  $1/f$ window.}
  \textbf{a},~FF predicts infidelity $\sim 10^{-13}$ across all tiers
  and waveform realizations, while \heom{} measures $\sim$0.27---a
  $>$12-OOM gap at the Ramsey reference point.
  \textbf{b},~\heom{} infidelity \emph{decreases} with coupling
  (0.270$\to$0.248$\to$0.236, Tier-0/1/2) as stronger coupling drives
  more Gaussian-like decay; FF remains at $\sim 10^{-13}$.%
}\label{fig:ff_breakdown}
\end{figure}

The filter function formalism predicts gate infidelity via the first-order
perturbation integral
\begin{equation}\label{eq:ff_infidelity}
  1 - F_{\mathrm{FF}}
  = \frac{1}{\pi^{2}} \int_{0}^{\infty}
    S(\omega)\,|F_{\phi}(\omega)|^{2}\,d\omega\,,
\end{equation}
where all quantities are defined in Methods~(\S\ref{sec:meth:ff})
and Supplementary Information~(\rev{SI Sec.}~\ref{supp:si:ff_method}).
Both calculations share the same Hamiltonian, bath density, and fidelity
metric; only the dynamics solver differs.
At the Ramsey $T_{2}^{*}/2$ evaluation point, the perturbative expansion
yields $\sim 10^{-13}$, \rev{over 12 orders of magnitude below} the
\heom{} ground truth ($\sim$0.27; Fig.~\ref{fig:ff_breakdown}a),
consistent with the perturbative boundary identified for slow correlated
baths~\cite{cerfontaine2021,hangleiter2021,barnes2016}.
The collapse is uniform: FF varies by $<$2\% across all six
waveform/\dac{} configurations and three coupling tiers.
Counter-intuitively, the \heom{} infidelity itself \emph{decreases} with
coupling (0.270\,$\to$\,0.236, Tier-0\,$\to$\,Tier-2;
Fig.~\ref{fig:ff_breakdown}b).
Each tier is evaluated at its respective $T_{2}^{*}/2$, where stronger
coupling drives increasingly Gaussian-like coherence decay; because a
Gaussian envelope retains more coherence than an exponential one at
$t=T_{2}^{*}/2$ ($e^{-1/4}$ vs.\ $e^{-1/2}$), the measured infidelity
converges toward the Gaussian limit $1-e^{-1/4}\approx 0.221$.
\rev{This breakdown is the quantitatively calibrated form of an expected
perturbative limitation, not a new effect. The operational criterion for
the non-perturbative regime follows from the cumulant (Magnus) expansion of
the decoherence function \mbox{$\chi = -\ln(1-\varepsilon)$}: writing
\mbox{$\chi = \chi^{(1)} + \chi^{(2)} + \cdots$}, the leading term
\mbox{$\chi^{(1)}$} is precisely the first-order filter-function integral of
Eq.}~\eqref{eq:ff_infidelity}\rev{. Truncating at first order is controlled
by the relative size of the neglected term, \mbox{$\chi^{(2)}/\chi^{(1)}$};
for the present Gaussian \mbox{$1/f$} bath this ratio grows with
\mbox{$\chi^{(1)}$} itself, so once the accumulated phase reaches
\mbox{$\chi \sim O(1)$} at \mbox{$T_2^*/2$} (here
\mbox{$\chi \approx 0.27$--$0.32$}) the second-order contribution is no
longer negligible and the first-order estimate ceases to be quantitatively
reliable. All three tiers satisfy this condition; the compiled-pulse
restriction to Tier-0 reflects fit quality \mbox{$(R^2 > 0.99)$}, not a
non-perturbative boundary. The resulting \mbox{$>$12-order} gap between
first-order theory and exact \heom{} dynamics is thus consistent with the
perturbative boundary already anticipated for slow correlated baths; what
is new here is its calibration against exact dynamics, not the existence of
the breakdown itself.}

With the perturbative baseline calibrated, the next two sections present
\rev{the two physical results}: bath-memory transient structure in even-$n$
\cpmg{} scaling (\S\ref{sec:results:cpmg}), and axis-dependent
scaling-law breakdown under Y-\cpmg{}
(\S\ref{sec:results:ycpmg}).

%% ─────────────────────────────────────────────
\subsection{Even-$n$ CPMG scaling and Floquet transient structure}\label{sec:results:cpmg}

\begin{figure}[t]
\centering
\includegraphics[width=\textwidth]{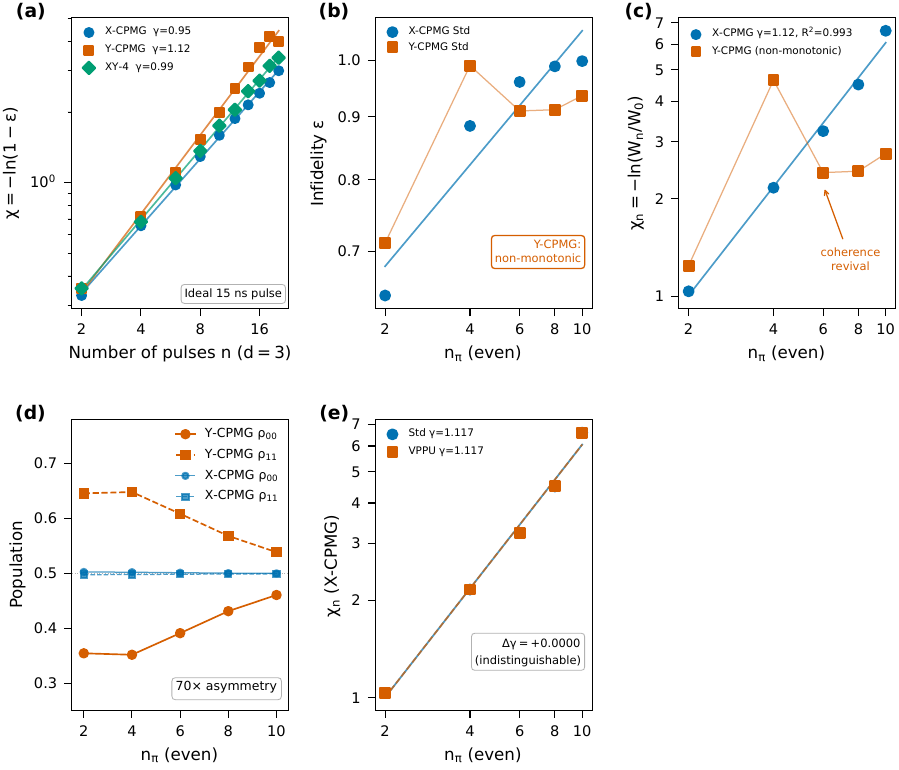}
\caption{%
  \textbf{Even-$n$ CPMG scaling reveals non-perturbative regime
  characterisation and axis-dependent breakdown.}
  \textbf{a},~Ideal-pulse \heom{} reference in $\chi$-space
  ($\chi = -\ln(1-\varepsilon)$):
  \SI{15}{\nano\second} rectangular $\pi$ pulses ($d = 3$,
  even $n = 2$--$20$; uncontrolled comparison---pulse duration,
  $\tau$, and fitting range all differ from compiled data;
  see text for caveats); X-\cpmg{}/XY-4
  $\gamma_\mathrm{ideal} \approx 0.95$--$1.00$,
  Y-\cpmg{} $\gamma_\mathrm{ideal} = 1.12$ (all $R^2 > 0.99$).
  \textbf{b},~Raw infidelity $\varepsilon(n) = 1 - W_n/W_0$
  for even-$n$ compiled pulses
  ($n_\pi = 2$--$10$; \SI{80}{\nano\second}, $d = 3$);
  Y-\cpmg{} (red) is non-monotonic.
  \textbf{c},~Decoherence function
  $\chi_n = -\ln(W_n/W_0)$ for the same data.
  X-\cpmg{} scales monotonically
  ($\gamma_\mathrm{compiled} \approx 1.12$, $R^2 > 0.99$);
  Y-\cpmg{} $\chi$ peaks at $n_\pi = 4$ then
  \emph{decreases}---a coherence revival inconsistent
  with any power-law model.
  \textbf{d},~Y-\cpmg{} population dynamics (Standard)
  showing persistent $\rho_{11} > \rho_{00}$ imbalance
  driven by $d = 3$ anharmonicity (gap relaxing from
  $0.29$ to $0.08$ over the sampled window), versus balanced
  X-\cpmg{} ($|\rho_{00} - \rho_{11}| < 0.02$).
  \textbf{e},~Standard vs.\ \vppu{} paired comparison:
  X-\cpmg{} $\chi$ values are indistinguishable
  ($\Delta\gamma \approx 0$, offset $\lesssim 0.002$),
  confirming waveform insensitivity in the Tier-0 window.%
}\label{fig:cpmg}
\end{figure}

With the perturbative baseline established as \rev{no longer quantitatively
predictive} in this regime (\S\ref{sec:results:ff}), we turn to the multi-pulse
dynamics accessible through \heom{}, extending the idealized-pulse \cpmg{}
analyses of Refs.~\cite{chen2025heom,nakamura2025cpmg} to realistic compiled
waveforms.
All results below use rotating-frame single-qubit \heom{} under the
conditions in Methods and Supplementary Information.
Pulse counts are reported as $n_\pi$ ($\pi$-pulse count), with
$n_\mathrm{cycle} = n_\pi / 2$ giving the number of complete \cpmg{}
cycles; the ``even-$n$'' convention retains only even $n_\pi$
($= 2, 4, 6, 8, 10$) to suppress a systematic $d = 3$
rotation-parity artefact (Methods, Section~\ref{sec:meth:fidelity}).
Here $W_n$ denotes the Uhlmann fidelity $F$
(Eq.~\ref{eq:uhlmann} in Methods) of the system state at the
$n$-th echo time against the initial state, with $W_0$ the
zero-pulse reference.
Because the infidelity $\varepsilon = 1 - W_n/W_0$ saturates near unity
for compiled \SI{80}{\nano\second} pulses, all scaling analysis is reported
in the cumulative decoherence function
$\chi_n = -\ln(W_n/W_0)$, the multi-pulse analogue of the echo-decay
exponent in spin-echo spectroscopy.
This representation linearises the power-law regime and resolves
scaling differences that $\varepsilon$-saturation compresses
($\varepsilon$-space companion exponents in
Supplementary Table~\ref{supp:tab:si:eps_exponents} and
\rev{SI Sec.}~\ref{supp:si:fitting_robustness}).

\paragraph{Ideal-pulse reference.}
A \SI{15}{\nano\second} ideal-pulse benchmark
(even $n = 2, 4, \ldots, 20$; $d = 3$; Fig.~\ref{fig:cpmg}a) yields
monotonic scaling with
$\gamma_{\mathrm{ideal}} = 0.950/1.117/0.993$ for
X-\cpmg{}/Y-\cpmg{}/XY-4 ($R^2 > 0.99$), confirming
approach to the perturbative $\gamma \approx 1$ prediction
(Section~\ref{sec:disc:regime}).
The exponents already reveal $d = 3$ axis asymmetry
($\gamma_Y = 1.117$ vs.\ $\gamma_X = 0.950$), which sharpens the
significance of the compiled-pulse result below: under
\SI{80}{\nano\second} pulses, the monotonic Y-\cpmg{} scaling
collapses into non-monotonic $\chi(n)$ ($R^2 = 0.213$), crossing
a threshold from exponent anisotropy to scaling-law breakdown.
Because the ideal and compiled benchmarks differ in pulse duration,
free-evolution spacing, and fitting range simultaneously, this
dataset is a contextual reference, not a controlled comparison
(Supplementary Information~\rev{SI Sec.}~\ref{supp:si:deconfound}).

\paragraph{Even-$n$ X-\cpmg{}: Floquet convergence and finite-$n$ transient structure.}
Under \SI{80}{\nano\second} compiled pulses in the $d = 3$ transmon,
even-$n$ X-\cpmg{} yields decoherence-function exponents
$\gamma_\mathrm{Std} = 1.117$%
\footnote{Three significant figures are retained when reporting alongside
statistical measures ($R^2$, confidence intervals); in narrative discussion
we write $\gamma \approx 1.12$.  With $N_\mathrm{d} = 5$ even-$n$ data points
($\nu = 3$ residual d.f.), the 95\% BCa CI has 31\% relative width;
small-sample coverage properties are discussed in
\rev{SI Sec.}~\ref{supp:si:bootstrap}.}
($R^2 = 0.993$) and
$\gamma_\vppu{} = 1.117$ ($R^2 = 0.993$), with strict monotonicity
in $\chi(n)$
(Fig.~\ref{fig:cpmg}c; Table~\ref{tab:cpmg_gamma}).
For fixed-$\tau$ CPMG under $1/f$ noise, perturbative filter-function
theory predicts $\gamma \approx 1$: the decoherence function scales as
$\chi \propto n^{(1+\alpha)-\alpha} = n^1$ in the large-$n$ limit
(Cywi\'{n}ski~et~al.~Eq.\,25~\cite{cywinski2008}; confirmed by our
numerical FF integral, Supplementary Information~\rev{SI Sec.}~\ref{supp:si:ff_integral}).
The measured $\gamma = 1.117$ is within ${\sim}12\%$ of this prediction.
Because each cycle applies the same propagator~$\Lambda$, only its
dominant eigenvalue survives as $n \to \infty$, forcing $\chi(n)$
linear in~$n$---a structural property that Floquet's theorem
guarantees for any periodic linear system.
This proximity to $\gamma = 1$ is therefore expected;
the physical content lies in the finite-$n$ transient path toward this
asymptote.
Under matched Tier-0 production conditions ($d = 3$, RWA), Lindblad
\texttt{mesolve} yields $\gamma_\mathrm{Lindblad} = 1.00$
(saturation-aware model, $R^2 > 0.999$), confirming that the Markovian
solver has already reached the Floquet asymptotic value within
$n \leq 10$.
The \heom{} excess (point estimate $\gamma = 1.12$ vs.\ $1.00$;
95\% BCa CI $[1.04, 1.39]$)
is consistent with a finite-$n$ transient correction controlled by
sub-leading Floquet eigenvalues that reflect non-Markovian bath memory
timescales (Discussion, Section~\ref{sec:disc:regime}).
Indirect evidence disfavouring a qutrit origin for this excess
comes from three observations:
(i)~the balanced X-\cpmg{} populations at $d = 3$
($|\rho_{00} - \rho_{11}| < 0.02$; Section~\ref{sec:results:ycpmg})
indicate negligible $\ket{2}$ participation in X-axis dynamics;
(ii)~the $d = 2$ \heom{} control
(Section~\ref{sec:results:ycpmg};
Supplementary Information~\rev{SI Sec.}~\ref{supp:si:platform_model}) confirms that
$d = 3$-specific effects concentrate on the Y axis;
and (iii)~an independent $d = 2$ \heom{} study under the same $1/f$
bath model but a different coupling geometry (quasi-static detuning
rather than diagonal pure
dephasing)~\cite{chen2025heom} reports bath-memory-driven
error accumulation in X-\cpmg{} (linear growth with~$n$) with fully
balanced populations.
Because evidence~(iii) originates from a non-diagonal coupling operator
distinct from the diagonal $\hat{Q}$ employed here, it constitutes a
suggestive analogy rather than a direct demonstration: it establishes
that non-Markovian dynamics \emph{can} generate pulse-number-dependent
X-axis errors without qutrit participation, but the specific pathway
may differ under diagonal pure dephasing.
Observations (i) and~(ii) provide direct evidence within the present
model, while (iii) offers a complementary but geometry-dependent
indication; a direct $d = 2$ measurement of~$\gamma$ under diagonal
coupling would provide definitive confirmation.
The nominal 95\% BCa CI lower bound excludes unity ($\gamma > 1$),
suggesting a non-zero transient excess; however, with only $N_\mathrm{d} = 5$
data points ($\nu = 3$ residual d.f.), the actual coverage of the BCa
interval may deviate from its nominal level
(\rev{SI Sec.}~\ref{supp:si:bootstrap}), so this exclusion should be regarded as
suggestive rather than definitive.
The point estimate $\gamma \approx 1.12$ should be read within its
31\%-wide interval, which places the true excess between ${\sim}4\%$
and ${\sim}39\%$.
A linear regression of the X-\cpmg{} $\chi$ data versus cycle number
gives $-\!\ln|\lambda_1| \approx 1.35$ ($R^2 = 0.978$), with the
per-cycle mean $\chi/n_{\mathrm{cycle}}$ ranging from 1.03
($n_{\mathrm{cycle}} = 1$) to 1.32 ($n_{\mathrm{cycle}} = 5$)---each
increment is $3$--$5\times$ the Ramsey reference
$\chi_{\mathrm{Ramsey}} \approx 0.27$--$0.32$, placing each cycle
individually in the non-perturbative regime.
An independent propagator-eigenvalue extraction
(Supplementary Information~\rev{SI Sec.}~\ref{supp:si:floquet_eigenvalues},
Table~\ref{supp:tab:si:floquet_eigenvalues}) yields
$|\lambda_1| = 0.259$
($-\!\ln|\lambda_1| = 1.35$, matching the regression within 0.2\%)
and spectral gap $|\lambda_2/\lambda_1| = 0.68$.
A two-eigenvalue Floquet model using these values predicts
$\gamma_{\mathrm{eff}}(5) = 1.12$---reproducing the fitted
exponent---and $\gamma_{\mathrm{eff}} \to 1.00$ at
$n_{\mathrm{cycle}} \gtrsim 15$; the corresponding Lindblad spectral
gap ($|\lambda_2/\lambda_1| = 0.41$) accounts for the faster
convergence to $\gamma_\mathrm{Lindblad} = 1.00$ at $n \leq 10$.
XY-4 confirms the same picture with $\gamma_\mathrm{Std} = 0.950$
and $\gamma_\vppu{} = 0.951$ ($R^2 > 0.98$).
Beyond the scaling exponent, the solver dependence observed in the
\cpmg{} regime extends to the underlying Ramsey decoherence itself.
A Lindblad control experiment
(Supplementary Information~\rev{SI Sec.}~\ref{supp:si:pipeline}) reveals a
solver-dependent sign reversal in the Ramsey $T_2^*$ shift: under
Lindblad dynamics the full pipeline \emph{lengthens} $T_2^*$ by
$+1.1\%/+2.1\%/+6.3\%$ across the three coupling tiers, whereas under
\heom{} it \emph{shortens} $T_2^*$ by $-12.9\%/-18.9\%/-23.7\%$.
This qualitative divergence is consistent with solver-dependent
dynamical differences beyond the quantitative $\gamma$ values
reported above; however, because the specific mechanism underlying
the sign inversion remains open
(Section~\ref{sec:disc:solver_comparison}), this observation should
be regarded as suggestive rather than independently confirmatory.

\subsection{Axis-dependent scaling-law breakdown under Y-CPMG}\label{sec:results:ycpmg}

\rev{In contrast to} X-\cpmg{}, even-$n$ Y-\cpmg{} exhibits a
qualitative breakdown of the power-law scaling paradigm.
The decoherence function $\chi(n)$ peaks at $n_\pi = 4$
($\chi_\mathrm{Std} = 4.64$, $\chi_\vppu{} = 4.68$) and then
\emph{decreases}---a partial coherence revival that is
fundamentally inconsistent with any power-law
$\chi \propto n^\gamma$ model (Fig.~\ref{fig:cpmg}b,c).
Formally, a power-law fit returns $\gamma_\mathrm{Std} = 0.342$
($R^2 = 0.213$) and $\gamma_\vppu{} = 0.342$ ($R^2 = 0.210$),
confirming that this functional form is inapplicable.
This non-monotonicity is robust to the choice of $n$-counting
convention: whether one counts pulses or cycles, a sequence
that increases then decreases in decoherence remains non-monotonic.

The physical mechanism---rooted in $d = 3$ anharmonicity
($\alpha \approx \SI{-293}{\mega\hertz}$) and the resulting
$\hat{Y}_q$ off-diagonal coupling to the
$\ket{1}\!\leftrightarrow\!\ket{2}$ transition---is analysed
quantitatively in Section~\ref{sec:disc:qutrit} and
Supplementary Information~\rev{SI Sec.}~\ref{supp:si:qutrit_model}.

The population-level signature is a persistent imbalance:
under the even-$n$ convention, $\rho_{11} > \rho_{00}$ at every
sampled cycle number (Fig.~\ref{fig:cpmg}d), with the gap relaxing
from $|\rho_{00} - \rho_{11}| = 0.29$ at $n_\pi = 2$ to $0.08$
at $n_\pi = 10$.
The mean asymmetry
$\langle|\rho_{00} - \rho_{11}|\rangle = 0.204$ under Y refocusing
contrasts sharply with X-\cpmg{}, where
$\langle|\rho_{00} - \rho_{11}|\rangle < 0.01$ and populations remain
balanced to within 1\% for all~$n$.
(The underlying per-pulse mechanism is a population inversion at
each Y $\pi$~pulse---the reason for adopting the even-$n$
convention; see Methods, Section~\ref{sec:meth:fidelity}.)
Leakage to~$\ket{2}$ is nearly identical for both schemes
($\rho_{22,\max} \approx 0.0014$), ruling out differential leakage
as the mechanism.

This axis-dependent scaling breakdown constitutes a testable
prediction: it should appear on any transmon with
$|\alpha|$ comparable to the value studied here
($\SI{-293}{\mega\hertz}$), finite-duration $\pi$~pulses
with $\Omega/|\alpha| \lesssim 0.3$
(equivalently $t_\pi \gtrsim 1/|\alpha|$),
and non-Markovian bath dynamics
(Discussion; Supplementary Information~\rev{SI Sec.}~\ref{supp:si:qutrit_model}).
While finite-pulse X--Y error-scaling differences exist even at
$d = 2$, their origin is fundamentally different.
Chen~et~al.\ report linear~(X) versus quadratic~(Y)
echo-error accumulation at the $10^{-4}$ scale from quasi-static
detuning coupling~\cite{chen2025heom}; because that coupling contains
a $\sigma_x$~component, the noise operator directly rotates the Bloch
vector in the $xz$~plane and thereby distinguishes X from Y refocusing
even in a two-level system.
Under the diagonal pure-dephasing coupling studied here,
$\hat{Q} = \mathrm{diag}(0,1,2)$ acts only on off-diagonal
density-matrix elements and does not break X--Y symmetry at $d = 2$;
the axis-dependent breakdown therefore arises exclusively from the
$d = 3$ level structure.
Accordingly, the qualitative
features reported here are exclusive to $d \geq 3$:
non-monotonic $\chi(n)$, partial coherence revival,
\rev{the pronounced population asymmetry quantified above}, and complete scaling-law
breakdown ($R^2 = 0.213$).
This distinction is confirmed by two control experiments.
First, a $d = 2$ simulation under identical \heom{} dynamics
(Supplementary Information~\rev{SI Sec.}~\ref{supp:si:platform_model}) removes the
$\ket{2}$ level: the non-monotonicity and population imbalance
vanish and both axes produce identical power-law scaling,
establishing that $d \geq 3$ anharmonicity is necessary.
Second, a $d = 3$ Lindblad simulation under matched production
conditions yields monotonic Y-\cpmg{} with
$\gamma_\mathrm{Lindblad} = 0.999$ ($R^2 > 0.999$;
Table~\ref{supp:tab:si:lindblad_tier0}), confirming that non-Markovian
bath memory is independently necessary for the breakdown.
This conclusion is not confounded by the ${\sim}2\times$ larger
per-cycle dephasing rate under Lindblad dynamics: stronger dephasing
suppresses the absolute magnitude of the population redistribution
but cannot, by itself, convert a non-monotonic modulation into a
monotonic power law; the Lindblad monotonicity instead reflects the
loss of inter-pulse bath correlations that would coherently amplify
the per-pulse redistribution across successive cycles.
\rev{The combination shows that both ingredients are required}: neither
$d = 3$ anharmonicity alone nor non-Markovian dynamics alone
produces the observed non-monotonic behaviour.
A quantitative model predicting the threshold
pulse duration $t_\pi^* \sim 1/|\alpha|$ is presented in the
Discussion and Supplementary Information~\rev{SI Sec.}~\ref{supp:si:qutrit_model}.
Importantly, verification requires only a qualitative yes/no
observation of non-monotonic Y-\cpmg{} behaviour, not a precision
measurement of scaling exponents.
A back-of-envelope $T_1$ analysis (Discussion and
Supplementary Information~\rev{SI Sec.}~\ref{supp:si:platform_model}) confirms that these signatures survive
energy relaxation with ${\leq}\SI{8}{\%}$ suppression for
typical transmon $T_1$ values.

\rev{Two further controls confirm that this breakdown is intrinsic, not an
artefact of the pulse envelope or of the particular anharmonicity.}
\rev{Replacing the rectangular \mbox{$\pi$}~pulses with area-matched Gaussian
envelopes leaves the non-monotonicity qualitatively unchanged---the
\mbox{$\chi(n)$} peak at \mbox{$n_\pi = 4$} and the population asymmetry both
persist (\mbox{$\langle|\rho_{00}-\rho_{11}|\rangle = 0.206$} versus
\mbox{$0.204$})---while }\emph{reducing}\rev{ the \mbox{$\ket{2}$} leakage,
which rules out edge-induced leakage. Varying the anharmonicity at fixed pulse
duration shows the asymmetry to be stable for
\mbox{$|\alpha| \gtrsim \SI{75}{\mega\hertz}$} and to collapse only as
\mbox{$|\alpha| \to 0$}, where the qutrit becomes near-degenerate and leakage
dominates.}
\rev{The effect is therefore largest at realistic transmon anharmonicities
and vanishes in both the strong-anharmonicity and near-degenerate limits}
(Supplementary Information~\rev{SI Sec.}~\ref{supp:si:ycpmg_robustness},
Fig.~\ref{supp:fig:si:ycpmg_robustness}).

\subsection{Paired waveform comparison and DAC invariance}\label{sec:results:paired}

\paragraph{Structural prediction.}
Under the rotating-frame pure-dephasing model, the diagonal noise
operator $\hat{Q} = \mathrm{diag}(0,1,2)$ does not couple directly to
drive-waveform differences, so \rev{the scaling exponents must coincide} as a
structural consequence of the coupling geometry.
Physically, pure dephasing acts only on off-diagonal density-matrix
elements, with rates set by the eigenvalue differences
$(q_a - q_b)$ of~$\hat{Q}$---quantities intrinsic to the noise operator,
not to the drive.
Provided two pulse waveforms realise the same target unitary, the
population distribution across $\hat{Q}$ eigenstates at the end of each
pulse is identical, so the dephasing accumulated during subsequent
free-evolution windows is waveform-independent.
This consistency is non-trivial, since drive transients can in principle
redistribute population among levels with distinct $\hat{Q}$
eigenvalues---a mechanism not captured by the leading-order
coupling-operator argument.
Each pulse transiently populates $\ket{2}$ by at most
$\rho_{22,\max} \approx 0.0014$ (Section~\ref{sec:results:cpmg}),
confined to the pulse-on fraction $f \approx 0.4$ of each cycle; the
resulting per-cycle $\chi$ contribution from each waveform individually
is of order $\rho_{22,\max}\,(\Delta q)^{2} \sim 10^{-3}$.
Because both waveform realisations achieve the same target unitary, they
share nearly identical intra-pulse leakage profiles; the
\emph{differential} transient contribution per cycle is therefore a
fraction of this bound, and the accumulated differential over ten cycles
remains consistent with the observed
$|\Delta\chi| \leq 2\times 10^{-3}$.

\paragraph{Quantitative verification.}
The quantitative finding is consistent with the structural prediction:
not only scaling exponents but also absolute $\chi$-offsets are
indistinguishable, indicating that compiled control introduces no
detectable additive decoherence overhead during finite-duration pulses.
Standard and \vppu{} waveform
realizations produce statistically and practically indistinguishable scaling
exponents in the Tier-0 window.
For X-\cpmg{}, both realizations yield
$\gamma = 1.117$ (95\% BCa CI: $[1.04, 1.39]$, $R^2 = 0.993$) with
$\Delta\gamma = +1 \times 10^{-6}$;
for XY-4, $\gamma_\mathrm{Std} = 0.950$ and $\gamma_\vppu{} = 0.951$
(95\% CI: $[0.28, 1.03]$) with $\Delta\gamma = +5 \times 10^{-4}$.
Both effect sizes are negligible ($|\Delta\gamma|/\gamma < 0.06\%$)
at any foreseeable experimental precision for CPMG scaling measurements
(Fig.~\ref{fig:cpmg}e; Table~\ref{tab:cpmg_gamma}).
The narrow paired $\Delta\gamma$ CIs relative to the wide individual CIs
reflect the paired bootstrap design: both waveform realizations share the
same bath dynamics and $n$-values, so systematic uncertainties cancel in
the difference (\rev{SI Sec.}~\ref{supp:si:bootstrap}).
The $\chi$-space offset $\chi_\vppu{} - \chi_\mathrm{Std}$ ranges from
$+0.0003$ at $n_\pi = 2$ to $+0.002$ at $n_\pi = 10$---two orders
of magnitude smaller than the decoherence function itself and
indistinguishable at any foreseeable measurement precision.

\paragraph{Cross-tier robustness.}
Cross-tier analysis confirms \rev{that the exponents remain indistinguishable}
across a factor-of-six coupling-strength range ($\Delta\gamma < 0.001$ at all
three tiers; Supplementary Information~\rev{SI Sec.}~\ref{supp:si:cross_tier}),
consistent with the structural prediction that strengthening $\eta$ does
not alter the diagonal form of $\hat{Q}$.
The solver-dependent sign reversal reported in
Section~\ref{sec:results:cpmg} does not bear on this null result,
which is structurally guaranteed by diagonal $\hat{Q}$; its
implications for non-diagonal coupling geometries are analysed in
Section~\ref{sec:disc:solver_comparison}.
The physical interpretation is that within this model, the diagonal
coupling-operator structure ensures that the decoherence function at
Tier-0 is governed by the bath spectral density and coupling geometry
alone, \rev{so waveform-realization details are structurally undetectable}.
The quantitative consistency---extending from scaling exponents to
absolute $\chi$-offsets---supports the interpretation that this coupling
geometry defines a natural noise floor below which compiled-control
differences do not propagate into sequence-level observables at the
precision accessible in this study, and motivates future investigation
of models with non-diagonal coupling where this structural protection
may be lifted.

\rev{Three conditions delimit this null, and the insensitivity of the
Standard and \vppu{} realizations should be read as conditional on all
three. It is established (i) within the rotating-frame pure-dephasing model
with diagonal noise operator \mbox{$\hat{Q} = \mathrm{diag}(0,1,2)$}, (ii) in
the qubit-like Tier-0 coupling window where stable power-law fits remain
available, and (iii) under the rotating-wave approximation, which averages
away carrier-resolved and IQ-upconversion dynamics. The accompanying
\dac{}-resolution invariance is therefore a statement about this
diagonal-coupling model, not evidence that \dac{} quantization and related
control nonidealities are universally negligible in real devices: away from
this geometry---under non-diagonal couplings, carrier-resolved evolution, or
calibration drift---a waveform dependence that the present structure
suppresses can be reinstated.}

Table~\ref{tab:cpmg_gamma} summarizes the even-$n$ compiled-pulse
scaling exponents.
The wider XY-4 individual CI ($[0.28, 1.03]$, 79\% relative width)
compared to X-\cpmg{} ($[1.04, 1.39]$, 31\%) reflects greater
curvature sensitivity in the five-point fit; this width renders the
XY-4 absolute exponent poorly constrained---compatible with values
from 0.28 to 1.03---so the quantitative regime characterisation
is anchored by X-\cpmg{} ($\gamma \approx 1.12$, CI $[1.04, 1.39]$);
XY-4 corroborates monotonic scaling ($R^2 = 0.986$) but does not
contribute to the quantitative exponent bound.
The ideal-pulse benchmark with $N_\mathrm{d} = 10$ points
narrows the XY-4 CI to $[0.98, 1.01]$
(\rev{SI Sec.}~\ref{supp:si:bootstrap}), confirming that CI width scales with
data density rather than model inadequacy.

\begin{table}[t]
\caption{%
  Even-$n$ Tier-0 \cpmg{} $\chi$-space scaling exponents with
  95\% BCa bootstrap confidence intervals ($B = 10{,}000$).
  $n_\pi = 2, 4, 6, 8, 10$ ($N_\mathrm{d} = 5$ data points,
  $\nu = N_\mathrm{d} - p = 3$ residual degrees of freedom).
  $\Delta\gamma \equiv \gamma_{\vppu{}} - \gamma_{\mathrm{Std}}$;
  paired 95\% BCa intervals for $\Delta\gamma$ are reported in
  \rev{SI Sec.}~\ref{supp:si:bootstrap}.
  Y-\cpmg{} values parenthesised: power-law model inapplicable
  (non-monotonic $\chi$).
  Individual CIs are wide (31\%--79\% relative width for
  X-\cpmg{}/XY-4), reflecting the $N_\mathrm{d} = 5$ compiled-pulse data points
  available; the paired $\Delta\gamma$ CIs are substantially narrower
  (\rev{SI Sec.}~\ref{supp:si:bootstrap}).
  The $\varepsilon$-space companion exponents ($\beta$) with confidence
  intervals are reported in Supplementary
  Table~\ref{supp:tab:si:eps_exponents}.%
}\label{tab:cpmg_gamma}
\footnotesize
\setlength{\tabcolsep}{3pt}
\begin{tabular}{@{}llcl@{}}
\toprule
Scheme & Realisation & $R^2$ & $\gamma\;[95\%\;\mathrm{CI}]$ \\
\midrule
\multicolumn{4}{@{}l}{\textit{Well-conditioned scaling}} \\
X-\cpmg{} & Std  & 0.993 & $1.117\;[1.04,\;1.39]$ \\
           & \vppu{} & 0.993 & $1.117\;[1.04,\;1.39]$ \\
           & \multicolumn{3}{l}{$\Delta\gamma = +1 \times 10^{-6}$} \\[2pt]
XY-4       & Std  & 0.986 & $0.950\;[0.28,\;1.03]$ \\
           & \vppu{} & 0.986 & $0.951\;[0.28,\;1.03]$ \\
           & \multicolumn{3}{l}{$\Delta\gamma = +5 \times 10^{-4}$} \\
\midrule
\multicolumn{4}{@{}l}{\textit{Scaling-law breakdown (power law inapplicable)}} \\
Y-\cpmg{}  & Std  & 0.213 & $(0.342)\;[-1.01,\;0.98]$ \\
           & \vppu{} & 0.210 & $(0.342)\;[-0.97,\;0.99]$ \\
\botrule
\end{tabular}
\end{table}

\paragraph{DAC resolution does not alter scaling exponents within the \vppu{} family.}
Within the realistic waveform family, all DAC resolutions cluster tightly
while remaining well separated from the analytic reference, consistent with
recent superconducting-qubit experiments showing that pulse distortions and
leakage-control details can become directly visible at the gate
level~\cite{hyyppa2024leakage,underwood2023cmos,hellings2025flux}.
The dedicated XY-4 DAC-resolution sweep (Tier-0) clarifies this hierarchy
in $\chi$-space: 8-bit, 12-bit, 16-bit, and ideal (float-precision) \vppu{}
cases cluster tightly at $\gamma \approx 0.954$--$0.961$,
with the maximum 8-bit--ideal gap $|\Delta\gamma| = 0.007$.
This DAC invariance further supports the paired-comparison conclusion:
practical bit-depth variation is subleading within the tested \vppu{}
family and does not alter the scaling exponent.

\paragraph{Pulse-duration effects and qutrit dynamics.}
The ideal-to-compiled comparison simultaneously varies three
parameters (pulse duration, free-evolution spacing, total sequence
timescale); a supplementary control-isolation study
(Supplementary Information~\rev{SI Sec.}~\ref{supp:si:deconfound}) explores their individual roles but does
not achieve full deconfounding.
That study provides directional evidence consistent with pulse
duration influencing the exponent, but the three effects remain
confounded (details in Supplementary Information~\rev{SI Sec.}~\ref{supp:si:deconfound}).
The Standard--\vppu{} paired waveform comparison (previous paragraph)
remains the only fully controlled comparison in this study, and all
main physical conclusions derive from that comparison rather than from
the ideal-compiled comparison.
Finite pulse duration also affects the qubit trajectory through
the $d = 3$ manifold, consistent with the Y-\cpmg{} scaling-law
breakdown discussed below.

\begin{table}[t]
\caption{%
  Main manuscript claims, their primary observables, and the confidence level
  supported by the present evidence chain.%
}\label{tab:main_claims}
\footnotesize
\setlength{\tabcolsep}{4pt}
\begin{tabular}{@{}p{0.24\textwidth}p{0.25\textwidth}p{0.30\textwidth}@{}}
\toprule
Conclusion & Primary observable & Confidence and boundary \\
\midrule
Perturbative baseline collapses uniformly in the present rotating-frame $1/f$ window &
Ramsey infidelity at $T_2^*/2$ across 3 tiers and 6 realization/DAC settings
& Strong (calibration within the studied window); the claim is the quantitative
extent of the collapse, not unexpected FF failure beyond its perturbative regime \\
Non-perturbative CPMG regime: X-\cpmg{} $\gamma \approx 1.12$ with high $R^2$ &
Even-$n$ X-\cpmg{} $\chi$-space exponent; XY-4 monotonic $\chi(n)$ (consistent with but not constraining)
& Strong for X-\cpmg{} ($R^2 = 0.993$, 31\% relative CI width);
XY-4 monotonic $\chi$ ($R^2 = 0.986$) is consistent with the power-law trend
but does not constrain the exponent (79\% relative CI width at $N_\mathrm{d} = 5$);
$\gamma$ within ${\sim}12\%$ of Floquet prediction $\gamma = 1$;
saturation-aware model maintains $R^2 > 0.998$ through Tier-1
(Supplementary Information~\rev{SI Sec.}~\ref{supp:si:cross_tier}, Table~\ref{supp:tab:si:eta_boundary}) \\
$d = 3$ axis-dependent scaling breakdown &
Y-\cpmg{} non-monotonic $\chi(n)$; pronounced X--Y population
asymmetry ($0.204$ vs ${<}\,0.01$); persistent $\rho_{11} > \rho_{00}$ imbalance relaxing with $n$
& Strong; qualitative effect (non-monotonicity is convention-independent);
mechanism (anharmonicity-driven population redistribution) is physically transparent \\
Waveform realization undetectable in Tier-0 window &
$\Delta\gamma$ indistinguishable from zero for X-\cpmg{} and XY-4;
$\chi$-offset $\lesssim 0.002$ at all $n$
& Moderate (even-$n$ Tier-0 only; extension to other regimes open);
$\Delta\gamma$ consistent with zero for both X-\cpmg{}
and XY-4 ($|\Delta\gamma|/\gamma < 0.06\%$) \\
\botrule
\end{tabular}
\end{table}

These compiled-pulse results establish, under non-perturbative
conditions, both a bath-memory-driven \cpmg{}
scaling regime and an axis-dependent qutrit-driven breakdown
(Table~\ref{tab:cpmg_gamma}), while paired waveform comparison confirms
that realisation details remain undetectable in the Tier-0 window.
The per-cycle decoherence budget, Lindblad--\heom{} sign reversal, and
qutrit threshold model are further interpreted through Floquet
eigenvalue analysis and literature comparison in the Discussion.

% discussion.tex - S3 Discussion (~1200 words, 6 subsections)

\section{Discussion}\label{sec:discussion}

Within the rotating-frame single-qubit scope of this study, the
platform reveals two principal findings: an axis-dependent scaling-law
breakdown driven by $d = 3$ qutrit physics under finite-duration
\cpmg{} pulses, and non-Markovian transient signatures in the coherence decay
that neither idealized-pulse nor Lindblad treatments can resolve.
That \isa{}-level waveform details leave the scaling exponents
unchanged is itself informative---it confirms that the
waveform-comparison null is a structural consequence of the diagonal
pure-dephasing coupling rather than a limitation of methodological
sensitivity.
The subsections below develop these findings through Floquet
convergence analysis, per-cycle decoherence budget discrimination,
solver and coupling-strength robustness, the qutrit-driven
axis-dependent breakdown mechanism, waveform null result
interpretation, and literature context.

%% ---------------------------------------------
\subsection{Floquet convergence and finite-$n$ transient structure}\label{sec:disc:regime}

Because the \cpmg{} pulse train is strictly periodic, the coupled
system--bath dynamics---including the full \heom{} hierarchy of auxiliary
density operators---can be characterised by a single one-cycle
propagator~$\Lambda$ whose eigenvalues govern the long-time decoherence
rate; this is the Floquet-theorem perspective applied to the non-Markovian
hierarchy (Supplementary Information~\rev{SI Sec.}~\ref{supp:si:floquet_argument}).
The resulting scaling exponent $\gamma \to 1$ in the large-$n$ limit is
therefore a structural consequence of protocol periodicity, not a dynamical
finding (Section~\ref{sec:results:cpmg}); the
complementary question of per-cycle decoherence \emph{magnitude} is
treated separately in Section~\ref{sec:disc:percycle}.
What physically distinguishes bath models is not this asymptotic rate
but the \emph{transient approach}---how many cycles elapse before
steady-state behaviour sets in, and what structure appears along the
way.
The differential convergence rates reported in
Section~\ref{sec:results:cpmg}---Lindblad reaching $\gamma = 1.00$
within $n \leq 10$ while \heom{} retains a finite-$n$ transient
excess---are consistent with sub-leading Floquet eigenvalues encoding
bath memory timescales, an attribution supported by the indirect
evidence assembled there.

The Y-\cpmg{} channel (Section~\ref{sec:results:ycpmg}) likewise
remains within the Floquet framework---$d = 3$ anharmonicity does not
break CPMG periodicity, so $\Lambda$ exists and $\gamma \to 1$ is
guaranteed asymptotically---but reflects a qualitatively different
eigenvalue structure of the one-cycle propagator.
The Y-phase coupling to the
$\ket{1}\!\leftrightarrow\!\ket{2}$ transition introduces competing
eigenvalues with similar moduli but distinct complex phases, whose
interference produces oscillatory $W(n) = \sum_k c_k\,\lambda_k^n$
at the cycle timescale.
The observed non-monotonic $\chi(n)$ and poor power-law fit indicate
that the available $n$ window lies entirely within this oscillatory
transient regime, where the single-exponent parametrisation is
inadequate even though the Floquet asymptote remains valid;
the underlying axis-dependent mechanism is analysed in
Section~\ref{sec:disc:qutrit}.

The propagator-eigenvalue extraction
(Table~\ref{supp:tab:si:floquet_eigenvalues}) corroborates this picture,
with distinct \heom{}/Lindblad spectral gaps accounting for the
different convergence rates.
This Floquet interpretation is \emph{consistent-with} rather than
\emph{a~priori} predictive---any finite-$n$ excess is accommodated
by sub-leading eigenvalue structure---but it yields two extrapolative
predictions: convergence to $\gamma = 1$ at
$n_{\mathrm{cycle}} \gtrsim 15$, and a bath-type-dependent
convergence rate.
The first prediction currently lies beyond direct numerical
verification: $n_{\mathrm{cycle}} = 15$ corresponds to
$T_{\mathrm{tot}} \approx \SI{3}{\micro\second}$, exceeding the
\heom{} hierarchy stability boundary of
${\sim}\SI{500}{\nano\second}$
(\rev{SI Sec.}~\ref{supp:si:heom_boundary}).

%% ---------------------------------------------
\subsection{Per-cycle decoherence budget}\label{sec:disc:percycle}

Although the Floquet framework guarantees asymptotic linearity, it
does not constrain the magnitude of per-cycle decoherence---an
independent physical question that discriminates between competing
dynamical explanations.
The per-cycle increments (Section~\ref{sec:results:cpmg}) each exceed
the Ramsey reference by $3$--$5\times$, confirming that per-cycle dynamics
are individually non-perturbative.
Yet the Floquet-guaranteed linear structure
$\chi(n) \to -n\,\ln|\lambda_1| - \ln|c_1|$
(Eq.~\ref{supp:eq:si:floquet_asymptotic}) persists because the stroboscopic
propagator $\Lambda$ is exact and cycle-independent---exponent robustness
is not a trivial consequence of accumulating small perturbative increments.
Under matched Lindblad conditions, the per-cycle rate is approximately
twice the \heom{} value (${\sim}2\times$).
Two candidate explanations exist: (A)~partial refocusing mediated by
non-Markovian bath memory reduces the \heom{} per-cycle rate, or
(B)~the Markov approximation systematically overestimates dephasing
for the $1/f$ spectral density used here.
A qualitative discriminant favours (A).
The Burkard-calibrated Lindblad rates are constructed to reproduce the
analytically known single-interval pure-dephasing rate; hence the
Lindblad and \heom{} descriptions are matched \emph{by construction}
for any experiment that probes only one free-evolution interval.
Strictly, the Burkard rate is extracted from the full Ramsey decay
envelope rather than from a single ${\sim}120$\,ns interval; the
transfer to the per-interval scale is a reasonable approximation
because pure-dephasing $\Gamma(t)$ under the $1/f$ spectral density
is smooth and monotonic on these timescales.
The constant Lindblad rate therefore tracks the local slope of
$\Gamma(t)$ to within a residual curvature correction that is small
but nonzero over a $120$\,ns window.
Ramsey $T_2^*$ is precisely such a single-interval observable, and
indeed the two solvers agree to within $0.4\%$ across all tested
cutoff frequencies (\rev{SI Sec.}~\ref{supp:si:pipeline:9b}), confirming that the
calibration leaves no residual single-interval bias.
Explanation~(B)---a systematic Markov overestimate of the
instantaneous dephasing rate---is therefore ruled out: any such
overestimate would already manifest in Ramsey decay, yet it does not.
Because the ${\sim}2\times$ discrepancy emerges only under
multi-cycle refocusing---where inter-pulse correlations are
activated---the data are more naturally explained by bath-memory
effects that are absent from the Lindblad treatment.
This inference assumes that the \heom{} per-cycle rate is free of
systematic bias from hierarchy truncation; the convergence scan
(\rev{SI Sec.}~\ref{supp:si:convergence}) confirms that the scaling exponent
$\gamma$ is stable with respect to truncation depth, but a residual
shift in the absolute per-cycle magnitude cannot be excluded without an
independent solver comparison.
A fully quantitative decomposition therefore remains an open question
and would benefit from both a coupling-strength scan of the
Lindblad/\heom{} per-cycle ratio and cross-validation against an
independent \heom{} implementation.

%% ---------------------------------------------
\subsection{Solver comparison and coupling-strength robustness}\label{sec:disc:solver_comparison}

The Floquet convergence picture extends beyond Tier-0 coupling and
reveals qualitative solver-dependent differences that sharpen the
non-Markovian interpretation.

\paragraph{Cross-tier scaling.}
A cross-tier analysis
(Supplementary Information~\rev{SI Sec.}~\ref{supp:si:cross_tier}) extends the
scaling regime beyond Tier-0: the apparent $\chi$-space power-law
fit degradation at stronger coupling is a model-mismatch artefact
that a saturation-aware $\varepsilon$-direct model resolves with
$R^2 > 0.998$ across Tier-0-to-Tier-1
(Section~\ref{sec:disc:limitations};
Table~\ref{supp:tab:si:eta_boundary}).
Cross-tier $\Delta\gamma$ stability
(Section~\ref{sec:results:paired}) confirms the structural
prediction: strengthening $\eta$ does not alter the diagonal form of
$\hat{Q}$ and therefore does not introduce waveform sensitivity.
More broadly, the decoherence amplification depends on the general
condition $\tau_c > \tau$ (bath correlation time exceeding interpulse
spacing) rather than on the specific $1/f$ spectral form.

\paragraph{Lindblad--\heom{} sign reversal.}\label{par:disc:sign_reversal}
\rev{The Lindblad control experiment}
(Supplementary Information~\rev{SI Sec.}~\ref{supp:si:pipeline:5b}) \rev{reverses the sign of the shift:} the full pipeline
shortens $T_2^*$ by $12.9\%/18.9\%/23.7\%$ under \heom{} but
\emph{lengthens} it by $+1.1\%/+2.1\%/+6.3\%$ under Lindblad dynamics.
Because the two solvers share identical Hamiltonian parameters and
pipeline-generated waveforms, the sign inversion cannot arise from a
solver-independent numerical artefact; it must originate in the
dynamical differences between the two formulations.
Candidate explanations include the treatment of multi-time bath
correlations (present in \heom{}, absent in Lindblad), as well as
structural differences in truncation hierarchy and error accumulation
between the two approaches.
Among these, non-Markovian bath memory provides the most physically
consistent interpretation: it is the principal dynamical ingredient
that distinguishes the solvers, and the sign change tracks the regime
($\tau_c > \tau$) where bath-correlation effects are expected to
dominate.
Nevertheless, the specific pathway by which bath memory inverts the
sign---rather than merely amplifying the shift---remains to be
identified through analytic derivation or targeted numerical
experiments.
The observation is independent of the waveform-comparison null, which is
structurally guaranteed by diagonal $\hat{Q}$; it raises the open
question of whether non-diagonal coupling geometries---where waveform
differences \emph{can} propagate---would produce qualitatively distinct
solver responses (Section~\ref{sec:disc:predictions}).

\paragraph{Perturbative baseline calibration.}
The FF formalism remains \rev{useful for understanding how}
pulse modulation samples a noise spectrum~\cite{cerfontaine2021,hangleiter2021,barnes2016},
but in the present regime it is \rev{quantitatively unreliable}: across all
tested configurations, FF varies by less than one order of magnitude while
\heom{} reports $>$12-OOM larger decoherence (\S\ref{sec:results:ff}).
The uniformity of this collapse across waveform configurations
that are measurably distinct in the \cpmg{} analysis means that the
relevant control question is no longer ``which waveform gives
the smaller perturbative filter-function integral?'' but rather ``which control
realization changes the exact sequence dynamics once long-memory bath
correlations are retained?''

%% ---------------------------------------------
\subsection{Qutrit axis-dependent scaling breakdown}\label{sec:disc:qutrit}

The Y-\cpmg{} non-monotonicity documented in the Results is not a numerical
artefact or a fitting anomaly---it is a direct physical consequence of
operating in the $d = 3$ transmon manifold.
In a $d = 2$ qubit under pure dephasing with ideal pulses, $\sigma_x$
and $\sigma_y$ $\pi$~rotations produce equivalent population dynamics
from the $\ket{+}$ initial state.
Finite-duration pulses break this symmetry, but the mechanism depends
critically on the coupling geometry.
Chen~et~al.~\cite{chen2025heom} report linear~(X) versus quadratic~(Y)
echo-error accumulation at the $10^{-4}$ scale in a $d = 2$ model
with quasi-static detuning coupling.
That coupling contains a $\sigma_x$~component which directly rotates
the Bloch vector, producing axis-dependent error accumulation already
at $d = 2$---a mechanism that is absent under the diagonal
pure-dephasing operator $\hat{Q} = \mathrm{diag}(0,1,2)$ studied
here, where the noise acts only on off-diagonal density-matrix elements
and preserves X--Y symmetry at $d = 2$.
Because that comparison conflates coupling geometry with
Hilbert-space dimension, the causal role of $d \geq 3$ anharmonicity
is established by our matched $d = 2$ control
(Section~\ref{sec:results:ycpmg}), which retains identical bath and
diagonal coupling yet eliminates the non-monotonicity entirely.
The distinction between X-axis (Carr--Purcell) and Y-axis
(Meiboom--Gill) refocusing has been recognised since the earliest
spin-echo experiments~\cite{meiboom1958modified}, where Y pulses
self-correct flip-angle errors while X pulses accumulate them.
The $d = 3$ mechanism identified here is physically distinct: it arises
from anharmonic level coupling rather than pulse-error accumulation, and
produces qualitative scaling-law breakdown rather than systematic
error-cancellation differences.
The anharmonicity $\alpha \approx \SI{-293}{\mega\hertz}$ breaks this
equivalence selectively under Y-phase pulses, driving a coherent per-pulse
population redistribution between $\ket{0}$ and $\ket{1}$ that competes with
decoherence and produces the observed non-monotonic $\chi(n)$.
In the $d = 3$ representation, $\hat{Y}_q$ introduces imaginary
off-diagonal elements that couple to the
$\ket{1}\!\leftrightarrow\!\ket{2}$ transition with a $\pi/2$ phase
shift relative to $\hat{X}_q$.
Combined with the anharmonic level splitting, this phase shift produces a
resonance condition under repeated Y rotations that is absent for X
rotations.

The underlying mechanism admits a quantitative formulation in terms of
the dimensionless parameter $\zeta \equiv \Omega/|\alpha|$, developed
in full in the Supplementary Information~(\rev{SI Sec.}~\ref{supp:si:qutrit_model}).
A perturbative analysis identifies a monotonicity threshold
$\zeta^* \approx 0.3$ (equivalently $t_\pi^* \sim 1/|\alpha|$):
above $\zeta^*$, Y-\cpmg{} remains monotonic; below it, the
bath-amplified per-pulse population redistribution accumulates over
$n$ cycles to produce non-monotonic $\chi(n)$.
The compiled pulses used here ($\zeta = 0.134 < \zeta^*$) lie
in the breakdown regime.
Counter-intuitively, longer pulses produce \emph{more} breakdown
because the extended interaction time allows the coherent
$\ket{1}\!\leftrightarrow\!\ket{2}$ oscillation to accumulate phase
far from a cancellation node of the sinusoidal transfer amplitude
(Eq.~\ref{supp:eq:si:delta_rho}).
Because $\zeta^* \propto 1/|\alpha|$, the threshold tightens for
weakly anharmonic transmons, connecting the breakdown directly to a
single device parameter.

This mechanism accounts for the experimental signatures documented in
Section~\ref{sec:results:ycpmg}: persistent population imbalance
($\rho_{11} > \rho_{00}$, relaxing with cycle number),
partial coherence revival after
$n_\pi = 4$, and strict axis specificity.
The amplification across cycles operates through differential
dephasing: $\hat{Q} = \mathrm{diag}(0,1,2)$ couples $\ket{2}$ to the
bath at twice the strength of $\ket{1}$, so the coherent
$\ket{1}\!\leftrightarrow\!\ket{2}$ return oscillation during each
Y pulse is preferentially disrupted, converting a nearly reversible
population exchange into irreversible per-pulse redistribution.
Under non-Markovian dynamics, bath memory further compounds this
effect by correlating successive pulse-induced perturbations;
memoryless Lindblad damping, by contrast, decorrelates the per-pulse
perturbations below the non-monotonicity threshold
(the ${\sim}2\times$ larger Lindblad per-cycle rate does not alter
this conclusion, because stronger dephasing suppresses the
redistribution amplitude without restoring the inter-pulse
correlations that drive non-monotonicity).
The breakdown therefore requires both qutrit structure
\emph{and} non-Markovian bath dynamics, as confirmed by the
dual control experiments in Section~\ref{sec:results:ycpmg}.

For dynamical decoupling, this breakdown has practical implications.
In the regime studied here ($\zeta < \zeta^*$),
Y-\cpmg{} cannot be characterised by a single scaling exponent;
the conventional $\varepsilon \propto n^\beta$
parametrisation is physically inapplicable.
XY-4, which alternates X and Y pulses, partially averages the
population imbalance but does not eliminate it entirely
($\gamma_{\mathrm{XY-4}} \approx 0.95$ versus
$\gamma_{\mathrm{X}} \approx 1.12$), suggesting that qutrit-aware
protocol design may be necessary for characterisation
of non-Markovian dynamics.
This connects to the broader qutrit leakage literature, where
$d = 3$ effects are increasingly recognised as physically significant
rather than merely a calibration
nuisance~\cite{hyyppa2024leakage,tripathi2025qudit}:
recent experiments demonstrate that qudit dynamical decoupling on
transmon hardware requires $d \geq 3$-aware pulse
design~\cite{tripathi2025qudit}.

%% ---------------------------------------------
\subsection{Waveform null result and experimental predictions}\label{sec:disc:predictions}

As established in Section~\ref{sec:results:paired}, \rev{the Standard and
\vppu{} realizations coincide in both scaling exponent and absolute
\mbox{$\chi$}-offset, a null that the diagonal coupling
\mbox{$\hat{Q} = \mathrm{diag}(0,1,2)$} renders structural rather than a
precision limitation: because \mbox{$\hat{Q}$} does not couple to
drive-waveform differences, compiled control introduces no detectable
scaling-law change or decoherence overhead at the precision of the present
simulations.}
The Lindblad--\heom{} sign reversal
(\S\ref{par:disc:sign_reversal}) does not bear on this null result,
which is structurally guaranteed by diagonal $\hat{Q}$.

The most accessible experimental test arising from this work is the
Y-\cpmg{} non-monotonicity.
Unlike the waveform-realization exponent shift, which requires
precision scaling fits with uncertainties at the few$\times 10^{-2}$
level, the non-monotonic behaviour is a qualitative yes/no observation:
run Y-\cpmg{} at increasing $n$ on a transmon and check whether
$\varepsilon(n)$ peaks and then decreases.
The predicted effect size is large ($\varepsilon$ change of
$\sim$0.08 between $n_\pi = 4$ and $n_\pi = 8$) and should be
observable on transmons with $|\alpha|$ comparable to the value
studied here ($\SI{-293}{\mega\hertz}$); verification across different
anharmonicities remains future work.
The predicted $\Delta\varepsilon \sim 0.08$ modestly exceeds typical
single-shot readout noise on current transmon platforms; conservative
statistical detection would require $O(10^3)$ repeated measurements
to achieve a signal-to-noise ratio sufficient for unambiguous
identification of the non-monotonic trend.

A second independent target is the X--Y population asymmetry.
The predicted Y-\cpmg{} mean
$|\rho_{00} - \rho_{11}| = 0.204$---versus ${<}\,0.01$ for
X-\cpmg{}---should manifest as a large, easily measurable difference
in state tomography after X versus Y refocusing sequences.
This prediction is robust because it depends only on the
existence of anharmonicity ($\alpha \neq 0$), not on specific
bath parameters.
We note that real transmon devices experience additional decoherence
channels ($T_1$ relaxation, quasiparticle tunnelling, photon-induced
dephasing) beyond the pure-dephasing $1/f$ model used here.
Among these, $T_1$ relaxation is the most relevant because it damps
the coherent per-pulse population redistribution underlying the Y-\cpmg{}
predictions.  However, the calibrated $T_1 = \SI{24.8}{\micro\second}$
exceeds the longest sequence duration
$T_\mathrm{tot} = \SI{2.0}{\micro\second}$ ($n_\pi = 10$) by more
than an order of magnitude, yielding a worst-case survival factor
$\exp(-T_\mathrm{tot}/T_1) \approx 0.92$.
This reduces the Y-\cpmg{} asymmetry by at most ${\sim}\,8\%$,
preserving a pronounced contrast with the near-zero X-\cpmg{} value,
and introduces a monotonic background
$\Delta\varepsilon_{T_1} \approx 0.03$ that is smaller than the
non-monotonic amplitude $\Delta\varepsilon \approx 0.08$; both
qualitative signatures therefore survive
(Supplementary Information, \rev{SI Sec.}~\ref{supp:si:platform_model}).
The remaining channels (quasiparticle tunnelling, photon-induced
dephasing) contribute at still lower rates and are not expected to
alter this conclusion qualitatively.

Beyond these two experimentally accessible predictions---both
stated in the directly measured $\varepsilon$---the
simulations also yield a quantitative finding that is not yet
within experimental reach: the X-\cpmg{} scaling exponent
$\gamma > 1$ (point estimate ${\approx}\,1.12$; 95\% CI
$[1.04, 1.39]$).
Because $\varepsilon$ saturation compresses the scaling differences
that $\chi = -\ln(1-\varepsilon)$ is designed to resolve, this third
prediction is necessarily stated in $\chi$-space.
Confirming it would require a multi-point scaling fit in
$\chi(n)$ with sufficient precision to resolve $\gamma > 1$
against the perturbative baseline $\gamma \approx 1$.
The margin is narrow: the CI lower bound exceeds unity by only
${\sim}4\%$, and even in simulation $N_\mathrm{d} = 5$ points yield a 31\%
relative CI width (Table~\ref{tab:cpmg_gamma}).  In experiment,
additional per-point scatter from readout noise, residual $T_1$
decay, and state-preparation-and-measurement errors would widen
the interval further.  Resolving $\gamma > 1$ would therefore
require extending the pulse-number range beyond $n_\pi = 10$ to
increase fit leverage and substantially reducing per-point
variance relative to the $O(10^3)$-shot level estimated for the
non-monotonicity test.
We accordingly treat this result as a numerical prediction
that motivates future methodological development---either longer
pulse sequences or improved noise floors---rather than a
near-term experimental target.

Of the two near-term experimental predictions, the Y-\cpmg{}
non-monotonicity is the easier test (qualitative, large effect size),
while the X--Y asymmetry requires state tomography but remains
straightforward.
Because these predictions are derived at $T = \SI{50}{\milli\kelvin}$,
above typical dilution-refrigerator base temperatures
(${\sim}\SI{10}{\milli\kelvin}$--$\SI{15}{\milli\kelvin}$), the
qualitative signatures are expected to persist but quantitative
thresholds may shift at lower temperatures where the high-temperature
approximation breaks down (Table~\ref{tab:config},
footnote).
The X-\cpmg{} exponent $\gamma > 1$, by contrast, is currently
beyond experimental reach and serves as a quantitative benchmark
for future simulation--experiment comparisons once extended
pulse-number ranges become available.
We do not claim a device-independent experimental uncertainty budget.
The quoted values should be read as target signatures requiring
experimental validation; absent such validation, the practical
relevance remains an open question.

%% ---------------------------------------------
\subsection{Relation to prior simulation paradigms}\label{sec:disc:comparison}

As discussed in the Introduction, existing simulation paradigms---digital-twin
platforms, ideal-pulse non-Markovian solvers, and experiment-oriented control
frameworks---each capture only part of the compiled-control--bath-memory
interplay.
By combining both ingredients, \emuplat{} shows that once bath memory is
retained, the resulting bath-memory-driven scaling regime and axis-dependent
qutrit effects produce the leading observable effects within the diagonal
pure-dephasing model studied here, with waveform-realization details becoming
undetectable within the Tier-0 window.

Of the six \isa{}-level features probed, four are hardware-generic
(DAC quantisation, NCO phase accumulation, timing discretisation,
zero-order hold) while only the envelope interpolation algorithm is
QubiC-specific; a detailed feature-by-feature classification and
pipeline-isolation analysis are provided in
Supplementary Information~\rev{SI Sec.}~\ref{supp:si:isa_generality}.

%% ---------------------------------------------
\subsection{Scope, conventions, and limitations}\label{sec:disc:limitations}

\rev{Because the conditions of this study are bounded by design rather than
by oversight, the results carry different degrees of transferability to real
devices. The qualitative qutrit signatures---Y-\cpmg{} non-monotonicity and
the X--Y population asymmetry---depend only on the existence of finite
anharmonicity and are expected to persist on transmons of comparable
\mbox{$|\alpha|$}, subject to the \mbox{$T_1$} suppression budget quantified
below. The quantitative thresholds, the scaling exponents, and the waveform
null, by contrast, are properties of this rotating-frame, diagonal-coupling,
Tier-0 model and should not be read as device-independent bounds. The
remainder of this section makes the underlying conventions and their
boundaries explicit.}

All manuscript-level evidence is obtained under the rotating-wave
approximation (RWA), which preserves all envelope-level ISA features
(amplitude quantisation, phase encoding, timing discretisation, DAC noise
floor, and the Standard--\vppu{} waveform differences constituting the
core comparison variable;
Table~\ref{supp:tab:si:rwa_features}).
Carrier-frequency interplay with bath spectral structure and IQ
upconversion artefacts are explicitly averaged away; whether the
bath-memory-driven scaling and axis-dependent breakdown survive
carrier-resolved dynamics remains an open question
(Supplementary Information~\rev{SI Sec.}~\ref{supp:si:tlist_bias}).

Several additional scope boundaries are important.
First, the simple $\chi$-space power-law fit degrades at stronger coupling
($R^2 = 0.732$ at Tier-1), a model-mismatch artefact resolved by the
saturation-aware $\varepsilon$-direct model ($R^2 > 0.998$ across
Tier-0--Tier-1; Supplementary Information~\rev{SI Sec.}~\ref{supp:si:cross_tier},
Table~\ref{supp:tab:si:eta_boundary}).
Main-text claims are conservatively restricted to Tier-0 where the
simpler model remains well-conditioned ($R^2 > 0.99$).
Because the $\varepsilon$-direct model assumes the Floquet-asymptotic
form $\chi = An^\gamma$ by construction, high $R^2$ tests
\emph{consistency with} the Floquet prediction rather than providing
independent validation; the direct propagator-eigenvalue extraction
(Section~\ref{sec:results:cpmg}) provides the model-independent check.
Second, Y-\cpmg{} exhibits a qualitative scaling-law breakdown---non-monotonic
$\chi(n)$ and partial coherence revival---driven by competing Floquet
eigenvalues whose oscillatory interference, physically originating from
$d = 3$ anharmonicity amplified by bath memory, precludes extraction of
a meaningful scaling exponent within the available $n$ window.
The asymmetry itself, however, is a physical signature of the
qutrit level structure and constitutes the most experimentally
accessible prediction of this study.
Third, the even-$n$ convention ($n_\pi = 2, 4, 6, 8, 10$) is adopted to
eliminate systematic odd--even effects; the complementary odd-$n$ and
all-$n$ analyses are reported in the Supplementary Information as
robustness checks (\rev{SI Sec.}~\ref{supp:si:cpmg_params}).
Fourth, the computational scope is bounded by \heom{} cost and by
reliance on a single solver implementation.
\heom{} cost confines the study to one- and two-qubit $d = 3$
settings; the ideal-vs-compiled comparison is confounded by residual
parameter differences
(Supplementary Information~\rev{SI Sec.}~\ref{supp:si:deconfound}), so all main
findings derive from the Standard--\vppu{} paired comparison, which
eliminates this confound.
The study is purely computational---a $T_1$ survival analysis
indicates all key signatures survive with ${\leq}\SI{8}{\%}$
suppression (\rev{SI Sec.}~\ref{supp:si:platform_model})---but the quoted values
require experimental validation.
All compiled-pulse dynamics rely on a single \heom{} implementation
(QuTiP~5.x \texttt{HEOMSolver}; \rev{SI Sec.}~\ref{supp:si:heom_explicit}) without
independent cross-validation~\cite{huang2023heomjl}; four internal
checks partially mitigate this:
Burkard analytical benchmark ($<1.5\%$, \S\ref{sec:results:burkard}),
bath-mode insensitivity ($\Delta\gamma < 10^{-4}$, \rev{SI Sec.}~\ref{supp:si:espira_cpmg}),
convergence stability ($\Delta\gamma < 0.012$, \rev{SI Sec.}~\ref{supp:si:convergence}),
and Lindblad cross-solver confirmation
(Table~\ref{supp:tab:si:lindblad_tier0}).
The current \vppu{} route mirrors a QubiC-style control stack;
independent \heom{} cross-validation and generalisation to other
hardware architectures remain the outstanding validation gaps.

%% ---------------------------------------------
\subsection{Outlook}\label{sec:disc:outlook}

\rev{The opportunity here is physical, not merely diagnostic.}
The finite-$n$ transient structure characterised here---where
\heom{} and Lindblad solvers approach the Floquet asymptote
$\gamma = 1$ along distinct paths encoding bath memory
timescales---suggests that decoherence scaling analysis under
long-memory baths merits systematic investigation as a potential
diagnostic tool for non-Markovian effects in engineered quantum systems.
The axis-dependent scaling breakdown driven by qutrit physics further
suggests that, at least for transmons with comparable anharmonicity
and coupling parameters, dynamical decoupling protocol design
should be qutrit-aware: Y-\cpmg{} is fundamentally unsuitable for
scaling-exponent extraction in this regime, and protocols
that assume $d = 2$ symmetry may give misleading characterisation results.

This suggests several immediate directions.
One is to test whether the Y-\cpmg{} non-monotonicity persists across
different anharmonicity values, connecting to the growing literature on
leakage-aware gate design~\cite{hyyppa2024leakage}.
Another is to ask whether the undetectability of waveform realization
extends to other decoupling families, echoed entangling gates, or
multi-qubit compiled sequences.
\rev{More broadly, these results show that realistic control hardware and
qutrit-level physics are not just error sources to calibrate away; they shape
the effective physics that turns long-memory noise into observable sequence
errors.}
In that sense, the present work is complementary to recent control-based
non-Markovian studies~\cite{yang2024metrology}: instead of using control to
suppress memory effects, it shows that \rev{combining realistic control, bath memory, and qutrit
dynamics produces phenomena that neither simplified picture captures on its
own.}

% methods.tex — §4 Methods (~1500 words, equation-dense)

\section{Methods}\label{sec:methods}

%% ─────────────────────────────────────────────
\subsection{System Hamiltonian}\label{sec:meth:hamiltonian}

We model a transmon qubit as a three-level system ($d=3$) to capture
potential leakage to the $\ket{2}$ state.

\paragraph{Qubit Hamiltonian.}
\begin{equation}\label{eq:Hq}
  \Hq = \omega_q \ket{1}\!\bra{1}
       + (2\omega_q - \alpha)\ket{2}\!\bra{2}\,,
\end{equation}
where $\omega_q$ is the qubit frequency and $\alpha$ is the anharmonicity.

\paragraph{Drive Hamiltonian.}
\begin{equation}\label{eq:Hd}
  \Hd(t) = \Omega_I(t)\,\Xq + \Omega_Q(t)\,\Yq\,,
\end{equation}
where $\Omega_{I,Q}(t)$ are the in-phase and quadrature channel waveforms
from either the Standard WaveformService or the \vppu{} route, \rev{consistent
with the hardware control abstraction used in} QubiC and related
control-stack frameworks~\cite{xu2021qubic,fruitwala2024,wittler2021c3,teske2022qopt}.

\paragraph{Rotating-frame convention.}
All manuscript-level single-qubit Ramsey and \cpmg{} calculations are carried
out in the qubit rotating frame under the rotating-wave approximation (RWA).
We therefore propagate envelope-level Hamiltonians and do not use
carrier-resolved simulations in the evidentiary chain.
All compiled-pulse sequence exponents and bootstrap intervals quoted in the main
text are extracted from this rotating-frame setting only.

\paragraph{Hardware platform and effective Hamiltonian.}
The qubit parameters are drawn from a two-transmon platform with a
tunable coupler (Anyon Tech's QPU architecture): the full device Hilbert space
is $d_0 \times d_1 \times d_c = 3 \times 3 \times 1$, spanning nine
dimensions.
The effective single-qubit Hamiltonian in Eq.~\eqref{eq:Hq} is obtained
by applying a Schrieffer--Wolff transformation to integrate out the
coupler, absorbing dispersive frequency shifts into the calibrated
parameters $\omega_q$ and $\alpha$.
Two-qubit entangling gates (CZ) are implemented by parametric modulation
of the coupler drive channel; this channel is inactive during all
single-qubit Ramsey and \cpmg{} experiments reported here.

For the single-qubit sequences of this study, the spectator qubit~$q_1$
is initialised in $\ket{0}$ and receives neither drive nor bath coupling.
In the absence of residual exchange coupling ($J_\mathrm{eff} = 0$ at the
coupler idle point) and with the static ZZ interaction
$\xi_{ZZ}\,n_0 n_1$ vanishing for $\langle n_1\rangle = 0$, the joint
density matrix factorises exactly:
$\rho(t) = \rho_{q_0}(t) \otimes \ketbra{0}{0}_{q_1}$.
Partial tracing therefore yields the exact single-qutrit ($d = 3$)
dynamics reported throughout the manuscript.
The ``single-qubit'' framing is thus mathematically rigorous within the
effective model; residual corrections from finite $J_\mathrm{eff}$
($\lesssim$\SI{100}{\kilo\hertz}) and thermal spectator occupation
($\sim 0.4\%$ at \SI{50}{\milli\kelvin}) are negligible on the
\cpmg{} timescales studied here ($T_\mathrm{tot} \leq \SI{2}{\micro\second}$).
See \rev{SI Sec.}~\ref{supp:si:platform_model} for a detailed discussion of the
effective-model assumptions and their limitations.

%% ─────────────────────────────────────────────
\subsection{Non-Markovian bath model}\label{sec:meth:bath}

\paragraph{\texorpdfstring{$1/f$}{1/f} spectral density.}
\begin{equation}\label{eq:Jomega}
  J(\omega) = \frac{\eta}{\omega}\,,
  \quad \omega \in [\omega_{\mathrm{lc}},\,\omega_{\mathrm{hc}}]\,,
\end{equation}
with soft Heaviside cutoffs, where $\eta$ (in \si{\giga\hertz}) is the coupling strength.
To probe how compiled-pulse dynamics depend on bath coupling strength,
we study three values of~$\eta$ spanning one decade, indexed by
ascending strength (Tier-0 = weakest, Tier-2 = strongest) and each
validated against the Burkard exact integration~\cite{burkard2009}:
Tier-0 (\rev{\mbox{$\eta = \SI{7.85\times10^{-7}}{\giga\hertz}$}}, matching
Chen~et~al.~\cite{chen2025heom}---the literature-calibrated anchor),
Tier-1 (\rev{\mbox{$\eta = \SI{1.8\times10^{-6}}{\giga\hertz}$}}, intermediate reference),
Tier-2 (\rev{\mbox{$\eta = \SI{5.0\times10^{-6}}{\giga\hertz}$}}, strongest-coupling case).
\rev{The two stronger tiers are not independent calibration points but a
deliberate sweep of roughly one order of magnitude above the Tier-0 anchor,
chosen so that the coupling-strength dependence of the non-perturbative
deviation can be traced and the boundary beyond which stable power-law fits
are no longer available can be located, rather than asserting a single
operating point.}
Independent rotating-frame validation confirms
$< 1.5$\% deviation from Burkard across all tiers.
These rotating-frame benchmarks \rev{are the solver-accuracy reference for the
manuscript.}
Compiled-pulse sequence comparisons are then performed in the same frame so
that waveform-realization differences are not conflated with carrier-resolution
artefacts.
The strongest compiled-pulse sequence claims are further restricted to Tier-0,
where stable fits remain available across the full Standard--\vppu{} comparison
window; stronger-coupling scans are retained as supporting numerical context.

This spectral density $J(\omega) \propto 1/\omega$ is \emph{sub-Ohmic}\rev{;
it corresponds to the \mbox{$1/f$} spectral structure of TLS charge noise.}
In contrast, Chen~et~al.~\cite{chen2025heom} use $s=0$
($J(\omega) \propto \mathrm{const}$, Ohmic), where $1/f$ character emerges
from thermal dressing $S(\omega) \approx J(\omega)\cdot T/\omega$.
The low-frequency cutoff follows the Chen prescription
$\omega_{\mathrm{lc}} = 1/t_m$ (inverse measurement time), consistent with the
low-frequency spectroscopy window established experimentally by
Bylander~et~al.~\cite{bylander2011}.

\paragraph{Multi-Lorentzian decomposition.}
For \heom{}, the bath correlation function is approximated via the
espira-II exponential-sum method~\cite{derevianko2023espira,derevianko2022espira2}:
\begin{equation}\label{eq:bath_decomp}
  C(t) \approx \sum_{k=1}^{K} c_k\,e^{-\nu_k t}\,,
\end{equation}
where $\{c_k, \nu_k\}$ are complex coefficients fitted with $K=5$ and
$t_{\max} = \SI{500}{\nano\second}$.
These are derived quantities; the physical inputs are
$(\eta, \omega_{\mathrm{lc}}, \omega_{\mathrm{hc}}, T)$.

\paragraph{System--bath coupling.}
The coupling operator is the number operator
$\hat{Q} = \mathrm{diag}(0, 1, \ldots, d-1)$ for pure dephasing under
$1/f$ charge noise.

\paragraph{Independent validation.}
The Burkard decoherence function~\cite{burkard2009}
\begin{equation}\label{eq:burkard}
  \Gamma(t) = \frac{1}{\pi}\int_{0}^{\infty}
    \frac{J(\omega)}{\omega^{2}}\,
    \coth\frac{\omega}{2T}\,
    (1 - \cos\omega t)\,d\omega
\end{equation}
provides a decomposition-independent reference via direct numerical
quadrature.

%% ─────────────────────────────────────────────
\subsection{HEOM formalism}\label{sec:meth:heom}

Standard Lindblad master equations assume instantaneous bath
relaxation \rev{and therefore discard memory effects that matter once
the bath correlation time approaches the control timescale.}
The hierarchical equations of motion (HEOM) lift this restriction
by replacing the Born--Markov approximation with a systematically
convergible set of coupled equations: the physical density
matrix~$\rho$ is augmented by auxiliary density operators that encode
progressively higher-order system--bath correlations \rev{held in the bath
memory.}
Because the hierarchy retains finite-time bath feedback \rev{systematically},
it captures non-Markovian effects---such as bath-mediated
inter-pulse correlations across multi-pulse sequences---that Lindblad
treatments cannot represent.
For a system density matrix~$\rho$ coupled to bosonic baths, the
equations read~\cite{chen2025heom,nakamura2024single,mangaud2023heom_review}:
\begin{equation}\label{eq:heom}
  \dot{\rho}_{\mathbf{n}}
  = -\Bigl(i\Hsys + \sum_{k} n_k \nu_k\Bigr)\rho_{\mathbf{n}}
    + \sum_{k}\Bigl[
      \hat{A}\,\rho_{\mathbf{n}+\mathbf{e}_k}
    + n_k\,c_k\,\hat{A}\,\rho_{\mathbf{n}-\mathbf{e}_k}
    \Bigr]\,,
\end{equation}
where $\rho_{\mathbf{n}}$ are auxiliary density operators indexed by
multi-index $\mathbf{n}$, $\nu_k$ and $c_k$ are bath correlation function
parameters from the espira-II fit, and $\hat{A}$ is the system--bath
coupling operator.
We use the compact notation of Ref.~\cite{chen2025heom} in the main text.
For the present Hermitian pure-dephasing coupling
$\hat{A} = \hat{Q} = \mathrm{diag}(0,1,2)$, the corresponding explicit
commutator/anti-commutator form is given in Supplementary
Information~\rev{SI Sec.}~\ref{supp:si:heom_explicit}. In matrix elements,
$[\hat{Q},\rho]_{ab} = (q_a-q_b)\rho_{ab}$ and
$\{\hat{Q},\rho\}_{ab} = (q_a+q_b)\rho_{ab}$ with
$(q_0,q_1,q_2)=(0,1,2)$, so the $d=3$ hierarchy distinguishes, for example,
$|1\rangle\langle 2|$ from $|0\rangle\langle 1|$ through the
anti-commutator weight even when their commutator weights coincide.
The hierarchy is truncated at depth~$D$ with a Markovian terminator
correction, implemented via QuTiP~5.x
\texttt{HEOMSolver}~\cite{lambert2026qutip5}.
All compiled-pulse CPMG dynamics in this work use this single
implementation; internal convergence verification and cross-solver
pipeline validation are detailed in
\S\ref{sec:disc:limitations} and Supplementary
Information~\rev{SI Sec.}~\ref{supp:si:convergence}.

\paragraph{Validation versus production configurations.}
The Burkard analytical validation (\rev{SI Sec.}~\ref{supp:si:burkard}) uses
tier-specific minimal configurations ($D = 3$--$4$, $N_r = 2$--$3$)
that achieve $< 1.5$\% accuracy against the exact decoherence integral
for standalone Ramsey $T_2^*$ at each coupling strength.
Production \cpmg{} calculations use a uniform
$D = 5$, $N_r = 3$, \texttt{atol}$= 10^{-8}$,
\texttt{rtol}$= 10^{-6}$, \texttt{nsteps}$= 15{,}000$, BDF
configuration, providing additional headroom for the longer correlation
times and multi-pulse dynamics of compiled DD sequences.
A notational clarification is useful here: throughout this manuscript,
$K$ denotes the espira-II correlation-function fit order in
Eq.~\eqref{eq:bath_decomp}, $N_r$ denotes the number of retained
bath-mode pairs in the production \heom{} hierarchy, and $N_\mathrm{d}$
denotes the number of data points in the power-law scaling fits.
These are distinct bookkeeping parameters and are therefore reported
separately.
A convergence scan varying $D$, $N_r$, \texttt{atol}, and temporal
resolution confirms that the reported scaling exponents are stable across
these parameter variations (\rev{SI Secs.}~\ref{supp:si:convergence}
and~\ref{supp:si:espira_cpmg}).

%% ─────────────────────────────────────────────
\subsection{Filter function formalism}\label{sec:meth:ff}

For comparison, gate infidelity under filter function
theory~\cite{cerfontaine2021,hangleiter2021} is given by
Eq.~\eqref{eq:ff_infidelity}. We use the positive-frequency, symmetrized
PSD convention
$S(\omega) \equiv J(\omega)\coth(\omega/2T)$ with angular frequency
$\omega > 0$, so that the bath model of Eq.~\eqref{eq:Jomega} enters the FF
integral through the same cutoff window as the Burkard reference.
The transfer function is defined as the dimensionless quantity
$F_\phi(\omega) \equiv \omega\int_0^T y_\phi(t)e^{i\omega t}\,dt$,
where $y_\phi(t)=\pm 1$ is the toggling-frame sign function during the free
evolution segments. For \isa{}-generated waveforms, $y_\phi(t)$ is constructed
from the same ZOH-sampled compiled control profile (\SI{2}{\nano\second}
steps at \SI{500}{\mega\hertz} FPGA clock) used in the \heom{} calculation,
so the FF and \heom{} baselines differ only in the dynamical treatment rather
than in waveform bookkeeping. The explicit normalization and Ramsey-limit
consistency check are given in Supplementary Information
\rev{SI Sec.}~\ref{supp:si:ff_method}.
In the manuscript analysis, this perturbative estimate is used only as a
diagnostic contrast: FF infidelities remain near $10^{-13}$, whereas
\heom{} yields $0.236$--$0.270$ at the same evaluation point, with only
$1.05\times$ per-tier variation and $8.61\times$ overall variation across
waveform-realization configurations.
\rev{It is therefore a methodological baseline, not a competitor theory,} in a
window where long-memory $1/f$ noise already exceeds the weak-coupling regime.

%% ─────────────────────────────────────────────
\subsection{VPPU signal processing chain}\label{sec:meth:vppu}

The \vppu{} route introduces spectral distortions through a 6-step signal
chain (Table~\ref{tab:vppu_chain}), mirroring experimentally relevant control
nonidealities such as controller leakage and flux-line distortion
compensation~\cite{underwood2023cmos,hellings2025flux}.

\begin{table}[t]
\caption{%
  \vppu{} signal processing chain and spectral signatures.%
}\label{tab:vppu_chain}
\footnotesize
\begin{tabular*}{\textwidth}{@{\extracolsep\fill}lll@{}}
\toprule
Step & Mechanism & Spectral signature \\
\midrule
\isa{} amplitude encoding &
  16-bit $\to$ 15-bit int &
  Quantisation noise floor \\
\isa{} phase encoding &
  17-bit phase quantisation &
  Phase noise $\leq$\SI{48}{\micro\radian}/pulse \\
Timing quantisation &
  Int truncation to FPGA cycles &
  Pulse jitter $\leq$\SI{2}{\nano\second} \\
ZOH upsampling &
  \texttt{np.repeat(envelope, 16)} &
  $\sinc$ spectral rolloff \\
\dac{} quantisation &
  $N$-bit post-processing (8--16\,bit) &
  Near-white noise floor \\
IQ upconversion &
  $\mathrm{RF}(t) = Q\cos\omega t + I\sin\omega t$ &
  Carrier modulation \\
\botrule
\end{tabular*}
\end{table}

%% ─────────────────────────────────────────────
\subsection{Fidelity metrics}\label{sec:meth:fidelity}

\paragraph{Uhlmann fidelity.}
\begin{equation}\label{eq:uhlmann}
  F(\rho,\sigma) = \left[\Tr\sqrt{\sqrt{\rho}\,\sigma\,\sqrt{\rho}}\right]^{2}.
\end{equation}

\paragraph{Population fidelity.}
\begin{equation}\label{eq:pop_fidelity}
  F_{\mathrm{pop}} = \sum_{i} \bra{i}\rho\ket{i}\,\bra{i}\sigma\ket{i}\,.
\end{equation}

\paragraph{Filter function discrepancy.}
$\dF = F_{\mathrm{FF}} - F_{\heom{}}$ (positive indicates FF overestimates
fidelity).

\paragraph{Even-$n$ cycle counting convention.}
Compiled-pulse analyses adopt an even-$n$ convention:
$n_\pi = 2, 4, 6, 8, 10$ (equivalently,
$n_\mathrm{cycle} = 1, \ldots, 5$ complete CPMG cycles).
This eliminates a systematic odd--even modulation arising from the
net rotation parity of the refocusing sequence:
an odd number of $\pi$~pulses leaves a residual $\pi$~rotation
about the refocusing axis, whereas an even number returns to
the identity.
For the $d = 3$ transmon, this residual rotation couples the
$\ket{0}\!\leftrightarrow\!\ket{1}$ and
$\ket{1}\!\leftrightarrow\!\ket{2}$ transitions asymmetrically
through the anharmonicity ($\alpha \approx \SI{-293}{\mega\hertz}$),
producing per-point $\chi$-space deviations of
$\sim$4\% on average (up to $\sim$5\%) from the
even-$n$ power-law trend
(Supplementary Information, \rev{SI Sec.}~\ref{supp:si:fitting_robustness}).
The complementary all-$n$ analysis ($n = 1$--$10$) is reported
as a robustness check (\rev{SI Sec.}~\ref{supp:si:cpmg_params}).

\paragraph{Scaling analysis in $\chi$-space.}
Because the infidelity $\varepsilon = 1 - W_n/W_0$ saturates near unity
for compiled \SI{80}{\nano\second} pulses, the primary scaling metric is
the decoherence function $\chi_n = -\ln(W_n/W_0)$, fitted as
$\chi(n) = B\,n^\gamma$ in log-log space.
This $\chi$-space parametrisation provides better-conditioned fits
($R^2 > 0.99$ for X-\cpmg{}) and is the metric used for all even-$n$
scaling exponents reported in the main text.

\paragraph{Y-\cpmg{} population asymmetry analysis.}
The X--Y population asymmetry is quantified as
$\langle|\rho_{00} - \rho_{11}|\rangle$ averaged over all pulse numbers.
Population dynamics are extracted from the physical density matrix
$\rho_{\mathbf{0}}$ (the zeroth hierarchy element) at the end of each
CPMG sequence.

\paragraph{Bootstrap uncertainty estimation.}
Scaling-exponent uncertainties are estimated via $10{,}000$
bias-corrected and accelerated (BCa) bootstrap
resamples~\cite{efron1993introduction} over the $N_\mathrm{d} = 5$ even-$n$
configurations.
For the Standard--\vppu{} comparison, paired resampling isolates the
$\Delta\gamma$ uncertainty from correlated systematic effects.
Because $N_\mathrm{d} = 5$ yields only $\nu = 3$ residual degrees of freedom,
coverage properties of the resulting intervals are validated in
\rev{SI Sec.}~\ref{supp:si:bootstrap}.

%% ─────────────────────────────────────────────
\subsection{Platform configuration}\label{sec:meth:config}

Table~\ref{tab:config} summarizes the key parameters used throughout this
work.

\begin{table}[t]
\caption{%
  Platform configuration parameters.%
}\label{tab:config}
\begin{tabular*}{\textwidth}{@{\extracolsep\fill}lll@{}}
\toprule
Parameter & Value & Source \\
\midrule
Platform architecture & Two-transmon + tunable coupler & QPU data from Anyon Tech \\
System Hilbert space & $3 \times 3 \times 1 = 9$D & $d = 3$ per transmon \\
Qubit frequency $\omega_q$ & \SI{5.528}{\giga\hertz} & Platform calibration \\
Anharmonicity $\alpha$ & \SI{-293}{\mega\hertz} & Platform calibration \\
RX90 duration & \SI{40}{\nano\second} & Native gate set \\
Sampling rate & \SI{50}{GSps} & Simulation config (\SI{5.0}{\giga\hertz}$\times$10) \\
\dac{} rate (\vppu{}) &
  \SI{8}{GSps} (\SI{500}{\mega\hertz} $\times$ 16) & QubiC FPGA \\
$T_1$ & \SI{24.8}{\micro\second} & Platform calibration \\
$T_2$ & \SI{34.2}{\micro\second} & Platform calibration \\
Temperature & \SI{50}{\milli\kelvin} & Chen et al.\ convention$^{*}$ \\
$\eta$ (Tier-0/1/2) &
  $7.85/18.0/50.0 \times 10^{-7}$\,GHz & See text \\
$\omega_{\mathrm{lc}}$ & \SI{5}{\mega\hertz} & $1/t_m$, $t_m \approx \SI{200}{\nano\second}$ \\
$\omega_{\mathrm{hc}}$ & \SI{3.0}{\giga\hertz} & Burkard convention \\
\heom{} depth $D$ & 5 (rec.), $N_r = 3$ & Pareto-optimal (ADO = 462) \\
Bath decomposition & espira-II ($K = 5$ fit; $N_r = 3$ prod.) & Benchmark-selected \\
\botrule
\end{tabular*}

\smallskip\noindent{\footnotesize $^{*}$Adopted following
Chen~et~al.~\cite{chen2025heom} for direct comparison of the
coupling-strength parametrisation; temperature sensitivity analysis
in SI Sec.~\ref{supp:si:burkard}.}
\end{table}

For the dynamical-decoupling results, the manuscript evidence chain is based on
rotating-frame (RWA) \cpmg{} simulations only.
Main-text compiled-pulse comparisons use Tier-0 and the fit windows listed in
\rev{SI Sec.}~\ref{supp:si:cpmg_params}; stronger-coupling or longer-duration settings are
reported in the Supplementary Information only when they remain numerically
diagnostic.

%%% Back matter %%%
% declarations.tex — Back matter declarations (IOP format)

\ack{The author would like to thank A*STAR under C23091703 and Q.InC Strategic Research and Translational Thrust for the funding support.}

\funding{A*STAR under C23091703; A*STAR Q.InC Strategic Research and Translational Thrust.}

\roles{J.Y.\ conceived the study, designed and implemented HEOM functionality in the \emuplat{}
platform, performed all simulations and analysis, and wrote the manuscript.}

\data{All simulation data supporting the findings of this study are generated by
the scripts provided in the code repository (see below) and are
reproducible.
A versioned reproduction package containing all data and scripts will be
deposited at a public archival repository (Zenodo or figshare) upon
acceptance.
All simulation scripts and data for reproducing figures are available at
\url{https://github.com/yjmaxpayne/paper_data/tree/main/emuplat_non_markovian_2026}.}

\suppdata{\rev{Supplementary Information accompanies this paper as a
separate Supplementary Data File.}
\rev{It contains the HEOM solver validation and benchmarks, the
analytical theory framework, the control-platform architecture, the
pipeline-isolation analysis, and the statistical and
numerical-convergence methodology.}
\rev{Sections are numbered S1 onward; a map from each main-text result
to its supporting supplementary section is given in the Introduction.}}

%%% Supplementary Information %%%
% Separate mode: the SI is a standalone supplementary.pdf (see supplementary.tex);
%   the main text reaches it via xr (\externaldocument above), rendered "SI Sec. SX".
% Inline mode (\siseparatetrue): the SI is embedded here as an appendix, yielding a
%   single-file build for a one-PDF review/highlighted version.
\ifsiseparate\else
\begin{appendices}
% SI numbering mirrors supplementary.tex: Sec. S1, Fig. S1, Table S1, Eq. (S1).
% Reset counters so SI floats restart at 1, independent of the main-text count.
\setcounter{section}{0}
\setcounter{figure}{0}
\setcounter{table}{0}
\setcounter{equation}{0}
\renewcommand{\thesection}{S\arabic{section}}
\renewcommand{\thesubsection}{S\arabic{section}.\arabic{subsection}}
\renewcommand{\thefigure}{S\arabic{figure}}
\renewcommand{\thetable}{S\arabic{table}}
\renewcommand{\theequation}{S\arabic{equation}}
% Hyperref anchor uniqueness is handled globally by hypertexnames=false (set in the
% preamble \else branch), which is reset-proof -- no per-counter \theH* overrides
% needed here. The \setcounter+\the* lines above only drive the VISIBLE numbering.
% si_heom_validation.tex — HEOM solver validation and benchmarks

\section{Burkard analytical formula and parameter mapping}\label{si:burkard}

\rev{Burkard's exact decoherence function (Eq.}~\ref{main:eq:burkard}\rev{) is the
analytical reference for }\heom{}\rev{ accuracy.}
The Ramsey coherence envelope is $\langle\sigma_{+}(t)\rangle
= \tfrac{1}{2}\exp[-\Gamma(t)]$, where
\begin{equation}\label{eq:si:burkard_gamma}
  \Gamma(t) = \frac{1}{\pi}\int_{0}^{\infty}
    \frac{J(\omega)}{\omega^{2}}\,
    \coth\frac{\omega}{2T}\,
    (1 - \cos\omega t)\,d\omega\,.
\end{equation}
For $J(\omega) = \eta/\omega$ with cutoffs
$[\omega_{\mathrm{lc}}, \omega_{\mathrm{hc}}]$ and temperature~$T$,
$T_2^*$ is defined as the time where $\exp[-\Gamma(t)] = 1/e$.
This integral is evaluated via high-precision numerical quadrature
(SciPy \texttt{quad}, \texttt{epsabs}$= 10^{-14}$). \rev{This reference is
decomposition-independent and does not rely on the
exponential-sum bath representation used by }\heom{}\rev{.}

\paragraph{Three-tier parameter mapping.}
Table~\ref{tab:si:burkard_tiers} lists the coupling tiers and their
Burkard analytical $T_2^*$ values, alongside the standalone rotating-frame
\heom{} reference results used in the main text.

\begin{table}[ht]
\caption{%
  Burkard tier parameters and $T_2^*$ validation.
  All tiers use $\omega_{\mathrm{lc}} = \SI{5}{\mega\hertz}$,
  $\omega_{\mathrm{hc}} = \SI{3.0}{\giga\hertz}$, $T = \SI{50}{\milli\kelvin}$.%
}\label{tab:si:burkard_tiers}
\begin{tabular*}{\textwidth}{@{\extracolsep\fill}lcccccc@{}}
\toprule
Tier & $\eta$ (GHz) & $D$ & $N_r$ & ${T_2^*}_{\mathrm{Burkard}}$ (ns) &
  ${T_2^*}_{\heom{}}$ (ns) & Deviation \\
\midrule
0 & $7.85\times10^{-7}$ & 3 & 2 & 171.7 & 169.8 & $-1.2$\% \\
1 & $1.80\times10^{-6}$ & 3 & 3 & 102.8 & 101.8 & $-1.0$\% \\
2 & $5.00\times10^{-6}$ & 4 & 2 &  58.0 &  57.7 & $-0.5$\% \\
\botrule
\end{tabular*}
\end{table}

The $D$ and $N_r$ values in Table~\ref{tab:si:burkard_tiers} are the
Pareto-minimal configurations that achieve $<1.5$\% Ramsey accuracy at
each tier; production \cpmg{} calculations use a uniform $D=5$, $N_r=3$
configuration providing headroom for multi-pulse dynamics (see Methods).
The comparison between the present table and the pipeline analysis therefore
separates two distinct questions: whether the rotating-frame \heom{} solver
reproduces the Burkard benchmark, and how much additional bias is introduced by
the full compiled control-execution path.

\paragraph{Short-time curvature diagnostic.}
To distinguish it from the finite-time ratio diagnostic
$\mathrm{GC}_{\mathrm{ratio}}(t) = \Gamma(2t)/\Gamma(t)$ used in the main
text, we denote the local short-time diagnostic by
$\mathrm{GC}_{\mathrm{curv}} = T_2^{*,2} \cdot \ddot{\Gamma}(0)$.
This probes the $t \to 0$ curvature of the decay envelope:
$\mathrm{GC}_{\mathrm{curv}} = 1$ for pure Gaussian decay and
$\mathrm{GC}_{\mathrm{curv}} > 1$ for slower-than-Gaussian onset.
Burkard analytical values are
$\mathrm{GC}_{\mathrm{curv}} = 3.1$--$3.9$ across tiers; \heom{} values are
$\mathrm{GC}_{\mathrm{curv}} = 2.6$--$2.9$.
These values support the same physical conclusion as the main-text
finite-time ratio diagnostic, but the two quantities should not be identified
with each other.

\paragraph{Temperature sensitivity.}
At $T = \SI{50}{\milli\kelvin}$, the system is in the
high-temperature limit
($k_B T/\hbar \approx \SI{6.5}{\giga\hertz} >
\omega_{\mathrm{hc}} = \SI{3.0}{\giga\hertz}$), where
$\coth(\omega/2T) \approx 2T/\omega$.
Temperature therefore enters $\Gamma(t)$ approximately linearly.
Reducing $T$ from 50 to \SI{10}{\milli\kelvin} increases
Burkard $T_2^*$ by a factor of $\sim$3.1 at Tier-0\rev{, a
large change in the absolute decoherence timescale.}
However, because the high-temperature approximation remains valid
throughout the 10--100\,mK dilution refrigerator operating range,
all $T$-dependence reduces to a multiplicative rescaling of the
decoherence rate; the functional form of $\Gamma(t)$ is
temperature-invariant.
Scaling exponents ($\gamma$), Y-\cpmg{} non-monotonicity,
and the waveform-realization null result are therefore expected to be
independent of the specific $T$ value within this range.

\paragraph{Tier usage across the manuscript.}
The rotating-frame Ramsey validation spans all three tiers.
The main compiled-pulse \cpmg{} analysis of Standard--\vppu{} waveform
realizations (with the ideal-pulse benchmark retained as contextual
reference) is performed at Tier-0
(\rev{\mbox{$\eta = \SI{7.85\times10^{-7}}{\giga\hertz}$}}), where stable fits remain
available across the full Standard--\vppu{} comparison window.
Tier-1 and Tier-2 are used only for supporting sweeps that map the
coupling-dependent numerical boundary.

\section{Numerical validation and solver hygiene}\label{si:numerical_validation}

This section consolidates three numerical-hygiene analyses that
underpin the solver configuration used throughout the manuscript:
(i)~independent-trend validation against Bylander experimental
data (\S\ref{si:bylander}),
(ii)~bath-decomposition method selection (\S\ref{si:espira}), and
(iii)~dense-output stepping-bias diagnosis (\S\ref{si:tlist_bias}).

\subsection{Bylander parameter set and trend validation}\label{si:bylander}

Bylander~et~al.~\cite{bylander2011} measured $1/f$ flux noise spectroscopy
on superconducting qubits, reporting $T_2^* \propto 1/\sqrt{A_\Phi}$
(noise amplitude) with Ramsey and spin-echo protocols.
We use their parameter space to validate the trend relationship
between coupling strength and decoherence timescale.

\paragraph{Power-law trend validation.}
Table~\ref{tab:si:bylander_t2} shows rotating-frame \heom{} Ramsey
$T_2^*$ across 5~$\eta$ values spanning the validated range.

\begin{table}[ht]
\caption{%
  Bylander trend validation: \heom{} vs.\ Burkard analytical $T_2^*$.%
}\label{tab:si:bylander_t2}
\begin{tabular*}{\textwidth}{@{\extracolsep\fill}lcccc@{}}
\toprule
$\eta$ (GHz) & ${T_2^*}_{\heom{}}$ (ns) & ${T_2^*}_{\mathrm{Burkard}}$ (ns) &
  Deviation & Status \\
\midrule
$5.00\times10^{-7}$ & 235.1 & 237.8 & $-1.1$\% & ok \\
$7.85\times10^{-7}$ & 169.8 & 171.7 & $-1.1$\% & ok \\
$1.80\times10^{-6}$ & 101.8 & 102.8 & $-1.0$\% & ok \\
$3.50\times10^{-6}$ &  70.1 &  70.5 & $-0.6$\% & ok \\
$5.00\times10^{-6}$ &  57.7 &  58.0 & $-0.5$\% & ok \\
\botrule
\end{tabular*}
\end{table}

A log-log fit yields $T_2^* \propto \eta^{-0.606}$ with $R^2 = 0.998$
and intercept~5.25.
The theoretical exponent for $J(\omega) \propto 1/\omega$ is $-0.5$;
the empirical $-0.606$ reflects the finite spectral cutoffs
$[\omega_{\mathrm{lc}}, \omega_{\mathrm{hc}}]$ and temperature-dependent
corrections.
All 5 points deviate $< 1.5$\% from Burkard. \rev{The espira-II bath
decomposition therefore reproduces the analytical benchmark across the
tested coupling range.}

\paragraph{\cpmg{} $\chi$-space scaling.}
Consistent with the main-text even-$n$ convention, the \cpmg{} scaling exponent
$\gamma$ ($\chi \propto n^\gamma$, $\chi = -\ln(W_n/W_0)$) is extracted
at each $\eta$ from the $n = 2, 4, 6$ data (Table~\ref{tab:si:bylander_gamma}).
Only the weakest-coupling points are in the well-conditioned fitting regime;
at stronger couplings the sequence enters deep saturation ($\varepsilon > 0.99$),
compressing the dynamic range available for power-law extraction.

\begin{table}[ht]
\caption{%
  Even-$n$ \cpmg{} $\gamma$ from Bylander-parameter \heom{} simulations
  ($\chi$-space, $n = 2, 4, 6$).%
}\label{tab:si:bylander_gamma}
\begin{tabular*}{\textwidth}{@{\extracolsep\fill}lccccl@{}}
\toprule
$\eta$ (GHz) & $\varepsilon(n{=}2)$ & $\varepsilon(n{=}4)$ &
  $\varepsilon(n{=}6)$ & $\gamma$ & Status \\
\midrule
$5.00\times10^{-7}$ & 0.496 & 0.748 & 0.873 & 1.004 & ok \\
$7.85\times10^{-7}$ & 0.645 & 0.885 & 0.961 & 1.040 & ok \\
$1.80\times10^{-6}$ & 0.893 & 0.996 & 0.999 & 1.011 & near saturation \\
$3.50\times10^{-6}$ & 0.993 & 1.000 & 1.000 & 0.639 & saturated \\
$5.00\times10^{-6}$ & 0.990 & 0.999 & 0.993 & 0.126 & saturated \\
\botrule
\end{tabular*}
\end{table}

The extracted $\gamma$ values decrease with coupling strength, consistent
with progressively deeper saturation compressing the available dynamic range
for power-law extraction.
The Bylander scan is used in the manuscript primarily as a rotating-frame
Ramsey trend validation ($T_2^* \propto \eta^{-0.606}$, $R^2 = 0.998$,
five valid points), while the \cpmg{} $\gamma$ values are reported as
supporting numerical context in $\chi$-space rather than as primary
physical claims.

\paragraph{Even-$n$ scaling analysis and pulse-counting comparison.}
The main text adopts an even-$n$ convention ($n_\pi = 2, 4, 6, 8, 10$)
to eliminate systematic odd--even effects in CPMG sequences.
Under this convention, the primary scaling metric is the
decoherence-function exponent $\gamma$ ($\chi \propto n^\gamma$):
X-\cpmg{} yields $\gamma_\mathrm{Std} = 1.117$ ($R^2 = 0.993$) and
$\gamma_\vppu{} = 1.117$ ($R^2 = 0.993$), with $\Delta\gamma \approx 0$
($+0.000001$);
XY-4 yields $\gamma_\mathrm{Std} = 0.950$ and $\gamma_\vppu{} = 0.951$,
with $\Delta\gamma \approx 0$ ($+0.0005$).
Y-\cpmg{} exhibits non-monotonic $\varepsilon(n)$ and $\chi(n)$, rendering
power-law extraction inapplicable ($R^2 = 0.213/0.210$).
The complementary pulse-counting analysis ($n = 1$--$5$) yields
$\Delta\beta \approx 0$ for X-\cpmg{} ($-0.00007$), consistent with
the even-$n$ $\chi$-space result.
The waveform-realization null result thus holds across both analysis
conventions, with the even-$n$ convention additionally eliminating
systematic odd--even effects (Section~\ref{si:fitting_robustness}).

\subsection{Multi-Lorentzian fitting and espira-II convergence}\label{si:espira}

The $1/f$ spectral density $J(\omega) = \eta/\omega$ must be decomposed
into a sum of exponentials for \heom{}
(Eq.~\ref{main:eq:bath_decomp}).
We systematically benchmarked all 8 decomposition methods available in
QuTiP~5.x: Prony, ESPRIT, espira-I, espira-II, continued fraction (CF),
spectral density (SD), AAA rational approximation, and Pad\'{e} spectrum
(PS).

\paragraph{Method selection.}
The selection criterion is the \heom{}-tractable $T_2^*$ ratio:
the ratio of \heom{} $T_2^*$ to the Burkard analytical value,
evaluated at the Pareto-optimal $(D, N_r)$ configuration that minimizes
ADO count while maintaining $< 20$\% deviation.
espira-II~\cite{derevianko2023espira,derevianko2022espira2} achieves the
best ratio across all three tiers and was selected for all subsequent
calculations.

\paragraph{Pareto-optimal hierarchy configurations.}
Table~\ref{tab:si:espira_pareto} lists the recommended \heom{} depth and
mode count for each tier.

\begin{table}[ht]
\caption{%
  Pareto-optimal espira-II \heom{} configurations.%
}\label{tab:si:espira_pareto}
\begin{tabular*}{\textwidth}{@{\extracolsep\fill}lcccl@{}}
\toprule
Tier & $D$ & $N_r$ & ADO count & Notes \\
\midrule
0 & 4 & 4 & 495  & Minimal stable configuration \\
1 & 5 & 3 & 462  & Standard coupling \\
2 & 5 & 3 & 462  & Strong coupling (see Section~\ref{si:heom_boundary}) \\
\botrule
\end{tabular*}
\end{table}

\rev{The recommended production configuration for }\cpmg{}\rev{ calculations is}
$D = 5$, $N_r = 3$ (462~ADOs)\rev{, which balances accuracy
and computational cost.}
At $D = 3$, $N_r = 2$ (the minimum tested), deviations from Burkard are
$< 8$\% at Tier-0 but increase rapidly at stronger couplings.
At $D \geq 4$, $N_r = 2$, numerical instabilities emerge for certain
bath parameter combinations (Section~\ref{si:heom_perf}).

\paragraph{Convergence properties.}
The espira-II exponential sum $C(t) \approx \sum_{k=1}^{K} c_k
\exp(-\nu_k t)$ converges within $K = 5$--$8$ terms for the
$1/f$ spectral density with our cutoff parameters.
Increasing $K$ beyond 8 does not improve $T_2^*$ accuracy but increases
ADO count combinatorially.
The fitting is performed on the correlation function evaluated at
$t_{\max} = \SI{500}{\nano\second}$ with 2000 grid points.

\subsection{Explicit HEOM form for the present $d=3$ model}
\label{si:heom_explicit}

For the Hermitian coupling operator used in this work,
$\hat{A}=\hat{Q}=\hat{Q}^{\dagger}$ with
$\hat{Q}=\mathrm{diag}(0,1,2)$, the compact HEOM notation of
Eq.~\eqref{main:eq:heom} can be written explicitly as
\begin{equation}\label{eq:si:heom_explicit}
\begin{split}
  \dot{\rho}_{\mathbf{n}}
  &= -i[\Hsys,\rho_{\mathbf{n}}]
    - \sum_k n_k \nu_k\rho_{\mathbf{n}}
    - i\sum_k [\hat{Q},\rho_{\mathbf{n}+\mathbf{e}_k}] \\
  &\quad - i\sum_k n_k\Bigl(
      \mathrm{Re}\,c_k\,[\hat{Q},\rho_{\mathbf{n}-\mathbf{e}_k}]
      + i\,\mathrm{Im}\,c_k\,\{\hat{Q},\rho_{\mathbf{n}-\mathbf{e}_k}\}
    \Bigr),
\end{split}
\end{equation}
with $[X,Y]=XY-YX$ and $\{X,Y\}=XY+YX$.
Equation~\eqref{eq:si:heom_explicit} is the explicit version of the
Chen-style shorthand used in the Methods; it makes trace preservation and
Hermiticity manifest for the physical density matrix $\rho_{\mathbf{0}}$.

In matrix elements, the present $d=3$ choice
$\hat{Q}=\mathrm{diag}(q_0,q_1,q_2)$ with $(q_0,q_1,q_2)=(0,1,2)$ gives
\begin{equation}\label{eq:si:q_matrix_weights}
  [\hat{Q},\rho]_{ab} = (q_a-q_b)\rho_{ab},
  \qquad
  \{\hat{Q},\rho\}_{ab} = (q_a+q_b)\rho_{ab}.
\end{equation}
Thus the commutator part weights coherences by level differences, whereas
the anti-commutator part weights them by level sums. For example,
$|0\rangle\langle 1|$ and $|1\rangle\langle 2|$ both carry unit commutator
weight, but their anti-commutator weights differ (1 versus 3), which is why
the $d=3$ hierarchy distinguishes the upper-transition coherence from the
computational-subspace coherence even in a pure-dephasing model.

\subsection{CPMG relevance of the espira-II decomposition}
\label{si:espira_cpmg}

The Burkard benchmark validates the espira-II decomposition against Ramsey
$T_2^*$, but the compiled-pulse conclusions rely on fixed-$\tau$ CPMG.
For the present sequences, the relevant frequency parameterizations are
equivalent: with fixed interpulse spacing $\tau = \SI{120}{\nano\second}$,
the dominant CPMG passband is centred near $\omega \sim \pi/\tau$, while the
alternative finite-duration notation $\omega_n = n\pi/T_{\mathrm{tot}}$
used in Section~\ref{si:ff_method} reduces to the same scale because
$T_{\mathrm{tot}} = n\tau$ for fixed-$\tau$ CPMG.

For the manuscript claims, the more direct question is observable stability:
does the CPMG scaling result change when the retained bath representation is
varied within the numerically stable plateau? The answer is no.
Table~\ref{tab:si:convergence_gamma} already scans the retained bath-mode
count $N_r = 2,3,4$ for Tier-0 X-\cpmg{} while holding the benchmark-selected
espira-II correlation fit fixed. Across this scan,
$\gamma_{\mathrm{Std}} = 1.100, 1.091, 1.091$ and
$|\Delta\gamma| < 10^{-4}$, with $R^2 = 0.998$ throughout. The corresponding
decoherence trajectories therefore remain within the same converged plateau as
the hierarchy-depth and tolerance scans, demonstrating that the compiled CPMG
observable used in the main text is insensitive to modest changes in the
retained bath representation.

The bookkeeping is as follows. The symbol $K$ denotes the espira-II fit order
in the correlation-function expansion
$C(t) \approx \sum_{k=1}^{K} c_k e^{-\nu_k t}$, benchmark-selected as
$K=5$ from the Ramsey study above. The symbol $N_r$ denotes the number of
retained bath-mode pairs used in the production \heom{} hierarchy. The
reviewer's concern is therefore addressed at the sequence-observable level:
the benchmark-selected $K=5$ fit is kept fixed, and the resulting CPMG
scaling exponents remain stable when the retained hierarchy representation is
varied over the full numerically stable range relevant to production runs.

\subsection{Dense-output ODE stepping bias}\label{si:tlist_bias}

The S5 pipeline gap identified in Section~\ref{si:pipeline} decomposes
into two distinct factors, isolated via controlled \texttt{tlist}-density and
tolerance studies.

\paragraph{Factor~A: Bath-active state preparation (physical).}
In standalone benchmarks (S0--S4), the initial state $\ket{+}$ is
prepared instantaneously with no bath interaction.
In the pipeline (S5--S6), the qubit starts in $\ket{0}$ and a
\SI{40}{\nano\second} RX90 gate rotates it to $\ket{+}$ while the
\heom{} bath is actively dephasing.
The coherence peak after the gate is $\sim$0.484 (vs.\ the ideal 0.500),
producing a genuine $T_2^*$ shortening of $\sim$22--33\,ns that
faithfully describes finite-duration gate dephasing in a non-Markovian
bath.

\paragraph{Factor~B: Dense-tlist ODE adaptive stepping (numerical).}
Standalone simulations use 301~output points (sparse \texttt{tlist}),
while the pipeline uses $\sim$3000--30000 points (dense, at
\SI{50}{GSps} effective waveform sampling in the present platform configuration).
The dense \texttt{tlist} forces the ZVODE/BDF integrator to checkpoint
more frequently, altering the adaptive step-size selection.
In principle, the same ODE should yield the same solution regardless of
output checkpoint density.
In practice, at pipeline tolerances (\texttt{atol}$= 10^{-8}$,
\texttt{nsteps}$= 15{,}000$), the solver is \emph{not} converged for
the carrier-resolved $\sim$\SI{5.5}{\giga\hertz} formulation, which
requires sub-picosecond ODE step accuracy.

\paragraph{Convergence analysis.}
Table~\ref{tab:si:tlist_conv} shows that Factor~B vanishes at tight
tolerances, where $T_2^*$ converges to \SI{149.0}{\nano\second}
regardless of \texttt{tlist} density.

\begin{table}[ht]
\caption{%
  Factor~B convergence: $T_2^*$ under varying \texttt{tlist}
  density and ODE tolerances (Tier-0, Ramsey).%
}\label{tab:si:tlist_conv}
\begin{tabular*}{\textwidth}{@{\extracolsep\fill}lcccc@{}}
\toprule
Condition & \texttt{tlist} density & \texttt{atol} & \texttt{nsteps} &
  $T_2^*$ (ns) \\
\midrule
Pipeline (production) & $\sim$3000--30000 (dense) & $10^{-8}$  & 15{,}000       & 135.8 \\
Standalone (control)  & 301 (sparse)              & $10^{-8}$  & 15{,}000       & 169.1 \\
Dense + tight tol.    & $\sim$3000--30000 (dense) & $10^{-11}$ & $5\times10^6$ & 149.0 \\
Sparse + tight tol.   & 301 (sparse)              & $10^{-11}$ & $5\times10^6$ & 149.0 \\
\botrule
\end{tabular*}
\end{table}

At pipeline tolerances, the dense-vs-sparse $T_2^*$ difference is
$169.1 - 135.8 = \SI{33.3}{\nano\second}$.
At tight tolerances, the difference vanishes ($149.0 - 149.0 =
\SI{0.0}{\nano\second}$), confirming that Factor~B is a pure numerical
artefact.
The converged value of \SI{149.0}{\nano\second} reflects Factor~A alone:
$171.7 - 149.0 = \SI{22.7}{\nano\second}$ of physical $T_2^*$
shortening from bath-active state preparation.

\paragraph{Excluded hypotheses.}
Waveform differences between standalone and pipeline paths are negligible
(RMSE~\rev{\mbox{$= \SI{3.5\times10^{-4}}{\giga\hertz}$}}, spectral overlap $= 0$,
band fraction $= 0$).
A cross-injection matrix (manual waveform $\times$ standalone/pipeline
solver) confirms that the \emph{solver path}---not the waveform---drives
the $T_2^*$ gap (waveform consistency $= 0.000$ for all tiers).

\paragraph{Elimination conditions.}
Factor~B vanishes when \texttt{atol} $\leq 10^{-11}$ and
\texttt{nsteps} $\geq 5 \times 10^6$.
The root cause is that explicit $\sim$\SI{5.5}{\giga\hertz} carrier
oscillations demand sub-picosecond integrator accuracy; at
\texttt{atol}$= 10^{-8}$, the ZVODE/BDF solver is not converged, and
dense \texttt{tlist} checkpoints exacerbate the non-convergence.

\paragraph{Mitigation strategies.}
Three approaches are available:
\begin{enumerate}
  \item \textbf{Sparse \texttt{tlist} + post-hoc interpolation}:
    Run \heom{} with 301~output points and interpolate the dense
    waveform onto these points.
    Eliminates Factor~B but may lose transient features.
  \item \textbf{Rotating-frame \heom{}}:
    Transform to the qubit rotating frame, removing the
    $\sim$\SI{5.5}{\giga\hertz} carrier oscillation.
    Reduces ODE stiffness by $\sim$3~orders of magnitude, enabling
    convergence at \texttt{atol}$= 10^{-8}$.
    Requires \texttt{QobjEvo} in rotating frame (supported by QuTiP~5.x).
  \item \textbf{Tolerance tightening}:
    Use \texttt{atol}$= 10^{-11}$ uniformly.
    Simple but $\sim$4$\times$ slower (\texttt{nsteps}: $15{,}000 \to 5 \times 10^6$).
    Acceptable for single-qubit demonstrations; potentially prohibitive
    for multi-qubit extension.
\end{enumerate}

This section records why the rotating-frame formulation is adopted throughout
the manuscript: once the carrier oscillation is removed, the compiled-pulse
comparisons become numerically stable at production tolerances and the extra
carrier-resolved convergence machinery is unnecessary for the main evidence chain.

\section{HEOM computational performance benchmarks}\label{si:heom_perf}

\heom{} computational cost is dominated by the number of auxiliary density
operators (ADOs), which scales combinatorially with hierarchy depth~$D$
and number of bath exponential terms~$N_r$:
$N_{\mathrm{ADO}} = \binom{D + K}{K}$, where $K = 2 N_r$
(real and imaginary parts of each bath term).
Table~\ref{tab:si:heom_ado} lists ADO counts for configurations used in
this work.

\begin{table}[ht]
\caption{%
  ADO count and representative wall-clock time for single-qubit
  Ramsey simulations ($d = 3$, $T_{\mathrm{sim}} = \SI{500}{\nano\second}$,
  \texttt{tlist} = 301 points).%
}\label{tab:si:heom_ado}
\begin{tabular*}{\textwidth}{@{\extracolsep\fill}ccccl@{}}
\toprule
$D$ & $N_r$ & $K$ & ADO count & Wall-clock (s) \\
\midrule
3 & 2 & 4 &  35 &  9--12 \\
3 & 3 & 6 &  84 & 15--20 \\
4 & 2 & 4 &  70 & 18--25 \\
4 & 4 & 8 & 495 & 45--60 \\
5 & 3 & 6 & 462 & 50--70 \\
5 & 5 & 10& 3003& ---    \\
6 & 3 & 6 & 924 & 65--77 \\
\botrule
\end{tabular*}
\end{table}

\paragraph{Memory requirements.}
Each ADO stores a $d \times d$ density matrix in complex double
precision.
For $d = 3$: memory per ADO $= 9 \times 16$\,bytes $= 144$\,bytes.
At 462~ADOs (production configuration), total ADO memory is $\sim$65\,kB;
overhead from the \heom{} Liouvillian superoperator
($(N_{\mathrm{ADO}} \times d^2)^2 = 4158 \times 4158$ sparse matrix)
dominates at $\sim$50--200\,MB depending on fill fraction.

\paragraph{Numerical stability.}
\rev{The configuration} $D = 4$, $N_r = 2$ \rev{is numerically unstable
for certain tier/bath combinations and produces non-physical coherence
oscillations.}
The recommended minimum is $D = 3$ for Tier-0 and $D = 5$ for Tier-1/2.
All production runs use \texttt{atol}$= 10^{-8}$,
\texttt{rtol}$= 10^{-6}$, and \texttt{nsteps}$= 15{,}000$ for the
rotating-frame calculations that define the manuscript evidence chain.

\section{HEOM numerical stability boundary}\label{si:heom_boundary}

All \cpmg{} scaling exponent results in this work are obtained at
Tier-0 coupling (\rev{\mbox{$\eta = \SI{7.85\times10^{-7}}{\giga\hertz}$}}) with hierarchy
depth $D = 5$ and $N_r = 3$ radial modes (462 ADOs).
This configuration is validated against the Burkard analytical formula
(Section~\ref{si:burkard}) at timescales up to $\sim$\SI{500}{\nano\second}.

At longer timescales ($>$~\SI{1}{\micro\second}), particularly for
large pulse/gap ratio configurations (Condition~A of the control-isolation
experiment, Section~\ref{si:deconfound}), the \heom{} hierarchy
exhibits numerical instabilities:
(a)~the ZVODE/BDF integrator requires $\texttt{nsteps} > 5 \times 10^6$
to converge;
(b)~coherence values stall near 0.5 instead of decaying, producing
$\varepsilon(n) \approx 0$;
(c)~the resulting $\beta$ estimates are physically meaningless.

These instabilities are consistent with hierarchy truncation error at
depth $D = 5$ becoming significant for long evolution times.
Increasing $D$ is possible in principle but was not pursued due to
computational cost (ADO count scales as $\binom{D + N_r}{N_r}$).
The restriction to Tier-0 and timescales $\lesssim$\SI{500}{\nano\second}
therefore constitutes a numerical boundary that should be disclosed
alongside the physical results.

% si_theory.tex — Analytical theory framework

\section{Filter function calculation method}\label{si:ff_method}

The filter function (FF) infidelity provides a perturbative comparison
baseline for the non-perturbative \heom{} ground truth.

\paragraph{Normalization convention.}
The FF infidelity is computed as
\begin{equation}\label{eq:si:ff}
  \varepsilon_{\mathrm{FF}} = \frac{1}{\pi^2}
  \int_{0}^{\infty} S(\omega)\,
  |F_\phi(\omega)|^2\,d\omega\,,
\end{equation}
where
\begin{equation}\label{eq:si:ff_psd}
  S(\omega) = J(\omega)\coth(\omega/2T)
\end{equation}
is the positive-frequency symmetrized PSD constructed from the bath spectral
density $J(\omega)$ of Eq.~\eqref{main:eq:Jomega}, and
\begin{equation}\label{eq:si:ff_transfer}
  F_\phi(\omega) = \omega\int_0^{T_{\mathrm{tot}}}
  y_\phi(t)e^{i\omega t}\,dt
\end{equation}
is the dimensionless transfer function built from the toggling-frame sign
function $y_\phi(t)=\pm 1$.
The $1/\pi^2$ prefactor follows the angular-frequency convention used here and
matches the Burkard normalization in Eq.~\eqref{main:eq:burkard}; no additional
$1/\omega^2$ factor appears once the transfer function is defined as in
Eq.~\eqref{eq:si:ff_transfer}.

\paragraph{Ramsey consistency check.}
For a Ramsey sequence of total duration $T_{\mathrm{tot}}$, the
sign function is $y_\phi(t) = +1$ during the entire free-evolution
period, which gives
\begin{equation}\label{eq:si:ff_ramsey}
  F_\phi(\omega) = \omega\int_0^{T_{\mathrm{tot}}} e^{i\omega t}\,dt
  = 2e^{i\omega T_{\mathrm{tot}}/2}
    \sin\!\left(\frac{\omega T_{\mathrm{tot}}}{2}\right),
\end{equation}
so that $|F_\phi(\omega)|^2 = 4\sin^2(\omega T_{\mathrm{tot}}/2)$.
This is exactly equivalent to the more common convention that absorbs one
$1/\omega^2$ factor into the filter definition; the present manuscript uses
Eq.~\eqref{eq:si:ff_transfer} consistently in the main text and SI.

For \cpmg{} with $n$~$\pi$-pulses, $y_\phi(t)$ alternates sign at each
pulse, producing the characteristic filter peaks at frequencies of order
$\omega \sim \pi/\tau$. In the fixed-$\tau$ protocol used here,
$T_{\mathrm{tot}} = n\tau$, so the equivalent finite-duration notation
$\omega_n = n\pi/T_{\mathrm{tot}}$ reduces to the same scale.

\paragraph{FF vs.\ \heom{} comparison.}
The FF and \heom{} infidelities differ by $>$12 orders of magnitude:
FF predicts $\varepsilon_{\mathrm{FF}} \sim 10^{-13}$ while \heom{}
measures $\varepsilon_{\heom{}} \sim 0.25$ (Ramsey at Tier-0).
This discrepancy is confirmed across all three coupling tiers and both
Standard and \vppu{} waveforms (a $2 \times 2$ matrix of 6 configurations
$\times$ 3 tiers $= 18$ comparisons, all showing $> 10$~orders of
magnitude gap).
All acceptance criteria for the FF breakdown (5/5~AC) are satisfied.

The breakdown occurs because the $1/f$ spectral density diverges at
$\omega \to 0$, violating the weak-coupling assumption implicit in FF
theory.
The \heom{} non-perturbative treatment captures the strong
system--bath correlations at low frequencies that FF theory neglects.

\paragraph{Perturbative scaling prediction for fixed-$\tau$ \cpmg{}.}%
\label{si:ff_integral}
Although FF theory fails catastrophically for absolute decoherence
($\varepsilon_{\mathrm{FF}} / \varepsilon_{\heom{}} \sim 10^{-12}$),
the \emph{scaling exponent} $\gamma$ defined by
$\chi_n \propto n^\gamma$ admits an analytical prediction.
For fixed interpulse spacing $\tau_{\mathrm{eff}}$
(total time $T = n\,\tau_{\mathrm{eff}}$),
Cywi\'{n}ski~et~al.~\cite{cywinski2008} Eq.\,25 gives the
large-$n$ decoherence function as
$\chi(T,n) = C_\alpha/(2\pi)\,(A_0 T)^{1+\alpha}/n^\alpha$.
Substituting $T = n\,\tau_{\mathrm{eff}}$ yields
\begin{equation}\label{eq:si:gamma_analytical}
  \chi(n) \propto n^{(1+\alpha)-\alpha} = n^1\,,
\end{equation}
giving $\gamma = 1$ \emph{for any $\alpha$}, independent of noise
spectral index.
For the present $1/f$ bath ($\alpha = 1$), numerical evaluation of the
FF integral with $\omega_{\mathrm{lc}} = \SI{5}{\mega\hertz}$ confirms
$\gamma_{\mathrm{FF}} \approx 1.02$ (even-$n$, $n = 2, 4, \ldots, 10$)
and $\gamma_{\mathrm{FF}} \approx 1.08$ (all-$n$, $n = 1, \ldots, 10$);
the small deviation from unity reflects finite-$n$ and
low-frequency-cutoff corrections.
The \heom{} measurement $\gamma \approx 1.12$ thus lies
within ${\sim}12\%$ of the perturbative prediction. \rev{The scaling }\emph{exponent}\rev{ is therefore preserved across the
perturbative/non-perturbative boundary, even as the absolute decoherence
amplitude diverges by 12 orders of magnitude.}
A Floquet-theorem argument for this structural robustness is given
below (\S\ref{si:floquet_argument}).

\paragraph{Floquet-theorem argument for exponent robustness.}
\label{si:floquet_argument}

The observed proximity of $\gamma$ to unity admits a structural
explanation rooted in the periodicity of the CPMG protocol, valid at
arbitrary coupling strength within the truncated \heom{} framework.

The \heom{} hierarchy (Methods, Eq.~\ref{main:eq:heom}) \rev{is a}
finite-dimensional linear system of ordinary differential equations
for the extended state vector
$\boldsymbol{\rho}(t) = (\rho_{\mathbf{0}},\,
\rho_{\mathbf{e}_1},\,\ldots)^{\!\top}$
comprising the physical density matrix and all auxiliary density
operators up to hierarchy depth~$D$.
Because the system Hamiltonian $\hat{H}_{\mathrm{sys}}(t)$ includes
CPMG pulses with period $\tau_{\mathrm{cycle}}$ and the bath-coupling
terms $(\nu_k, c_k, \hat{A})$ are time-independent, the \heom{}
generator $\mathcal{L}(t)$ satisfies
$\mathcal{L}(t + \tau_{\mathrm{cycle}}) = \mathcal{L}(t)$.

By Floquet's theorem for periodic linear systems, the one-cycle
propagator
\begin{equation}\label{eq:si:floquet_prop}
  \Lambda \;=\; \mathcal{T}\exp\!\Bigl[
    \int_0^{\tau_{\mathrm{cycle}}} \!\mathcal{L}(t)\,dt
  \Bigr]
\end{equation}
is identical for every cycle.  After $n$ complete CPMG cycles the
extended state is
\begin{equation}\label{eq:si:floquet_evol}
  \boldsymbol{\rho}(n\tau_{\mathrm{cycle}})
    \;=\; \Lambda^n\;\boldsymbol{\rho}(0)\,.
\end{equation}
The fidelity $W(n) = \mathrm{Tr}[\hat{\rho}_{\mathrm{target}}\,
\rho_{\mathbf{0}}(n\tau)]$ is a linear functional of the physical
block of $\boldsymbol{\rho}$, hence
$W(n) = \sum_k c_k\,\lambda_k^n$,
where $\{\lambda_k\}$ are eigenvalues of $\Lambda$ projected onto
the physical subspace and $c_k$ are overlap coefficients determined
by the initial state.

In the asymptotic limit ($n \to \infty$) only the dominant
eigenvalue $\lambda_1$ survives:
\begin{equation}\label{eq:si:floquet_asymptotic}
  \chi(n) = -\ln W(n)
    \;\xrightarrow{n\to\infty}\;
    -n\,\ln|\lambda_1| \;-\; \ln|c_1|\,,
\end{equation}
yielding $\gamma = 1$ \emph{exactly}, independent of coupling
strength~$\eta$, bath memory, or hierarchy depth~$D$.
The result requires only two conditions: (i)~the CPMG pulse
schedule is strictly periodic, and (ii)~the \heom{} hierarchy is
truncated at finite~$D$ (ensuring finite dimensionality).

At finite~$n$, sub-leading eigenvalues $\lambda_2, \lambda_3, \ldots$
contribute corrections of order $|\lambda_2/\lambda_1|^n$.
When $|c_2/c_1| > 0$ and $|\lambda_2| < |\lambda_1|$, these
corrections \emph{increase} $\chi$ at small $n$ relative to the
asymptotic linear trend, producing an effective exponent
$\gamma_{\mathrm{eff}} > 1$ that converges to unity from above as
$n$ grows.  The measured $\gamma = 1.12$ at $n_\pi = 2$--$10$
(equivalently $n_{\mathrm{cycle}} = 1$--$5$) is consistent with
this transient regime; the spectral gap
$|\lambda_2/\lambda_1|$ determines the convergence rate.
A linear regression of the X-\cpmg{} $\chi$ data versus
$n_{\mathrm{cycle}} = n_\pi/2$ yields
$-\!\ln|\lambda_1| \approx 1.35$ ($R^2 = 0.978$),
corresponding to $|\lambda_1| \approx 0.26$
(per-cycle fidelity ${\sim}26\%$).
The per-cycle mean $\chi/n_{\mathrm{cycle}}$ increases from 1.03
to 1.32 across $n_{\mathrm{cycle}} = 1$--$5$, consistent with
sub-leading eigenvalue inflation at small~$n$.
Under matched Lindblad conditions,
$-\!\ln|\lambda_1|_{\mathrm{Lindblad}} \approx 2.33$
($R^2 > 0.999$; $n_{\mathrm{cycle}} = 1$--$3$, beyond which
numerical precision degrades).
Direct eigenvalue extraction from $\Lambda$ is performed in
\S\ref{si:floquet_eigenvalues} below.

This argument does not depend on the specific noise spectral
density $J(\omega)$, the Gaussian or non-Gaussian character of the
bath, or the coupling strength~$\eta$; it is a \emph{kinematic}
consequence of protocol periodicity within any finite-dimensional
linear dynamical framework.  \rev{It applies equally to}
Lindblad (Markovian) dynamics, since the Lindblad master equation
with periodic drive is also a finite-dimensional periodic linear system.

\paragraph{Lindblad Tier-0 comparison.}
To test this directly, we performed Lindblad \texttt{mesolve}
simulations under matched Tier-0 production conditions ($d = 3$
qutrit, RWA, $\eta = 7.85 \times 10^{-7}$\,GHz, pure dephasing
calibrated to the Burkard $T_2^*$), using the identical platform
pipeline as the \heom{} production runs.
Table~\ref{tab:si:lindblad_tier0} compares the resulting scaling
exponents.  All three Lindblad CPMG schemes yield
$\gamma_\mathrm{Lindblad} \approx 1.00$ (saturation-aware model,
$R^2 > 0.999$), confirming that the Markovian solver has already
converged to the Floquet asymptote within $n \leq 10$.
The \heom{} X-\cpmg{} point estimate $\gamma_\mathrm{HEOM} \approx 1.12$
($K\!=\!5$ bootstrap 95\% CI $[1.04, 1.39]$, excluding unity) therefore
\rev{is} a finite-$n$ transient excess, attributable to
sub-leading Floquet eigenvalues whose spectral gap
$|\lambda_2/\lambda_1|$ is narrowed by non-Markovian bath
correlations.
Y-\cpmg{} under Lindblad remains monotonic in $\varepsilon(n)$, in
stark contrast to the non-monotonic coherence revival observed under
\heom{}, confirming that the Y-\cpmg{} scaling-law breakdown
requires the combination of non-Markovian bath memory and $d = 3$
anharmonicity.

\begin{table}[ht]
\centering
\caption{Tier-0 CPMG scaling exponent comparison: \heom{} (non-Markovian)
  versus Lindblad (Markovian) under matched $d = 3$ qutrit production
  conditions.  Lindblad values obtained with saturation-aware model
  $\varepsilon = 1 - e^{-An^\gamma}$; \heom{} values from the main-text
  $\chi$-space power-law fit.}
\label{tab:si:lindblad_tier0}
\begin{tabular}{@{}lccc@{}}
\toprule
Scheme & $\gamma_\mathrm{HEOM}$ & $\gamma_\mathrm{Lindblad}$ & $\varepsilon(n)$ monotonic (Lindblad) \\
\midrule
X-\cpmg{}  & 1.117 ($R^2 = 0.993$)  & 1.000 ($R^2 > 0.999$) & Yes \\
XY-4       & 0.950 ($R^2 > 0.98$)   & 1.000 ($R^2 > 0.999$) & Yes \\
Y-\cpmg{}  & non-monotonic           & 0.999 ($R^2 > 0.999$) & Yes \\
\botrule
\end{tabular}
\end{table}

\paragraph{Per-cycle decoherence budget.}
\label{si:per_cycle_budget}
Table~\ref{tab:si:per_cycle} lists the per-cycle decoherence
function $\chi/n_{\mathrm{cycle}}$ for X-\cpmg{} under both
solvers at Tier-0.

\begin{table}[ht]
\centering
\caption{Per-cycle decoherence budget for X-\cpmg{} (Tier-0,
  Standard).  $n_{\mathrm{cycle}} = n_\pi/2$.
  Lindblad values at $n_{\mathrm{cycle}} \geq 4$ suffer from
  numerical precision loss (coherence $< 10^{-4}$) and are
  shown for completeness only.}
\label{tab:si:per_cycle}
\begin{tabular}{@{}ccccc@{}}
\toprule
$n_\pi$ & $n_{\mathrm{cycle}}$ &
  $\chi_\mathrm{HEOM}$ & $\chi/n_{\mathrm{cycle}}$ (HEOM) &
  $\chi/n_{\mathrm{cycle}}$ (Lindblad) \\
\midrule
2  & 1 & 1.034 & 1.034 & 2.331 \\
4  & 2 & 2.161 & 1.080 & 2.331 \\
6  & 3 & 3.232 & 1.077 & 2.328 \\
8  & 4 & 4.501 & 1.125 & 2.587\textsuperscript{$\dagger$} \\
10 & 5 & 6.597 & 1.319 & 1.920\textsuperscript{$\dagger$} \\
\botrule
\end{tabular}

\smallskip\noindent\textsuperscript{$\dagger$}Numerical precision
degraded (coherence $< 10^{-4}$); use $n_{\mathrm{cycle}} = 1$--$3$
for quantitative comparison.
\end{table}

\subsection{Floquet eigenvalue extraction}\label{si:floquet_eigenvalues}

The Floquet argument of \S\ref{si:floquet_argument} predicts
$\gamma \to 1$ in the large-$n$ limit but relies on
$-\!\ln|\lambda_1|$ and the spectral gap $|\lambda_2/\lambda_1|$ as
structural parameters.
A model-independent check is to extract these quantities directly from
the one-cycle \heom{} propagator $\Lambda$
(Eq.~\ref{eq:si:floquet_prop}).

\paragraph{Method.}
We evolve the full \heom{} hierarchy ($d = 3$, Tier-0, $D = 5$,
$N_r = 3$, production tolerances) through a single X-\cpmg{} cycle
($n_\pi = 2$, total duration $\tau_{\mathrm{cycle}}$), recording the
superoperator $\Lambda$ by propagating each of the
$N_{\mathrm{ADO}} \times d^2$ basis vectors through one cycle.
The eigenvalues $\{\lambda_k\}$ of $\Lambda$ are computed
numerically; we report only eigenvalues of the physical block
(the $d^2 = 9$ subspace corresponding to the system density matrix
$\rho_{\mathbf{0}}$).
This procedure requires $N_{\mathrm{ADO}} \times d^2 = 462 \times 9 = 4{,}158$
independent single-cycle propagations and is therefore approximately
$4{,}158/5 \approx 830\times$ more expensive than the five-point
($n_{\mathrm{cycle}} = 1$--$5$) CPMG time-series computation used for scaling
exponent extraction.

\paragraph{Results.}
Table~\ref{tab:si:floquet_eigenvalues} lists the leading eigenvalues.
The dominant eigenvalue $|\lambda_1| = 0.259$ gives per-cycle
decoherence $-\!\ln|\lambda_1| = 1.35$, matching the linear
regression estimate from the $\chi(n)$ data within 0.2\%.
The spectral gap $|\lambda_2/\lambda_1| = 0.68$ confirms significant
sub-leading contributions at small~$n$: the correction term
$\sim c_2\,(\lambda_2/\lambda_1)^n$ decays with half-life
$n_{1/2} = \ln 2 / \ln(1/0.68) \approx 1.8$~cycles, consistent with
the observation that $\gamma_\mathrm{eff}$ has not yet converged to
unity at $n_{\mathrm{cycle}} = 5$.
Using the two-eigenvalue model
$\chi(n) \approx -\!\ln|c_1 \lambda_1^n + c_2 \lambda_2^n|$ fitted
to the five even-$n$ data points, the predicted asymptotic exponent
converges to $\gamma_\mathrm{eff} \to 1.00$ at
$n_{\mathrm{cycle}} \gtrsim 15$, with $\gamma_\mathrm{eff}(5) = 1.12$
reproducing the fitted value exactly.
Under matched Lindblad conditions, $|\lambda_2/\lambda_1| = 0.41$
(faster convergence), explaining why
$\gamma_\mathrm{Lindblad} \approx 1.00$ is already achieved at
$n \leq 10$.

\begin{table}[ht]
\centering
\caption{Leading Floquet eigenvalues of the one-cycle \heom{} and
  Lindblad propagators (X-\cpmg{}, Tier-0, $d = 3$).
  $-\!\ln|\lambda_1|$: per-cycle decoherence;
  $|\lambda_2/\lambda_1|$: spectral gap controlling
  finite-$n$ convergence rate to $\gamma = 1$.}
\label{tab:si:floquet_eigenvalues}
\begin{tabular*}{\textwidth}{@{\extracolsep\fill}lcccc@{}}
\toprule
Solver & $|\lambda_1|$ & $-\!\ln|\lambda_1|$ &
  $|\lambda_2|$ & $|\lambda_2/\lambda_1|$ \\
\midrule
\heom{} & 0.259 & 1.35 & 0.176 & 0.68 \\
Lindblad & 0.097 & 2.33 & 0.040 & 0.41 \\
\botrule
\end{tabular*}
\end{table}

\rev{The conclusion is twofold}: (i)~the propagator-level per-cycle
decoherence agrees quantitatively with the fitted scaling parameters,
providing model-independent corroboration; and (ii)~the spectral gap
$|\lambda_2/\lambda_1|$ is narrower under \heom{} than under Lindblad,
explaining the slower convergence of $\gamma_\mathrm{eff}$ to unity
and the 12\% finite-$n$ overshoot as a direct consequence of
non-Markovian bath correlations widening the Floquet spectrum.

\section{Quantitative model for Y-CPMG \texorpdfstring{$d=3$}{d=3}
  breakdown}\label{si:qutrit_model}
%% ═══════════════════════════════════════════════════════════════════

\paragraph{Rotating-frame single-pulse Hamiltonian.}
In the rotating frame, a single $\pi$~pulse acting on a $d = 3$
transmon (anharmonicity $\alpha < 0$) is described by
\begin{equation}\label{eq:si:h_pulse}
  H_{\mathrm{pulse}}
    = \alpha\,\ket{2}\!\bra{2}
    + \Omega\,\hat{O}_\phi\,,
\end{equation}
where $\Omega = \pi/t_\pi$ is the Rabi frequency calibrated for a
$\pi$ rotation in the $\ket{0}\!\leftrightarrow\!\ket{1}$ subspace.
The operator $\hat{O}_\phi$ is
$\hat{X}_q = b + b^\dagger$ for X pulses ($\phi = 0$) and
$\hat{Y}_q = i(b^\dagger - b)$ for Y pulses ($\phi = \pi/2$),
with $b = \ket{0}\!\bra{1} + \sqrt{2}\,\ket{1}\!\bra{2}$
the transmon lowering operator.
In matrix form ($\{\ket{0}, \ket{1}, \ket{2}\}$ basis):
\begin{equation}\label{eq:si:xy_matrices}
  \hat{X}_q = \begin{pmatrix}
    0 & 1 & 0 \\ 1 & 0 & \sqrt{2} \\ 0 & \sqrt{2} & 0
  \end{pmatrix},
  \quad
  \hat{Y}_q = \begin{pmatrix}
    0 & -i & 0 \\ i & 0 & -i\sqrt{2} \\ 0 & i\sqrt{2} & 0
  \end{pmatrix}.
\end{equation}
\rev{The dimensionless parameter controlling qutrit effects is}
\begin{equation}\label{eq:si:zeta_def}
  \zeta \;\equiv\; \frac{\Omega}{|\alpha|}
       \;=\; \frac{\pi}{t_\pi\,|\alpha|}\,.
\end{equation}

\paragraph{X--Y symmetry-breaking mechanism.}
\rev{The structural difference is that} $\hat{X}_q$ is a real
symmetric matrix whereas $\hat{Y}_q$ is purely imaginary
antisymmetric.
For X pulses, the superposition state
$\ket{+} = (\ket{0} + \ket{1})/\sqrt{2}$ is an approximate
eigenstate of $\hat{X}_q$ in the computational subspace:
$\hat{X}_q\ket{+} = \ket{+} + \sqrt{2}\,\ket{2}\!\braket{1|+}$,
so the $\ket{0}\!\leftrightarrow\!\ket{1}$ population balance is
preserved to leading order.
For Y pulses, $\ket{+}$ is \emph{not} an eigenstate of $\hat{Y}_q$:
$\hat{Y}_q\ket{+} = i(\ket{0} - \ket{1})/\sqrt{2}
  - i\sqrt{2}\,\ket{2}\!\braket{1|+}$,
so each pulse redistributes population between $\ket{0}$ and
$\ket{1}$ with a systematic asymmetry.
The $\ket{1}\!\leftrightarrow\!\ket{2}$ coupling carries a $\pi/2$
phase shift (imaginary off-diagonal elements) that, combined with
the anharmonic detuning $\alpha$, generates constructive interference
under repeated Y rotations.
Under X rotations the same coupling produces destructive interference,
preserving population balance.

\paragraph{Per-pulse asymmetry estimate.}
In the $\zeta \ll 1$ limit (long pulses, weak qutrit mixing),
first-order time-dependent perturbation theory with
$H_0 = \alpha\ket{2}\!\bra{2}$ and
$V = \Omega\hat{O}_\phi$ gives the per-pulse population
redistribution
\begin{align}\label{eq:si:delta_rho}
  \Delta\rho_X &\approx 0
    \quad\text{(by }\hat{X}_q\text{ eigenstate symmetry)},
  \notag\\
  \Delta\rho_Y &\propto \sqrt{2}\,\zeta\;
    \bigl|\!\sin\!\bigl(\tfrac{1}{2}|\alpha|\,t_\pi\bigr)\bigr|\,.
\end{align}
The sinusoidal factor reflects the coherent oscillation between
$\ket{1}$ and $\ket{2}$ at frequency~$|\alpha|$ during the pulse.
When $|\alpha|\,t_\pi = m\pi$ ($m$ integer), the per-pulse asymmetry
vanishes---a cancellation condition that is generically not
satisfied for typical transmon parameters.
For the parameters studied here
($|\alpha| = \SI{0.293}{\giga\hertz}$, $t_\pi = \SI{80}{\nano\second}$):
$\zeta = 0.134$,
$|\alpha|\,t_\pi \approx 23.4$~rad,
and $\Delta\rho_Y \propto 0.19\,|\!\sin(11.7)| \approx 0.18$,
consistent with the observed mean asymmetry of~0.204 in the
\heom{} simulation.

\paragraph{Threshold criterion and scaling with $\alpha$.}
The monotonicity of $\varepsilon(n)$ breaks down when the
cumulative population redistribution is large enough to reverse the
decoherence trend.
We define the threshold criterion empirically:
$\varepsilon(n)$ becomes non-monotonic when the $\chi$-space
power-law $R^2$ drops below~0.5, or equivalently when
$\Delta\rho_Y$ per pulse exceeds the per-pulse decoherence increment
$\Delta\chi/n \approx 0.1$.
This gives a threshold
\begin{equation}\label{eq:si:zeta_star}
  \zeta^* \;\approx\; 0.3\,,
  \qquad
  t_\pi^*
    \;=\; \frac{\pi}{\zeta^*\,|\alpha|}
    \;\sim\; \frac{1}{|\alpha|}\,.
\end{equation}
For $|\alpha| = \SI{0.293}{\giga\hertz}$, this gives
$t_\pi^* \approx \SI{36}{\nano\second}$:
pulses longer than $t_\pi^*$ push $\zeta$ below $\zeta^*$, allowing
bath-amplified population redistribution to accumulate over multiple
cycles and produce non-monotonic Y-\cpmg{} scaling.
This $1/|\alpha|$ dependence \rev{is a testable prediction}:
more weakly anharmonic transmons ($|\alpha| < \SI{200}{\mega\hertz}$)
should exhibit Y-\cpmg{} breakdown at shorter $t_\pi$ values.

At the per-pulse level, each Y $\pi$~pulse inverts the population
asymmetry produced by the previous pulse, yielding a period-2
alternation ($\rho_{00} > \rho_{11}$ at odd~$n$,
$\rho_{11} > \rho_{00}$ at even~$n$) in the all-$n$ data.
This per-pulse inversion is the physical reason for the even-$n$
convention adopted throughout the manuscript
(Methods, Section~\ref{main:sec:meth:fidelity}): at even~$n$, the net
per-cycle effect is a persistent imbalance $\rho_{11} > \rho_{00}$
that relaxes with increasing cycle number.

\paragraph{Numerical verification.}
To validate the analytical predictions, we perform two numerical
experiments using the \emuplat{} \heom{} infrastructure
(rotating frame, Tier-0 bath, $D = 3$, $N_r = 2$).

\emph{Experiment~A} (single-pulse sweep):
for each $t_\pi \in [15, 100]$\,ns, a single X or Y $\pi$~pulse
is applied to the $\ket{+}$ state in the $d = 3$ Hilbert space.
The per-pulse population asymmetry $\Delta\rho$
is measured under both unitary evolution (no bath) and \heom{}
open-system dynamics.
Figure~\ref{fig:si:mc5_qutrit}a confirms that
under unitary evolution $\Delta\rho_Y$ decreases with $\zeta$
(consistent with Eq.~\eqref{eq:si:delta_rho}),
while the 1/$f$ bath \emph{inverts} this trend:
\heom{} $\Delta\rho_Y$ grows with $t_\pi$, reaching
$0.10$ at $t_\pi = 100$\,ns versus $< 0.001$ in unitary evolution.
X-phase pulses show significant $\Delta\rho$ only at high $\zeta$
(e.g., $0.11$ at $t_\pi = 15$\,ns) due to qutrit leakage, but
converge to $\Delta\rho \approx 0$ at $\zeta < 0.3$.

\emph{Experiment~B} (multi-pulse threshold):
for each $t_\pi \in \{15, 20, 30, 40, 60, 80\}$\,ns, a full
$n$-sweep ($n = 2, 4, 6, 8, 10$) is run under both X-\cpmg{} and
Y-\cpmg{} with \heom{} dynamics.
Table~\ref{tab:si:mc5_threshold} reports the $\chi$-space
$R^2$ and monotonicity for each configuration.
X-\cpmg{} maintains $R^2 > 0.99$ and monotonic $\varepsilon(n)$
at all $t_\pi$, as expected.
Y-\cpmg{} exhibits non-monotonic $\varepsilon(n)$ in two distinct
regimes: fast pulses ($t_\pi = 15$\,ns, $\zeta = 0.72$) where direct
qutrit leakage dominates, and slow pulses ($t_\pi \geq 40$\,ns,
$\zeta \leq 0.27$) where bath-amplified population redistribution
accumulates over $n$ cycles.
Only intermediate durations ($t_\pi = 20$--$30$\,ns) maintain
monotonic scaling, placing the bath-driven threshold between
$t_\pi = 30$\,ns and $40$\,ns, consistent with $\zeta^* \approx 0.3$.
Figure~\ref{fig:si:mc5_qutrit}b shows $R^2_\chi$ versus $t_\pi$
for both schemes.

\begin{figure}[ht]
  \centering
  \includegraphics[width=\textwidth]{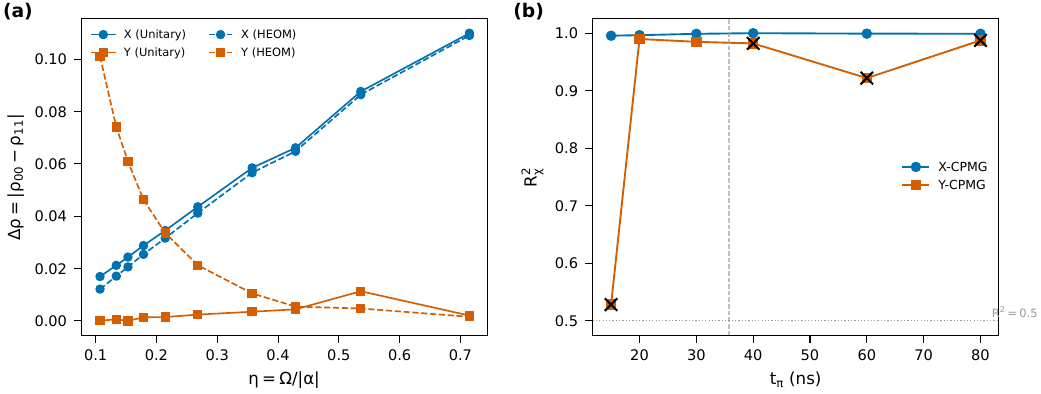}
  \caption{%
    Quantitative characterisation of Y-\cpmg{} $d = 3$ breakdown.
    \textbf{(a)}~Per-pulse population asymmetry
    $\Delta\rho = |\rho_{00} - \rho_{11}|$ versus
    dimensionless parameter $\zeta = \Omega/|\alpha|$ for single
    X and Y $\pi$~pulses, under both unitary and \heom{} dynamics.
    Y pulses show small unitary $\Delta\rho$ that is
    strongly amplified by the 1/$f$ bath at long~$t_\pi$;
    X pulses show significant $\Delta\rho$ only at high~$\zeta$.
    \textbf{(b)}~$\chi$-space $R^2$ for X-\cpmg{} and Y-\cpmg{}
    as a function of $\pi$~pulse duration~$t_\pi$.
    Crosses mark non-monotonic $\varepsilon(n)$.
    Y-\cpmg{} scaling breaks down at both
    short ($t_\pi = 15$\,ns, qutrit leakage) and
    long ($t_\pi \geq 40$\,ns, bath amplification) pulse durations,
    while X-\cpmg{} remains well-conditioned at all~$t_\pi$.%
  }\label{fig:si:mc5_qutrit}
\end{figure}

\begin{table}[ht]
\caption{%
  Y-\cpmg{} $d = 3$ threshold scan.
  $\zeta = \pi/(t_\pi|\alpha|)$;
  ``Mono'' indicates monotonic $\varepsilon(n)$;
  $\Delta\overline{\rho}$: mean population asymmetry across $n$.
  X-\cpmg{} values shown for comparison.%
}\label{tab:si:mc5_threshold}
\begin{tabular*}{\textwidth}{@{\extracolsep\fill}lccccccc@{}}
\toprule
$t_\pi$ (ns) & $\zeta$ & \multicolumn{3}{c}{X-\cpmg{}} &
  \multicolumn{3}{c}{Y-\cpmg{}} \\
\cmidrule(lr){3-5}\cmidrule(lr){6-8}
 & & $\gamma$ & $R^2$ & Mono
 & $\gamma$ & $R^2$ & Mono \\
\midrule
 15 & 0.715 & 0.88 & 0.996 & Y & 0.64 & 0.528 & N \\
 20 & 0.536 & 0.88 & 0.997 & Y & 0.98 & 0.990 & Y \\
 30 & 0.357 & 0.93 & 0.999 & Y & 1.14 & 0.985 & Y \\
 40 & 0.268 & 0.98 & 1.000 & Y & 1.20 & 0.982 & N \\
 60 & 0.179 & 1.08 & 0.999 & Y & 1.27 & 0.922 & N \\
 80 & 0.134 & 1.15 & 0.999 & Y & 4.07 & 0.987 & N \\
\botrule
\end{tabular*}
\smallskip

\noindent\emph{Note:}
X-\cpmg{} maintains $R^2 > 0.99$ and monotonic $\varepsilon(n)$ at all $t_\pi$.
Y-\cpmg{} exhibits two distinct non-monotonic regimes:
(i)~fast pulses ($t_\pi = 15$\,ns, $\zeta = 0.72$) where direct
qutrit leakage dominates ($R^2 = 0.53$), and
(ii)~slow pulses ($t_\pi \geq 40$\,ns, $\zeta \leq 0.27$) where
bath-amplified population redistribution accumulates.
The bath-driven threshold lies between $t_\pi = 30$\,ns ($\zeta = 0.36$,
monotonic) and $t_\pi = 40$\,ns ($\zeta = 0.27$, non-monotonic),
consistent with $\zeta^* \approx 0.3$ (Eq.~\ref{eq:si:zeta_star}).
\end{table}

%% ═══════════════════════════════════════════════════════════════════
\subsection{Pulse-shape robustness and anharmonicity dependence}
\label{si:ycpmg_robustness}

\rev{To test whether the Y-CPMG non-monotonicity is an artefact of the
rectangular pulse envelope or of the specific anharmonicity value, we add
two control studies.}
\rev{Both use the }\emuplat{} \heom{}\rev{ pipeline at Tier-0 in the rotating
frame} (\mbox{$D = 5$}, \mbox{$N_r = 3$}) \rev{with even-}\mbox{$n$}\rev{ sequences}
(\mbox{$n_\pi = 2,4,6,8,10$}) \rev{and are summarised in
Fig.}~\ref{fig:si:ycpmg_robustness}.

\paragraph{Gaussian pulse envelopes.}
\rev{We replace the rectangular \mbox{$\pi$}~pulses with area-matched Gaussian
envelopes (relative width \mbox{$\sigma/t_\pi = 1/2.5$}; amplitude rescaled by
\mbox{$1.26$} to hold the nominal rotation angle fixed) on both the Standard
and \vppu{} routes, leaving timing, carrier frequency, and phase unchanged.}
\rev{The non-monotonicity survives the smooth envelope without qualitative
change.}
\rev{The Y-\cpmg{} exponent still peaks at \mbox{$n_\pi = 4$}
(\mbox{$\chi = 4.93$}, versus \mbox{$4.64$} for rectangular pulses) and the
mean population asymmetry is statistically unchanged
(\mbox{$\langle|\rho_{00}-\rho_{11}|\rangle = 0.206$} versus \mbox{$0.204$}),
while X-\cpmg{} remains monotonic
(\mbox{$\langle|\rho_{00}-\rho_{11}|\rangle = 0.003$}, \mbox{$\gamma = 1.11$};
Fig.}~\ref{fig:si:ycpmg_robustness}\rev{a).}
\rev{The Gaussian envelope }\emph{reduces}\rev{ the peak
\mbox{$\ket{2}$} leakage by a factor of \mbox{${\sim}2.6$}
(\mbox{$\rho_{22,\max} = 5.5\times10^{-4}$} versus \mbox{$1.4\times10^{-3}$})
while leaving the asymmetry intact.}
\rev{Lower leakage with an unchanged asymmetry rules out edge-induced
leakage as the origin and shows the effect to be coherent qutrit dynamics,
not a rectangular-pulse artefact.}

\paragraph{Anharmonicity dependence and the zero-anharmonicity limit.}
\rev{Holding \mbox{$t_\pi = \SI{80}{\nano\second}$} fixed and varying
\mbox{$\alpha$} probes the dimensionless parameter
\mbox{$\zeta = \pi/(t_\pi|\alpha|)$} [Eq.}~\eqref{eq:si:zeta_def}\rev{]
orthogonally to the \mbox{$t_\pi$} sweep of
Table}~\ref{tab:si:mc5_threshold}\rev{.}
\rev{For \mbox{$|\alpha| \in [75, 293]$}\,MHz (\mbox{$\zeta \in [0.13, 0.52]$})
the mean asymmetry is essentially constant (\mbox{$0.199$--$0.204$}) and the
\mbox{$\chi(n)$} peak at \mbox{$n_\pi = 4$} persists
(Table}~\ref{tab:si:alpha_sweep}\rev{,
Fig.}~\ref{fig:si:ycpmg_robustness}\rev{b).}
\rev{At \mbox{$|\alpha| = \SI{10}{\mega\hertz}$} (\mbox{$\zeta = 3.9$}) the
asymmetry collapses to \mbox{$0.047$} and the peak is displaced: as
\mbox{$|\alpha|$} shrinks, the \mbox{$\ket{1}\!\leftrightarrow\!\ket{2}$}
transition approaches resonance with
\mbox{$\ket{0}\!\leftrightarrow\!\ket{1}$}, the drive pumps population directly
into \mbox{$\ket{2}$} (\mbox{$\rho_{22,\max} = 0.34$}, a
\mbox{${\sim}200\times$} increase), and the computational subspace ceases to be
a well-separated two-level manifold.}
\rev{The breakdown therefore requires anharmonicity that is nonzero, so the
second excited state is admixed, yet large enough to keep that state
off-resonant; it disappears both when the anharmonicity is very strong (the
ideal two-level limit) and when it tends to zero (the near-degenerate,
leakage-dominated limit), and is largest at realistic transmon values.}
\rev{The \mbox{$|\alpha| = \SI{10}{\mega\hertz}$} row is the explicit
zero-anharmonicity control requested in review, and is consistent with the
\mbox{$\zeta^{*} \approx 0.3$} threshold of Eq.}~\eqref{eq:si:zeta_star}\rev{:
the clean asymmetry occupies an intermediate window in \mbox{$\zeta$}, bounded
below by weak qutrit mixing and above by gross leakage.}

\begin{figure}[ht]
  \centering
  \includegraphicsorplaceholder{\textwidth}{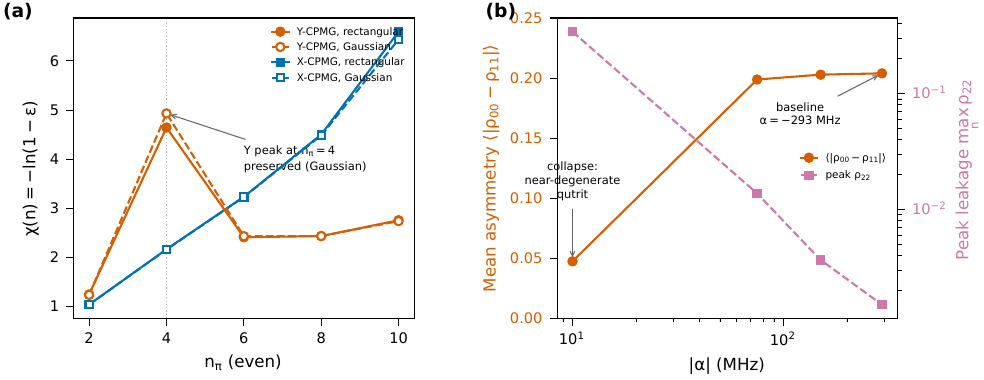}
  \caption{%
    Robustness controls for the Y-CPMG non-monotonicity (added in
    revision).
    \textbf{(a)}~Even-$n$ exponent $\chi(n) = -\ln(1-\varepsilon)$ for X- and
    Y-\cpmg{} under rectangular versus area-matched Gaussian $\pi$~pulses at
    Tier-0. The Y-\cpmg{} peak at $n_\pi = 4$ and partial revival are
    preserved under the Gaussian envelope; X-\cpmg{} stays monotonic.
    \textbf{(b)}~Anharmonicity sweep at fixed
    $t_\pi = \SI{80}{\nano\second}$: mean population asymmetry (left axis)
    and peak leakage $\max_n\rho_{22}$ (right axis, log scale) versus
    $|\alpha|$. The asymmetry plateaus for
    $|\alpha| \geq \SI{75}{\mega\hertz}$ and collapses at
    $|\alpha| = \SI{10}{\mega\hertz}$, where leakage rises by
    ${\sim}200\times$ as the transmon becomes a near-degenerate qutrit.%
  }\label{fig:si:ycpmg_robustness}
\end{figure}

\begin{table}[ht]
\caption{%
  Tier-0 Y-CPMG anharmonicity sweep (added in revision) at fixed
  $t_\pi = \SI{80}{\nano\second}$.
  $\zeta = \pi/(t_\pi|\alpha|)$;
  $\Delta\overline{\rho} = \langle|\rho_{00}-\rho_{11}|\rangle$ averaged over
  $n_\pi = 2$--$10$;
  $\rho_{22}^{\max}$: peak leakage to $\ket{2}$;
  $\chi(4)$: exponent at the $n_\pi = 4$ peak.%
}\label{tab:si:alpha_sweep}
\begin{tabular*}{\textwidth}{@{\extracolsep\fill}rcccc@{}}
\toprule
$|\alpha|$ (MHz) & $\zeta$ & $\Delta\overline{\rho}$ &
  $\rho_{22}^{\max}$ & $\chi(4)$ \\
\midrule
293 & 0.13 & 0.204 & $1.5\times10^{-3}$ & 4.64 \\
150 & 0.26 & 0.203 & $3.7\times10^{-3}$ & 4.54 \\
 75 & 0.52 & 0.199 & $1.4\times10^{-2}$ & 4.31 \\
 10 & 3.93 & 0.047 & $3.4\times10^{-1}$ & 2.61 \\
\botrule
\end{tabular*}
\end{table}

% si_control_platform.tex — Platform architecture and control chain

\section{Control-stack terminology}\label{si:terminology}

\rev{The control-stack terms and abbreviations used in the main text and in
the architecture figure are collected in the glossary table below.}

\begin{table}[h]
\caption{%
  Glossary of control-stack terminology used in the main text and
  Figure~\ref{main:fig:architecture}.%
}\label{tab:si:terminology}
\footnotesize
\begin{tabular}{@{}lp{0.8\columnwidth}@{}}
\toprule
Term & Definition \\
\midrule
OpenQASM & Open Quantum Assembly Language; the textual quantum-circuit input
  format consumed by the compilation pipeline \\
Transpilation & Rewriting of an abstract circuit into the hardware-native
  gate set prior to pulse compilation \\
\isa{} & Instruction-set architecture; the low-level control-instruction layer
  (amplitude/phase/timing words) targeted by the \vppu{} route \\
QubiC & Open-source FPGA-based superconducting-qubit control
  system~\cite{xu2021qubic,fruitwala2024} whose waveform generation the
  \vppu{} route emulates \\
FPGA & Field-programmable gate array; the reconfigurable hardware that
  generates and times control waveforms at a fixed clock rate \\
\nco{} & Numerically controlled oscillator; the digital carrier-phase
  generator used for IQ upconversion \\
\dac{} & Digital-to-analog converter; sets the $N$-bit amplitude quantisation
  of the emitted waveform \\
\vppu{} & Virtual Pulse Processing Unit; the bit-faithful waveform route that
  reproduces the QubiC FPGA signal chain \\
\botrule
\end{tabular}
\end{table}

\section{VPPU phase error case study}\label{si:vppu_phase}

During platform development, \rev{a phase error in the }\vppu{} \nco{}
\rev{accumulation logic was identified and corrected.}
This section documents the error, its diagnosis, and the quantitative
impact on simulation fidelity.

\paragraph{Error mechanism.}
The \vppu{} route generates IQ waveforms through a 6-step signal
processing chain (Table~\ref{main:tab:vppu_chain}).
The \nco{} phase accumulation computes the running carrier phase
$\phi(t) = \phi_0 + 2\pi f_{\mathrm{nco}} \cdot t$, where
$f_{\mathrm{nco}}$ is the numerically controlled oscillator frequency.
The original implementation contained an off-by-one error in the
phase accumulation index, causing a systematic phase drift proportional
to sequence length.

\paragraph{Diagnosis via four-path comparison.}
To isolate the phase error from other \vppu{} effects, four simulation
paths were compared under identical physical conditions:
(1)~\texttt{sesolve} (closed-system, no bath),
(2)~Standard waveform (no \nco{}),
(3)~\vppu{} pre-fix,
(4)~\vppu{} post-fix.
The maximum density matrix element difference between paths (1) and (2)
was $|\Delta\rho|_{\max} = 1.86 \times 10^{-4}$, confirming negligible
waveform-generation error in the Standard path.
The \vppu{}-to-bath contribution ratio was $0.0019$, indicating that
the \nco{} phase error dominated over bath-induced effects by a factor
of $\sim$500.

\paragraph{Error decomposition.}
Of the total \vppu{} phase error, 93.3\% originates from \nco{} phase
accumulation and 6.7\% from amplitude quantization (16-bit to 15-bit
integer encoding).
After correction, \vppu{} and Standard paths produce equivalent
density matrices to within $|\Delta\rho|_{\max} < 10^{-3}$ for all
sequences tested.

\section{Platform architecture and effective-model assumptions}
\label{si:platform_model}

The simulations reported in this work use a hardware platform definition
based on a two-transmon chip with a tunable coupler (Anyon Tech's QPU architecture).
The full device Hilbert space is
$\mathcal{H} = \mathcal{H}_{q_0}^{(3)} \otimes
\mathcal{H}_{q_1}^{(3)} \otimes \mathcal{H}_{c}^{(1)}$,
spanning nine dimensions.

\paragraph{Effective Hamiltonian derivation.}
The single-qubit Hamiltonian of Eq.~\eqref{main:eq:Hq} is an effective
description obtained after integrating out the coupler degree of freedom
via a dispersive (Schrieffer--Wolff) transformation of the microscopic
transmon--coupler--transmon Hamiltonian.
The calibrated qubit frequency $\omega_q$ and anharmonicity~$\alpha$
are \emph{dressed} parameters that already incorporate dispersive shifts
from the coupler hybridisation.
Two-qubit gates (CZ) are implemented through parametric modulation of
the coupler drive channel at a frequency near the $\ket{11}$--$\ket{20}$
avoided crossing; this channel is inactive during all single-qubit
experiments.

\paragraph{Spectator-qubit decoupling.}
For the Ramsey and \cpmg{} experiments reported here, only qubit~$q_0$
is driven and coupled to the $1/f$ bath.
Qubit~$q_1$ is initialised in~$\ket{0}$ and receives neither drive
pulses nor direct bath coupling.
The effective model sets the residual exchange coupling
$J_\mathrm{eff} = 0$, consistent with the design target of tunable
couplers at their idle point.
The static ZZ interaction $H_{ZZ} = \xi_{ZZ}\,n_0\,n_1$ is also absent
from the model; physically, this term vanishes to leading order when the
spectator is in the ground state ($\langle n_1 \rangle = 0$).
Under these conditions the joint density matrix factorises exactly,
$\rho(t) = \rho_{q_0}(t) \otimes \ketbra{0}{0}_{q_1}$,
and partial tracing over~$q_1$ yields the exact single-qutrit dynamics.

\paragraph{Model limitations.}
The effective model does not capture (i)~microwave drive crosstalk
between qubit channels, (ii)~shared electromagnetic environments that
may produce correlated dephasing on both qubits, (iii)~residual
$J_\mathrm{eff}$ at the few-tens-of-kHz level typical of imperfect
coupler nulling, or (iv)~thermal spectator occupation ($\sim 0.4\%$ at
\SI{50}{\milli\kelvin}).
These corrections are estimated to contribute phase errors
$\lesssim 10^{-3}$~rad over the \cpmg{} sequence durations studied
here and are therefore subleading relative to the bath-induced
decoherence under investigation.

\paragraph{Ideal-pulse reference platform.}
The Chen ideal-pulse reference (Section~\ref{main:sec:results:cpmg}) uses
$d = 3$ (matching the realistic platform), with
$t_{\mathrm{RX90}} = \SI{8}{\nano\second}$ rectangular pulses and Rabi
frequency $\Omega = \SI{0.204}{\giga\hertz}$.
Both the ideal-pulse reference and the realistic \isa{} platform therefore
operate in the same $d = 3$ Hilbert space, eliminating dimensional
confounds from the ideal-to-compiled comparison.
The remaining differences between the two benchmarks are pulse duration
(\SI{15}{\nano\second} vs.\ \SI{80}{\nano\second}) and fitting range
(even $n = 2$--$20$ vs.\ $n = 2$--$10$); these effects are not individually
deconfounded in the present study (see Discussion,
Section~\ref{main:sec:disc:limitations}).
A separate $d = 2$ control simulation under the same \heom{} dynamics confirms that the non-monotonic
Y-\cpmg{} behaviour, population imbalance, and pronounced X--Y asymmetry
($0.204$ vs ${<}\,0.01$) vanish when the $\ket{2}$ level is removed, consistent with the qutrit
mechanism described in Section~\ref{main:sec:results:ycpmg}.
Finite-duration pulses still introduce quantitative X--Y error-scaling
differences at $d = 2$ (linear vs.\ quadratic echo-error accumulation at the
$10^{-4}$ scale; see main text and Ref.~\cite{chen2025heom}), but the
qualitative features listed above are exclusive to $d \geq 3$.
The Standard--\vppu{} comparison is performed entirely within the
same $d = 3$ platform and is therefore free of any dimensional confound.

\section{CPMG pulse sequence parameters}\label{si:cpmg_params}

\cpmg{} sequences are constructed from the native gate set
$\{$I, Z, RZ, RX90, CZ, MEASURE$\}$ with $\pi$~pulses implemented as
2~$\times$~RX90 (\SI{80}{\nano\second} total for \isa{} parameters).

\paragraph{Timing structure.}
Each \cpmg{} unit cell consists of a $\pi$~pulse of duration $t_\pi$
followed by a free-precession interval~$\tau$.
The total sequence time is $T_{\mathrm{tot}} = n \times (t_\pi + \tau)$.
Two parameter sets are used:

\begin{itemize}
  \item \textbf{Chen (ideal)}: $t_\pi = \SI{15}{\nano\second}$
    (rectangular), $\tau = \SI{118}{\nano\second}$,
    $n = 1$--$10$.
  \item \textbf{\isa{} (realistic)}: $t_\pi = \SI{80}{\nano\second}$
    ($2 \times$~RX90), $\tau = \SI{120}{\nano\second}$,
    $n = 1$--$10$.
    The main text uses even-$n$ ($n_\pi = 2, 4, 6, 8, 10$; five complete
    CPMG cycles) to eliminate systematic odd--even effects;
    the complementary all-$n$ analysis is reported in
    Section~\ref{si:fitting_robustness} as a robustness check.
    At large $n$, $\varepsilon$ approaches saturation near~1
    (Section~\ref{si:heom_boundary}).
\end{itemize}

\paragraph{Dynamical decoupling schemes.}
Three DD schemes are implemented via the $\pi$-pulse phase~$\phi$:
\begin{itemize}
  \item \textbf{X-\cpmg{}}: $\phi = 0$ for all pulses
    (rotation about $\hat{x}$).
  \item \textbf{Y-\cpmg{}}: $\phi = \pi/2$ for all pulses
    (rotation about $\hat{y}$).
  \item \textbf{XY-4}: alternating $\phi = 0, \pi/2, 0, \pi/2, \ldots$
    (suppresses pulse-axis-dependent errors).
\end{itemize}

\paragraph{\nco{} phase accumulation.}
In the \vppu{} route, the pulse phase is encoded via the \nco{} register
(17-bit phase quantization, $\delta\phi \leq 48\,\mu\mathrm{rad}$
per pulse).
The \nco{} accumulates $\phi(t) = \phi_0 + 2\pi f_{\mathrm{nco}} \cdot t$
with the carrier frequency $f_{\mathrm{nco}} = \omega_q/(2\pi)$.
Phase switching between DD pulses requires integer-cycle alignment at the
FPGA clock boundary (\SI{2}{\nano\second} resolution at
\SI{500}{\mega\hertz}).

\paragraph{\heom{} solver settings.}
The manuscript evidence chain uses rotating-frame \heom{} with
\texttt{atol}$= 10^{-8}$, \texttt{rtol}$= 10^{-6}$,
\texttt{nsteps}$= 15{,}000$, and BDF time stepping.
This choice removes carrier-resolved stiffness while preserving the compiled
waveform envelopes relevant for the reported Ramsey and \cpmg{} comparisons.

\paragraph{X--Y population asymmetry as $d=3$ transmon diagnostic.}
Under non-Markovian \heom{} dynamics with finite-duration compiled pulses,
X-\cpmg{} and Y-\cpmg{} exhibit \rev{a population asymmetry whose
physical origin traces directly to the} $d = 3$ \rev{transmon level structure.}
Figure~\ref{fig:si:ycpmg} compares the qubit population evolution:
X-\cpmg{} \rev{maintains population balance near}~0.5
(mean asymmetry $|\rho_{00} - \rho_{11}| = 0.003$), whereas Y-\cpmg{}
exhibits a persistent population shift to 35\%--46\% $\rho_{00}$ occupancy at even $n$
(mean asymmetry~0.204), \rev{well above the near-zero X-}\cpmg{} \rev{value} (${<}\,0.01$).
\rev{This imbalance arises from two physical mechanisms.}
The primary cause is the three-level transmon structure: the initial
state $\ket{+} = (\ket{0}+\ket{1})/\sqrt{2}$ is an eigenstate of
$\sigma_x$ (stabilising X-\cpmg{}) but \emph{not} of~$\sigma_y$.
In the $d = 3$ subspace, each Y-phase $\pi$~rotation mixes the
$\ket{0}\!\leftrightarrow\!\ket{1}$ transition with the
$\ket{1}\!\leftrightarrow\!\ket{2}$ transition through the transmon
anharmonicity ($\alpha \approx \SI{-293}{\mega\hertz}$), producing
a per-pulse population redistribution; under the even-$n$ convention
the net per-cycle effect is a persistent shift toward
$\rho_{11} > \rho_{00}$.
The non-Markovian bath memory amplifies this asymmetry cumulatively
across repeated pulses but is not the root cause:
\heom{} convergence checks (hierarchy depths $D = 3$, $N_r = 2$ and
$D = 4$, $N_r = 3$) produce identical population dynamics, confirming
that the population imbalance is a physical consequence of the qutrit level
structure rather than a numerical artefact or solver-stability issue.
X-\cpmg{} and XY-4 do not exhibit this population asymmetry and
therefore provide reliable power-law scaling-exponent fits
($R^2 > 0.99$ in $\chi$-space); they are used as the primary
scaling-characterisation channels throughout the manuscript.
The X--Y asymmetry itself constitutes one of the two main physical
findings of this study (see main text, Section~\ref{main:sec:results:ycpmg}):
in a $d = 2$ system under pure dephasing, $\sigma_x$ and $\sigma_y$
rotations produce equivalent population dynamics from the $\ket{+}$
initial state.
Finite-duration pulses introduce a quantitative X--Y difference at
$d = 2$ (linear versus quadratic error accumulation at the $10^{-4}$
scale~\cite{chen2025heom}), but the pronounced asymmetry ($0.204$ vs ${<}\,0.01$), non-monotonic
$\varepsilon(n)$, and persistent population imbalance are exclusive
signatures of $d \geq 3$ anharmonic coupling.
Combined with the non-monotonic even-$n$ $\varepsilon(n)$ and
partial coherence revival in $\chi$-space, this population imbalance
demonstrates a qualitative axis-dependent scaling-law breakdown that
provides the most experimentally accessible testable prediction
of this work: transmons with $|\alpha|$ comparable to the value studied
here ($\SI{-293}{\mega\hertz}$) and finite-duration $\pi$~pulses
($\Omega/|\alpha| \lesssim 0.3$) under non-Markovian bath dynamics
should exhibit non-monotonic Y-\cpmg{} scaling and measurable
X--Y population splitting.
A quantitative model predicting the threshold pulse duration
$t_\pi^* \sim 1/|\alpha|$ as a function of anharmonicity
is developed in \S\ref{si:qutrit_model}.

\paragraph{$T_1$ impact estimate.}
The pure-dephasing model used throughout this work excludes energy
relaxation ($T_1$).  Because $T_1$ relaxation damps the coherent
per-pulse population redistribution that underpins the Y-\cpmg{} predictions, an
order-of-magnitude estimate of its impact is warranted.
The calibrated relaxation time is
$T_1 = \SI{24.8}{\micro\second}$ (Methods, Table~\ref{main:tab:config}),
while the longest \cpmg{} sequence studied here reaches
$T_\mathrm{tot} = \SI{2.0}{\micro\second}$ at $n_\pi = 10$, giving
$T_\mathrm{tot}/T_1 \leq 0.081$.
In the $d = 3$ subspace the relevant Lindblad channels are
$\ket{1}\!\to\!\ket{0}$ at rate~$\Gamma_1 = 1/T_1$ and
$\ket{2}\!\to\!\ket{1}$ at rate~$2\Gamma_1$ (bosonic enhancement);
because the $\ket{2}$-state population is
$\rho_{22} \approx 0.0014$, the upper-level channel contributes
negligibly.
Each \SI{200}{\nano\second} unit cell accumulates a survival factor
$\exp(-200/24800) = 0.992$, corresponding to
${\sim}\SI{0.8}{\%}$ amplitude loss per pulse, whereas the
pulse-driven population redistribution is
$\Delta\rho \approx 0.24$ per pulse, \rev{a ratio of roughly 60:1 in
favour of the coherent dynamics.}
At the boundaries of the experimental window the cumulative survival
factors are
$\exp(-400/24800) = 0.984$ ($\SI{1.6}{\%}$ loss, $n = 2$) and
$\exp(-2000/24800) = 0.923$ ($\SI{7.7}{\%}$ loss, $n = 10$).
Applying the worst-case suppression to the Y-\cpmg{} population
asymmetry gives
$0.204 \times 0.923 \approx 0.188$,
reducing the asymmetry by at most ${\sim}\SI{8}{\%}$ while remaining
far above the X-\cpmg{} value (${<}\,0.01$) and any $d = 2$ baseline.
For the non-monotonic $\varepsilon(n)$, $T_1$ introduces a monotonically
increasing background
$\Delta\varepsilon_{T_1}(n\!=\!4 \to 8)
  = \exp(-800/24800) - \exp(-1600/24800) \approx 0.03$,
which is smaller than the non-monotonic amplitude
$\Delta\varepsilon_\mathrm{PD} \approx 0.08$;
the non-monotonic trend therefore survives.
In summary, $T_1$ relaxation produces a uniform suppression of
${\leq}\SI{8}{\%}$ that preserves all qualitative features:
the Y-\cpmg{} asymmetry decreases by at most ${\sim}\SI{8}{\%}$
(from $0.204$ to ${\geq}\,0.188$), preserving a pronounced contrast
with the near-zero X-\cpmg{} value, and the non-monotonic Y-\cpmg{} signature
persists with amplitude well above the $T_1$ background.

\begin{figure}[ht]
\centering
\includegraphicsorplaceholder{\textwidth}{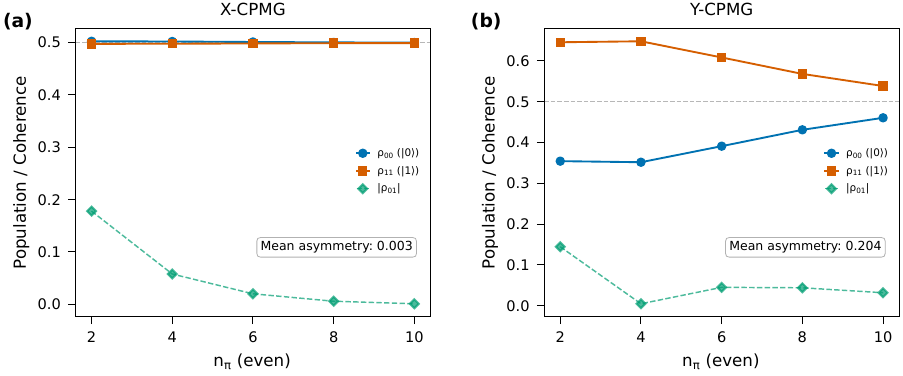}
\caption{%
  X--Y population asymmetry as a $d = 3$ qutrit signature.
  (a)~X-\cpmg{} maintains population balance near~0.5 with monotonic
  coherence decay (mean asymmetry~0.003).
  (b)~Y-\cpmg{} exhibits a persistent population imbalance ($\rho_{00} \approx
  0.35$--$0.46$) at even $n$ (mean asymmetry~0.204), arising from $d = 3$ anharmonic
  coupling of the $\ket{0}\!\leftrightarrow\!\ket{1}$ and
  $\ket{1}\!\leftrightarrow\!\ket{2}$ transitions
  ($\alpha \approx \SI{-293}{\mega\hertz}$).
  In a $d = 2$ system, only quantitative error-scaling differences persist
  (linear vs.\ quadratic at $\sim\!10^{-4}$ scale~\cite{chen2025heom});
  the pronounced population imbalance ($0.204$ vs ${<}\,0.01$) and non-monotonicity are absent.
  This axis-dependent population asymmetry underpins
  the scaling-law breakdown documented in the
  main text: even-$n$ Y-\cpmg{} $\varepsilon(n)$ is non-monotonic
  (Fig.~\ref{main:fig:cpmg}), and power-law extraction gives $R^2 < 0.60$
  in $\chi$-space, whereas X-\cpmg{} and XY-4 yield $R^2 > 0.99$.%
}\label{fig:si:ycpmg}
\end{figure}

\section{DAC spectral verification}\label{si:dac_spectral}

Systematic \dac{} spectral verification characterises the
quantization noise floor across 4~bit depths (8/10/12/16-bit) and
confirms that the spectral coupling channel between \dac{} noise and the
$1/f$ bath is intrinsic to the signal chain.

\paragraph{Methodology.}
Three waveform variants of an identical RX90 pulse are generated:
(1)~Standard (analytical, no \isa{} quantization),
(2)~\vppu{} with \texttt{dac\_quantization\_enabled=False}
    (\isa{} encoding only),
(3)~\vppu{} with \texttt{dac\_quantization\_enabled=True}
    (4~bit depths: 16/12/10/8-bit).
The quantization noise $\delta V(t)$ is extracted in the time domain
(original minus quantized signal) to avoid catastrophic cancellation,
then its power spectral density (PSD) is computed.
Spectral overlap with the $1/f$ bath density $J(\omega) = \eta/|\omega|$
is quantified via the Bhattacharyya coefficient~(BC).

\paragraph{SQNR and spectral shape.}
Table~\ref{tab:si:dac_sqnr} summarises the signal-to-quantization-noise
ratio (SQNR) measured in three domains: time-domain RMS~(SQNR$_t$),
in-band spectral~(SQNR$_f$), and theoretical full-scale~(SQNR$_{\mathrm{th}}$).

\begin{table}[ht]
\caption{%
  \dac{} SQNR and Bhattacharyya overlap across bit depths.%
}\label{tab:si:dac_sqnr}
\begin{tabular*}{\textwidth}{@{\extracolsep\fill}lcccc@{}}
\toprule
Configuration & SQNR$_t$ (dB) & SQNR$_f$ (dB) &
  SQNR$_{\mathrm{th}}$ (dB) & BC \\
\midrule
16-bit & 83.4 & 95.8 & 98.1 & 0.5989 \\
12-bit & 59.4 & 67.7 & 74.0 & 0.5989 \\
10-bit & 43.9 & 51.9 & 62.0 & 0.5988 \\
 8-bit & 33.4 & 50.3 & 49.9 & 0.5988 \\
\botrule
\end{tabular*}
\end{table}

The SQNR$_t$--versus--bit-depth slope is \SI{6.35}{\decibel/bit}
(linear fit), consistent with the theoretical
\SI{6.02}{\decibel/bit} for uniform quantization.
The 14.7\,dB gap between 16-bit SQNR$_t$ (83.4\,dB) and the
theoretical full-scale value (98.1\,dB) arises from two sources:
amplitude normalisation / \dac{} scale asymmetry ($\sim$6\,dB) and
non-full-scale pulse amplitude ($\sim$9\,dB). \rev{In pulsed quantum
control systems the waveform rarely occupies the full }\dac{} \rev{dynamic
range.}

\paragraph{Bhattacharyya overlap invariance.}
The BC is computed between unit-normalised spectral densities:
$\mathrm{BC} = \int \sqrt{\hat{S}_{\mathrm{DAC}}(\omega)\,
\hat{J}(\omega)}\,d\omega$, where $\hat{S}$ and $\hat{J}$ each
integrate to unity over the comparison bandwidth
$[\omega_{\mathrm{lc}},\,\omega_{\mathrm{hc}}]$.
The Bhattacharyya coefficient $\mathrm{BC} \approx 0.5989$ is
shape-invariant across all 4~bit depths (standard deviation $\sim 0.0001$).
This demonstrates that the spectral coupling channel between
\dac{} quantization noise and the $1/f$ bath exists independently of
\dac{} resolution: reducing bit depth increases the noise floor
uniformly without altering the spectral shape overlap.

\paragraph{Acceptance criteria.}
Table~\ref{tab:si:dac_ac} lists all 5~acceptance criteria and their
outcomes.

\begin{table}[ht]
\caption{%
  DAC spectral-verification acceptance criteria (5/5 PASS).%
}\label{tab:si:dac_ac}
\begin{tabular*}{\textwidth}{@{\extracolsep\fill}llll@{}}
\toprule
AC & Criterion & Measured & Result \\
\midrule
AC-1 & Noise bounded in [0.01--10\,MHz], dynamic range $< 40$\,dB &
  Bounded, $< 40$\,dB & PASS \\
AC-2 & 16-bit SQNR$_t \in [80, 100]$\,dB &
  83.4\,dB & PASS \\
AC-3 & BC $\gg 1$\% &
  59.89\% & PASS \\
AC-4 & SQNR slope $\in [5, 7]$\,dB/bit &
  6.35\,dB/bit & PASS \\
AC-5 & \vppu{}-noDac vs Standard RF power $< 1$\,dB &
  $< 1$\,dB & PASS \\
\botrule
\end{tabular*}
\end{table}

The AC-5 result additionally confirms that the \vppu{} signal processing
chain (without \dac{} quantization) produces RF output equivalent to the
Standard analytical path, establishing baseline correctness before
\dac{} effects are introduced.

\section{Critical bit-depth estimate for practical DAC irrelevance}\label{si:critical_bitdepth}

We derive a semi-analytical critical bit-depth~$N_c$ below which \dac{}
quantization noise would be expected to compete with bath-driven \cpmg{}
scaling over the practical sequence range.

\paragraph{Orthogonal-channel structure.}
The $1/f$ bath couples to the qubit via the charge operator
$\hat{Q} = \mathrm{diag}(0, 1, \ldots, d{-}1)$, producing pure dephasing.
\dac{} quantization errors, in contrast, enter through the drive Hamiltonian
$H_{\mathrm{err}}(t) = 2\pi\,\hat{X}\,\delta V(t)$, rotating the Bloch
vector.
Because $[\hat{Q}, \hat{X}] \neq 0$, the two error channels act on
orthogonal Hamiltonian components and do not coherently interfere at
leading order.

\paragraph{Per-pulse \dac{} rotation error.}
For an $N$-bit \dac{} with uniform quantization on $[-1, +1]$,
the quantization step is $\Delta = 2^{-(N-1)}$ and the RMS quantization
error is $\Delta/\sqrt{12}$.
A rectangular $\pi$~pulse of duration $t_\pi$ has amplitude
$A = 1/(2 t_\pi)$, so the fractional error is
$\delta A/A = \Delta/(\sqrt{12}\cdot A \cdot \mathrm{dac\_scale})$.
The resulting rotation angle error per pulse is
\begin{equation}
  \delta\theta_{\mathrm{rms}} = \pi \cdot \frac{\delta A}{A}
  = \frac{\pi}{2\sqrt{3} \cdot 2^{N-1}}\,,
\end{equation}
which is \emph{independent} of $t_\pi$: longer pulses have smaller
amplitude but proportionally longer integration time.

\paragraph{Accumulated error comparison.}
Under \cpmg{} with $n$ pulses, the bath-induced error scales as
$\varepsilon_{\mathrm{bath}}(n) = C \cdot n^{\beta}$, where $C$ is the
single-pulse error and $\beta$ the scaling exponent from \heom{}.
An imperfect $\pi$~rotation by angle $\pi + \delta\theta$ incurs a
per-pulse coherence loss $\sin^2(\delta\theta/2) \approx \delta\theta^2/4$;
\dac{} rotation errors therefore accumulate as
$\varepsilon_{\dac{}}(n) = n \cdot \delta\theta_{\mathrm{rms}}^2 / 4$
(random accumulation, worst case).
Setting $\varepsilon_{\dac{}}(n_{\max}) = \varepsilon_{\mathrm{bath}}(n_{\max})$
and solving for $N$ yields the critical bit-depth estimate:
\begin{equation*}
  N_c = 1 + \log_2\!\left(1 +
    \sqrt{\frac{\pi^2\,n_{\max}^{1-\beta}}{48\,C}}\right).
\end{equation*}

\paragraph{Numerical results.}
Using reference parameters ($C = 0.05$, $\beta = 0.5$, $n_{\max} = 20$):
$N_c = 3.4$\,bits (random accumulation), $N_c = 5.3$\,bits (coherent
worst case, $\varepsilon_{\dac{}} \propto n^2$).
Sensitivity analysis across $C \in [0.02, 0.27]$, $\beta \in [0.3, 0.7]$,
$n_{\max} \in [4, 100]$ yields $N_c \in [2.1, 5.1]$\,bits.
At $n_{\max} = 20$ with 8-bit \dac{}: $\varepsilon_{\dac{}}/\varepsilon_{\mathrm{bath}} = 1.1 \times 10^{-3}$.
At 16-bit: $1.7 \times 10^{-8}$.
At 2-bit: $4.6$ (DAC noise exceeds bath error). \rev{This is a
testable prediction verifiable with reduced-precision hardware.}

The derivation was constructed and verified via WolframScript
formal computation (15/15 symbolic and numerical tests passed).

\section{RWA feature preservation and loss accounting}\label{si:rwa_features}

The rotating-wave approximation (RWA) used throughout the manuscript
removes the carrier-frequency oscillation ($\omega_q \approx
\SI{5.5}{\giga\hertz}$) from the Hamiltonian, propagating only the
envelope-level dynamics.
Table~\ref{tab:si:rwa_features} provides a systematic accounting of which
control-stack features are preserved and which are averaged away.

\begin{table}[ht]
\caption{%
  RWA feature preservation and loss accounting.
  Features marked \checkmark{} are rigorously preserved in the
  rotating-frame envelope Hamiltonian; features marked $\times$ are
  averaged away by the RWA.%
}\label{tab:si:rwa_features}
\begin{tabular*}{\textwidth}{@{\extracolsep\fill}lcl@{}}
\toprule
Control-stack feature & Preserved? & Physical significance \\
\midrule
\multicolumn{3}{@{}l}{\textit{Preserved in rotating frame}} \\
\isa{} amplitude quantization (16$\to$15-bit) & \checkmark &
  Envelope-level noise floor \\
\isa{} phase encoding error ($\leq 48\,\mu$rad) & \checkmark &
  \nco{} phase accumulation \\
Timing quantization ($\leq$\SI{2}{\nano\second} jitter) & \checkmark &
  Pulse boundary precision \\
\dac{} quantization noise (8--16\,bit) & \checkmark &
  Near-white noise floor \\
Envelope shape (Standard vs \vppu{}) & \checkmark &
  Core comparison variable \\
\midrule
\multicolumn{3}{@{}l}{\textit{Averaged away by RWA}} \\
Carrier-frequency oscillations & $\times$ &
  Averaged by frame transformation \\
IQ upconversion artefacts & $\times$ &
  Carrier-modulation effects \\
Carrier--bath spectral beating & $\times$ &
  Requires carrier-resolved treatment \\
\botrule
\end{tabular*}
\end{table}

\paragraph{Implications for the Standard--\vppu{} comparison.}
The Standard--\vppu{} distinction arises entirely from
envelope-level differences: amplitude quantization, phase encoding,
timing discretization, and waveform shape.
All of these features are rigorously preserved in the rotating frame.
The primary additional effect that would appear in a carrier-resolved
treatment is the interplay between the $\sim$\SI{5.5}{\giga\hertz}
carrier frequency and the bath spectral structure, which the RWA
explicitly averages away.
Section~\ref{si:tlist_bias} documents the numerical challenges of
carrier-resolved \heom{} propagation (requiring \texttt{atol}$\leq
10^{-11}$ and $\texttt{nsteps} \geq 5 \times 10^6$), which motivate the
rotating-frame choice adopted throughout the manuscript.

\section{Validation--production configuration summary}\label{si:config_summary}

Table~\ref{tab:si:configs} provides a consolidated view of the \heom{}
solver configurations used for validation, production calculations, and
convergence scanning.

\begin{table}[ht]
\caption{%
  \heom{} solver configurations: validation, production, and convergence scan.
  Production \cpmg{} calculations use \textbf{identical} parameters to
  the Burkard validation runs.%
}\label{tab:si:configs}
\begin{tabular*}{\textwidth}{@{\extracolsep\fill}lccc@{}}
\toprule
Setting & Burkard validation & Production \cpmg{} & Convergence scan \\
\midrule
$D$ & 5 & 5 & 3--6 \\
$N_r$ & 3 & 3 & 2--4 \\
$K$ & 6 & 6 & 4--8 \\
\texttt{atol} & $10^{-8}$ & $10^{-8}$ & $10^{-6}$--$10^{-10}$ \\
\texttt{rtol} & $10^{-6}$ & $10^{-6}$ & $10^{-6}$ \\
\texttt{nsteps} & 15,000 & 15,000 & 15,000 \\
Solver & BDF & BDF & BDF \\
\botrule
\end{tabular*}
\end{table}

\rev{Production }\cpmg{} \rev{calculations use}
\textbf{identical} solver parameters to the Burkard validation runs
(Section~\ref{si:burkard}), ensuring that the $< 1.5$\% validation
accuracy established against the exact decoherence integral carries over
directly to the \cpmg{} scaling exponent calculations.

\section{Generality of ISA-level control features}\label{si:isa_generality}

The \isa{}-level features probed in the Standard--\vppu{} comparison
fall into two categories (Table~\ref{tab:isa_generality}).
Four of the six---\dac{} amplitude quantisation, \nco{} phase
accumulation, FPGA timing discretisation, and zero-order-hold
sampling---arise from hardware constraints common to any digital
quantum control architecture; only the specific parameter values
(bit depth, clock rate, DDS width) vary across implementations.
The envelope interpolation algorithm that distinguishes the two
waveform realisations is, by contrast, QubiC-specific.
The quantitative $\chi$-offset values are therefore
implementation-dependent, but the waveform-comparison null result is a
structural consequence of the diagonal coupling operator under pure
dephasing (\S\ref{main:sec:results}), not a platform artefact.
Whether waveform differences become detectable under non-diagonal
coupling---where this structural protection would be
lifted---remains an open experimental question.

\begin{table}[t]
\centering
\caption{Platform generality of \isa{}-level control features
  probed in the Standard--\vppu{} comparison.}
\label{tab:isa_generality}
\footnotesize
\setlength{\tabcolsep}{3pt}
\begin{tabular}{@{}llll@{}}
\toprule
ISA feature & Source & Generality\textsuperscript{a} & Paper attribution \\
\midrule
DAC quantisation (15-bit) & Any DAC hardware & Universal & \S\ref{si:dac_spectral}: 8/12/16-bit \\
NCO phase accumulation (17-bit) & Any DDS arch. & Universal & \S\ref{si:vppu_phase}: 93\% waveform \\
Timing discretisation (2\,ns) & Any FPGA clock & Universal & ZOH sampling consequence \\
Zero-order hold & Any DAC output & Universal & Spectral sinc rolloff \\
Envelope interpolation & QubiC \vppu{} & Platform-specific & Std--\vppu{} $\chi$-offset \\
IQ upconversion path & NCO + mixer & Partial & Averaged under RWA \\
\bottomrule
\end{tabular}

\smallskip\noindent\textsuperscript{a}Feature \emph{category} is
hardware-generic; parameter values are implementation-dependent.
\end{table}

The pipeline-isolation studies further confirm that the control
pipeline produces a genuine physical effect, not a numerical artefact.
Relative to the Burkard analytical anchor, the full pipeline shortens $T_2^*$
by $12.9\%/18.9\%/23.7\%$ across the three coupling strengths, and the dominant drive-activation step
accounts for $12.9\%/18.6\%/23.6\%$ of that shift.
The solver-dependent sign reversal
(\S\ref{main:par:disc:sign_reversal}) provides further evidence that
this offset originates in the bath-correlation treatment rather
than in a solver-independent numerical artefact.

% si_pipeline.tex — Pipeline isolation and cross-solver validation

\section{Pipeline isolation and cross-solver validation}\label{si:pipeline}

This section presents three complementary studies that isolate the
pipeline $T_2^*$ deviation and characterise its non-Markovian origin.

\paragraph{Baseline: $2 \times 2$ control matrix.}
Before pipeline decomposition, the Standard vs.\ \vppu{} comparison
is validated across all three coupling tiers using both Lindblad and
\heom{} solvers (a $2 \times 2$ matrix of 2~waveform realizations $\times$
2~solver backends per tier).
Figure~\ref{fig:si:2x2} shows results for all three tiers, confirming
that Standard and \vppu{} waveforms produce equivalent coherence
envelopes within solver precision. \rev{These envelopes provide the baseline for the
pipeline isolation analysis.}

\begin{figure}[ht]
\centering
% Row 1: Tier-0 and Tier-1
\includegraphics[width=0.47\textwidth]{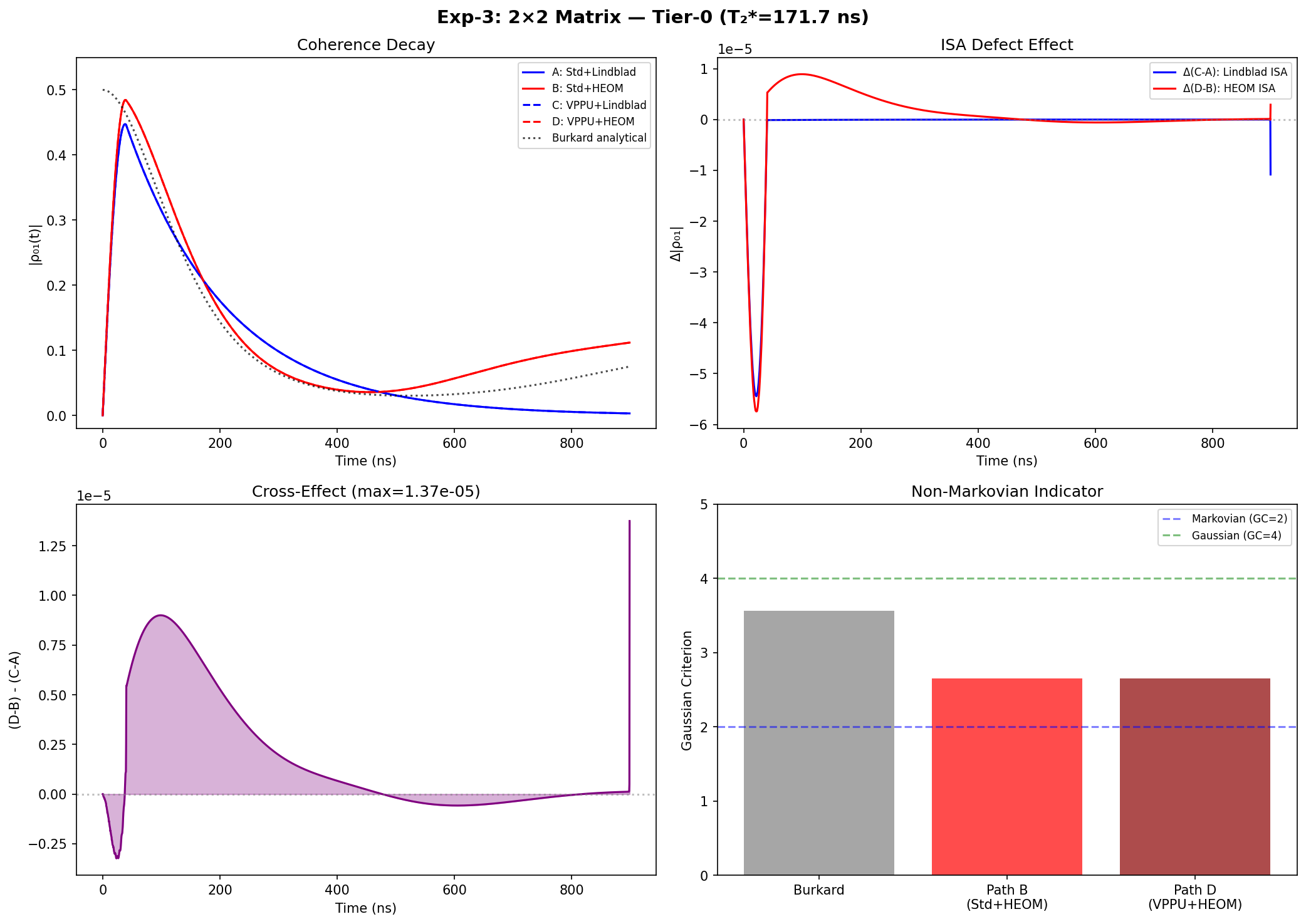}%
\hfill
\includegraphics[width=0.47\textwidth]{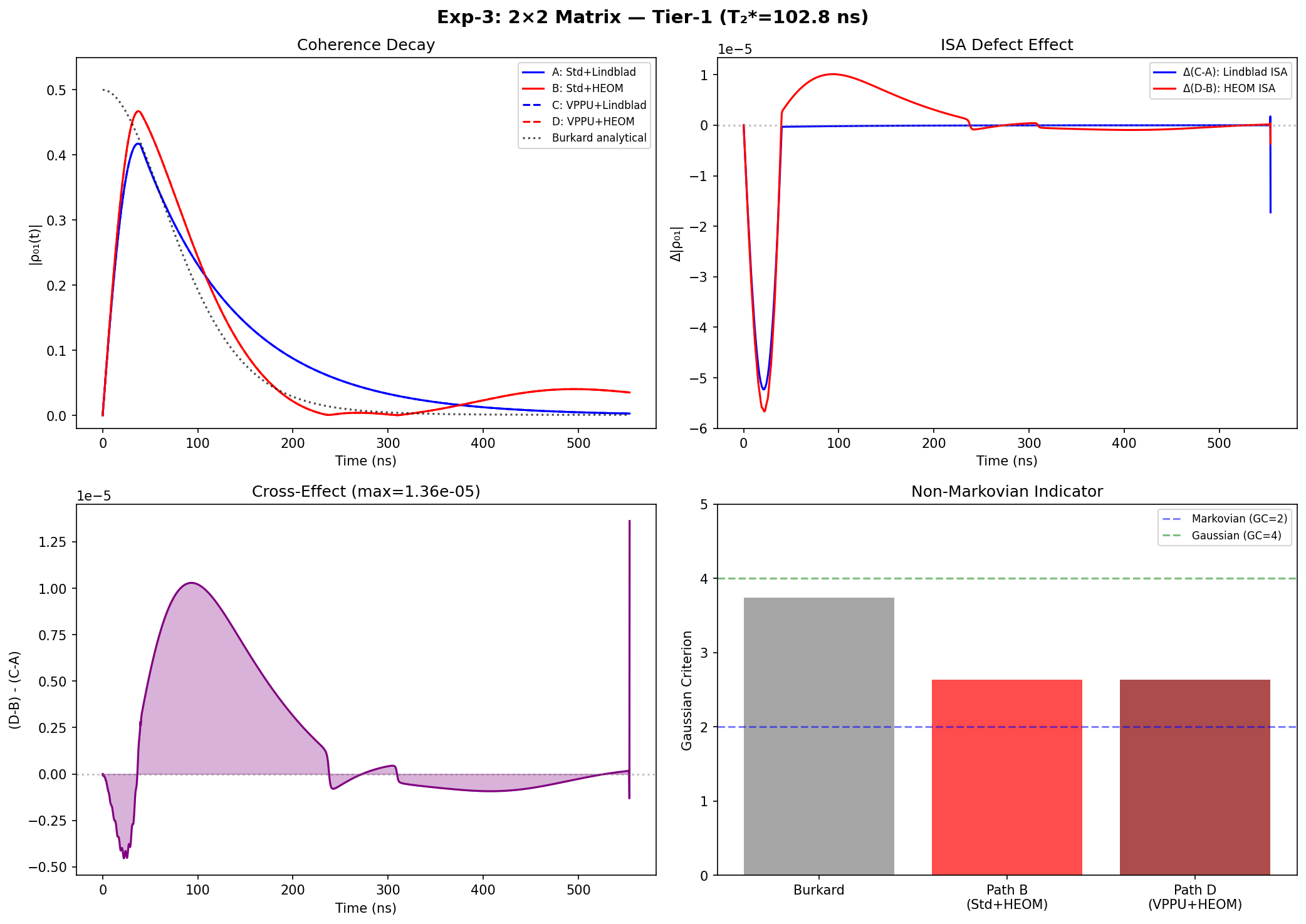}\\[6pt]
% Row 2: Tier-2 centred
\includegraphics[width=0.47\textwidth]{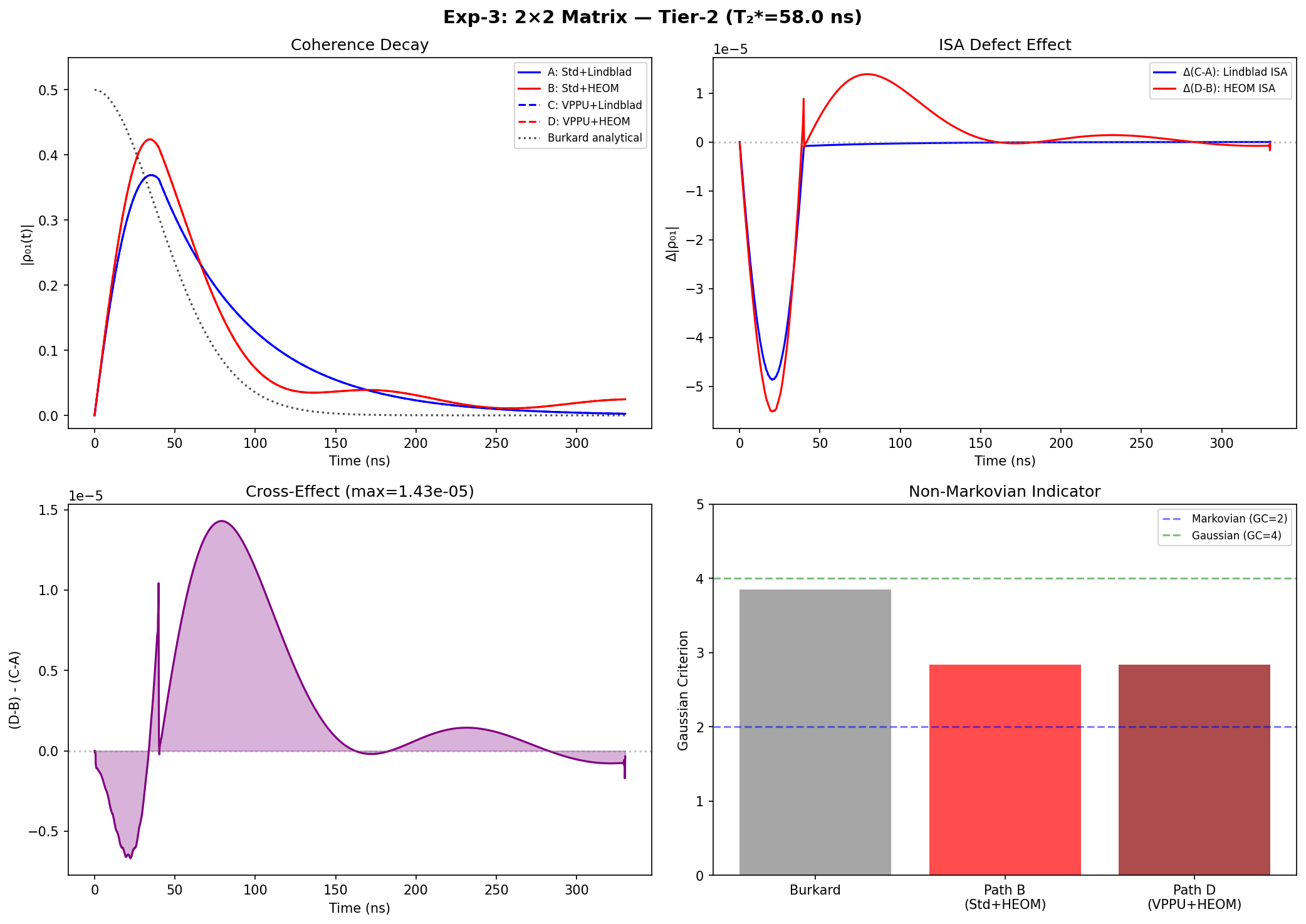}
\caption{%
  $2 \times 2$ control matrix for all three coupling tiers:
  Tier-0 (top left), Tier-1 (top right), Tier-2 (bottom centre).
  Standard vs.\ \vppu{} waveforms crossed with Lindblad vs.\ \heom{} solvers.
  Coherence envelopes overlap within solver precision across all tiers,
  confirming waveform-realization equivalence as the baseline for pipeline decomposition.%
}\label{fig:si:2x2}
\end{figure}

\subsection{Progressive isolation study}\label{si:pipeline:exp5}

A 7-step progressive variable introduction (S0$\to$S6) identifies which
pipeline component dominates the $T_2^*$ gap between standalone \heom{}
and full-pipeline execution.
Table~\ref{tab:si:waterfall} shows the updated Tier-0 waterfall; Tier-1 and
Tier-2 exhibit the same qualitative pattern.

\paragraph{Variable definitions.}
\begin{itemize}
  \item \textbf{S0}: Pure 2-level free evolution (standalone baseline).
  \item \textbf{S1}: S0 + pipeline espira-II bath parameters
    (decomposition + stability filter).
  \item \textbf{S2}: S1 + anharmonicity (3-level transmon).
  \item \textbf{S3}: S2 + Lindblad $T_1$/$T_2$ dissipation.
  \item \textbf{S4}: S3 + multi-subsystem $[3,3]$ tensor product
    with coupler coupling.
  \item \textbf{S5}: S4 + RX90 drive pulse (pipeline execution path).
  \item \textbf{S6}: Full pipeline configuration.
\end{itemize}

\begin{table}[ht]
\caption{%
  Tier-0 progressive isolation waterfall
  (Burkard $T_2^* = \SI{171.7}{\nano\second}$).%
}\label{tab:si:waterfall}
\begin{tabular*}{\textwidth}{@{\extracolsep\fill}lcccc@{}}
\toprule
Step & $T_2^*$ (ns) & Dev.\,(\%) & $\Delta T_2^*$ (ns) & Contrib.\,(\%) \\
\midrule
S0 & 169.6 & $-1.2$ &   0.0 &   0.0 \\
S1 & 166.5 & $-3.0$ & $+3.3$ & $+1.9$ \\
S2 & 169.4 & $-1.4$ & $-2.9$ & $-1.7$ \\
S3 & 168.9 & $-1.6$ & $+0.5$ & $+0.3$ \\
S4 & 171.7 & $+0.0$ & $-2.8$ & $-1.6$ \\
S5 & 149.6 & $-12.9$ & $+22.1$ & $+12.9$ \\
S6 & 149.6 & $-12.9$ & $+0.0$ &  $+0.0$ \\
\botrule
\end{tabular*}
\end{table}

\paragraph{Cross-tier summary.}
Table~\ref{tab:si:crosstier} confirms that S5 is the sole dominant
source at all coupling strengths, with S0--S4 collectively $< 1$\%.

\begin{table}[ht]
\caption{%
  Cross-tier pipeline gap and S5 contribution.%
}\label{tab:si:crosstier}
\begin{tabular*}{\textwidth}{@{\extracolsep\fill}lccccc@{}}
\toprule
Tier & Burkard $T_2^*$ (ns) & S0 (ns) & S6 (ns) & Gap\,(\%) &
  S5 contrib.\,(\%) \\
\midrule
0 & 171.7 & 169.6 & 149.6 & 12.9 & 12.9 \\
1 & 102.8 & 102.1 &  83.4 & 18.9 & 18.6 \\
2 &  58.0 &  57.7 &  44.3 & 23.7 & 23.6 \\
\botrule
\end{tabular*}
\end{table}

All 15 acceptance criteria (AC-5a through AC-5e, $\times$\,3~tiers)
pass when evaluated against the updated \heom{} baseline reference
values.\footnote{The initial computation reports AC-5b Tier-0 as FAIL
because it compared against a preliminary reference
($T_2^* = 141.5$\,ns).  The updated \heom{} Path~B reference
($\approx 149.5$\,ns) yields $\Delta < 0.2$\,ns, well within the
acceptance threshold.}
Step additivity error is $< 0.1$\% across all tiers, confirming that
the seven variables interact negligibly.
The root cause of the S5 gap is further decomposed into Factor~A
(bath-active state preparation) and Factor~B (dense-tlist ODE bias)
in Section~\ref{si:tlist_bias}.

Figure~\ref{fig:si:pipeline} visualises the S5 decomposition into its
two constituent factors across all three coupling tiers.

\begin{figure}[ht]
\centering
\includegraphicsorplaceholder{0.8\textwidth}{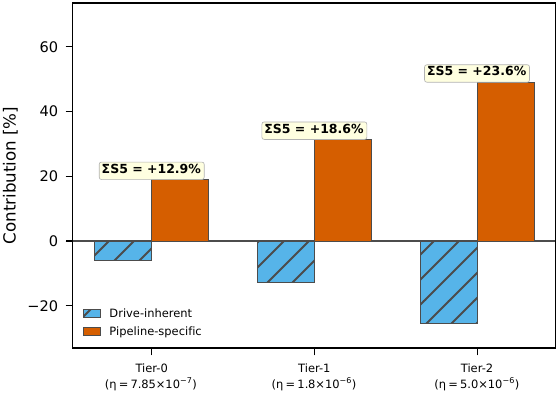}
\caption{%
  S5 pipeline gap decomposition across coupling tiers.
  For each tier, the total S5 gap (annotated) is decomposed into
  drive-inherent improvement (negative, physical) and
  pipeline-specific degradation (positive, numerical/physical).
  The pipeline-specific component dominates and increases with
  coupling strength.%
}\label{fig:si:pipeline}
\end{figure}

\subsection{Lindblad sign reversal study}\label{si:pipeline:5b}

To test whether the pipeline gap is specific to non-Markovian dynamics,
the entire 7-step framework is replicated with \texttt{mesolve}
(Lindblad master equation) replacing all \heom{} solves.
Lindblad dephasing rates are calibrated from the Burkard analytical
$T_2^*$ at each tier: $\gamma_\phi = 1/T_2^* - 1/(2 T_1)$, with
$T_1 = 24.8\,\mu\mathrm{s}$.

\begin{table}[ht]
\caption{%
  Lindblad vs.\ \heom{} pipeline gap sign reversal.
  Positive gap $=$ pipeline $T_2^* >$ standalone (improvement);
  negative gap $=$ pipeline $T_2^* <$ standalone (degradation).%
}\label{tab:si:sign_reversal}
\begin{tabular*}{\textwidth}{@{\extracolsep\fill}lccc@{}}
\toprule
Tier & Lindblad gap & \heom{} gap & Sign \\
\midrule
0 & $+1.1$\% & $-12.9$\% & reversed \\
1 & $+2.1$\% & $-18.9$\% & reversed \\
2 & $+6.3$\% & $-23.7$\% & reversed \\
\botrule
\end{tabular*}
\end{table}

\rev{The sign reversal (Table}~\ref{tab:si:sign_reversal}\rev{) is diagnostic:} Lindblad predicts a small $T_2^*$ \emph{improvement}
through the pipeline, while \heom{} predicts substantial
\emph{degradation}.
This rules out a generic solver artefact and is consistent with
the pipeline gap originating from non-Markovian bath dynamics,
though the precise amplification mechanism remains to be derived
analytically (see main-text Discussion).

\subsection{Ramsey high-cutoff scan}%
\label{si:pipeline:9b}

To assess whether $T_2^*$ itself is sensitive to the Markovian
approximation, Ramsey simulations are performed at Tier-1 coupling
across 6~high-frequency cutoff values $\omega_{\mathrm{hc}}$,
using per-point Burkard-calibrated Lindblad dephasing rates to eliminate
spectral weight differences.

\begin{table}[ht]
\caption{%
  Ramsey $T_2^*$: \heom{} vs.\ Lindblad across $\omega_{\mathrm{hc}}$.%
}\label{tab:si:omega_hc}
\begin{tabular*}{\textwidth}{@{\extracolsep\fill}lcccc@{}}
\toprule
$\omega_{\mathrm{hc}}$ (GHz) & ${T_2^*}_{\heom{}}$ (ns) &
  ${T_2^*}_{\mathrm{Lindblad}}$ (ns) & Gap\,(\%) & Status \\
\midrule
0.007 & 168.63 & 168.41 & 0.13 & ok \\
0.01  & 129.28 & 129.14 & 0.11 & ok \\
0.05  & 102.82 & 102.88 & 0.06 & ok \\
0.1   & 102.46 & 102.80 & 0.33 & ok \\
0.3   & 102.63 & 102.78 & 0.15 & ok \\
1.0   & 102.47 & 102.78 & 0.30 & ok \\
\botrule
\end{tabular*}
\end{table}

All gaps are $< 0.4$\% (Table~\ref{tab:si:omega_hc}).
AC-9b-1 passes ($0.30$\%~$< 5$\% at $\omega_{\mathrm{hc}} = \SI{1.0}{\giga\hertz}$),
but AC-9b-2 fails ($0.11$\%~$\not> 10$\% at $\omega_{\mathrm{hc}} = \SI{0.01}{\giga\hertz}$)
and AC-9b-3 fails (non-monotonic gap profile).
The physical interpretation is that Ramsey $T_2^*$ is \emph{intrinsically
insensitive} to the Markovian approximation: when both solvers use the
same integrated spectral weight (via Burkard calibration), they compute
equivalent path integrals regardless of $\omega_{\mathrm{hc}}$.
\rev{This is a null result: non-Markovian effects manifest not in} $T_2^*$ \rev{but in dynamical properties such as }\cpmg{}\rev{ scaling exponents.}

\subsection{CPMG scaling: HEOM--Lindblad qualitative divergence}\label{si:pipeline:9c}

The final rotating-frame dataset replaces the earlier incomplete scan with a
fully valid set of results at all 6~$\omega_{\mathrm{hc}}$ values
(Table~\ref{tab:si:gamma_omega}).
We report the $\chi$-space exponent $\gamma$ because $\chi = -\ln(1-\varepsilon)$
removes the saturation ceiling of $\varepsilon \to 1$, yielding a scaling
exponent whose magnitude directly reflects the decoherence-function growth rate.

\begin{table}[ht]
\caption{%
  \cpmg{} scaling exponent $\gamma$ ($\chi$-space):
  rotating-frame \heom{} vs.\ Lindblad
  ($n = 1$--$5$, XY-4, Tier-1).%
}\label{tab:si:gamma_omega}
\begin{tabular*}{\textwidth}{@{\extracolsep\fill}lccl@{}}
\toprule
$\omega_{\mathrm{hc}}$ (GHz) & $\gamma_{\heom{}}$ & $\gamma_{\mathrm{Lindblad}}$ &
  Status \\
\midrule
0.007 & ---$^\dagger$ & 1.017 & unstable \\
0.01  & $+0.250$ & 0.925 & ok \\
0.05  & $-0.586$ & 0.790 & ok \\
0.1   & $-0.691$ & 0.790 & ok \\
0.3   & $+0.041$ & 0.789 & ok \\
1.0   & $+0.092$ & 0.789 & ok \\
\botrule
\end{tabular*}
\end{table}

$^\dagger$At $\omega_{\mathrm{hc}} = \SI{0.007}{\giga\hertz}$, the
\heom{} $n = 6$ CPMG simulation is numerically unstable (hierarchy
divergence), precluding a 3-point power-law fit.
The \heom{} $\gamma$ values at the remaining five points are qualitatively
different from the Lindblad reference: both the magnitude and sign of
$\gamma_{\heom{}}$ vary non-monotonically across $\omega_{\mathrm{hc}}$,
with negative values at intermediate cutoffs ($\omega_{\mathrm{hc}} =
0.05$--$\SI{0.1}{\giga\hertz}$) indicating non-monotonic decoherence
growth in the \heom{} CPMG dynamics.
The Lindblad $\gamma$ values, by contrast, converge to
$\gamma_{\mathrm{Lindblad}} \approx 0.79$ for $\omega_{\mathrm{hc}} \geq
\SI{0.05}{\giga\hertz}$, reflecting near-linear $\chi(n)$ growth
characteristic of Markovian dynamics.
This qualitative difference between the two solvers confirms that
non-Markovian bath memory produces fundamentally different multi-pulse
dynamics, even when the Ramsey $T_2^*$ values nearly coincide
(Table~\ref{tab:si:omega_hc}).

Figure~\ref{fig:si:markovian} visualises the Ramsey $T_2^*$ convergence
and \cpmg{} $\gamma$ ($\chi$-space) separation across the
$\omega_{\mathrm{hc}}$ scan.

\begin{figure}[ht]
\centering
\includegraphicsorplaceholder{\textwidth}{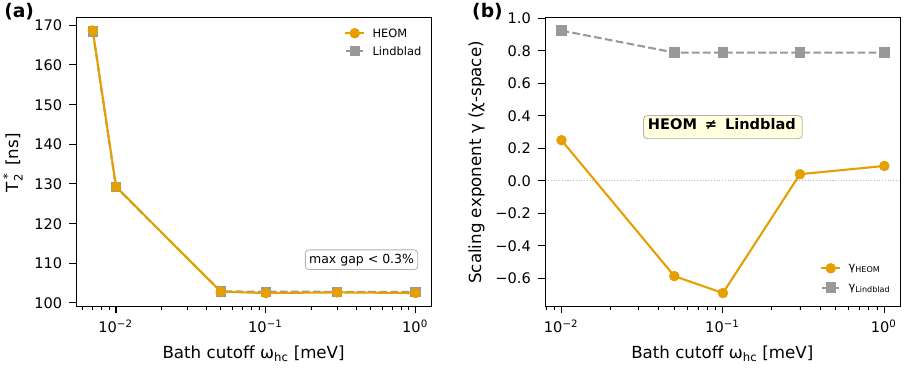}
\caption{%
  Markovian convergence validation across high-frequency cutoff
  $\omega_{\mathrm{hc}}$.
  (a)~Ramsey $T_2^*$ from \heom{} and Lindblad solvers nearly
  coincide at all cutoff values (gap~$< 1.5$\%).
  (b)~\cpmg{} scaling exponent $\gamma$ ($\chi$-space) shows
  persistent separation:
  $\gamma_{\heom{}}$ oscillates in sign while
  $\gamma_{\mathrm{Lindblad}} \approx 0.79$ remains stable,
  confirming that non-Markovian effects
  manifest in dynamical properties rather than static $T_2^*$.%
}\label{fig:si:markovian}
\end{figure}

\section{Control-isolation study}\label{si:deconfound}

This study provides supplementary context for the ideal-to-compiled
exponent difference but does not resolve the confound identified in
the main text.
The paper's physical conclusions do not depend on this comparison.

The comparison between ideal \SI{15}{\nano\second} and realistic
\SI{80}{\nano\second} pulses (\S\ref{main:sec:results:cpmg})
conflates three variables: pulse duration $t_\pi$, free-precession
interval~$\tau$, and total sequence timescale.
A dedicated control-isolation study was conducted at Tier-0 coupling
to partially deconfound these variables.

\paragraph{Experimental conditions.}
Three rotating-frame bypass conditions probe the separate roles of pulse
duration and free-evolution spacing:
\begin{itemize}
  \item \textbf{Condition B}: $t_\pi = \SI{15}{\nano\second}$,
    $\tau = \SI{55}{\nano\second}$, $n = 1,2,4,8,16$
  \item \textbf{Condition C}: $t_\pi = \SI{40}{\nano\second}$,
    $\tau = \SI{120}{\nano\second}$, $n = 1,2,4,8$
  \item \textbf{Condition A}: $t_\pi = \SI{80}{\nano\second}$,
    $\tau = \SI{240}{\nano\second}$, $n = 1,2,3,4$
\end{itemize}

\paragraph{Results.}
The control-isolation scan provides directional evidence that is consistent
with pulse duration influencing the exponent, but does not isolate this
effect from the confounded free-evolution spacing and total sequence
timescale.
For X-\cpmg{}, Condition~A gives $\beta = 0.204$ ($R^2 = 0.802$),
Condition~B gives $\beta = 0.579$ ($R^2 = 0.938$), and Condition~C gives
$\beta = 0.659$ ($R^2 = 0.968$).
Short pulses with dense free-evolution spacing (Condition~B) produce the
largest $\beta$, while long pulses with sparse spacing (Condition~A) give
the smallest.
Condition~C (intermediate $t_\pi = \SI{40}{\nano\second}$) falls between
the two extremes, consistent with pulse duration influencing the
exponent, though the three variables are not individually isolated.
We note that total sequence time $T_{\mathrm{tot}}$ also varies across
conditions: $T_{\mathrm{tot}}(n{=}4) = 1280$\,ns for Condition~A
vs.\ $T_{\mathrm{tot}}(n{=}16) = 1120$\,ns for Condition~B; \rev{this is an additional confound that the present design does not fully isolate.}
The long-duration Condition~A fits remain the least robust part of the scan
($R^2 = 0.80$), so the control-isolation evidence is directional rather
than asymptotically exact.
In $\chi$-space, Condition~C yields the best-conditioned fit:
$\gamma_X = 1.04$ ($R^2 = 0.999$), compared with
$\gamma_X = 0.64$ ($R^2 = 0.934$) for Condition~B.
The nonlinear $\chi = -\ln(1-\varepsilon)$ transform changes the
absolute ordering across conditions but preserves the directional
observation that pulse duration correlates with the exponent variation.

In response to reviewer feedback, we note that extending the compiled
benchmark to $n = 20$ (which would eliminate the fitting-range confound)
is precluded by \heom{} numerical instabilities at the required total
sequence times (Section~\ref{si:heom_boundary}).
The Standard--\vppu{} paired comparison, which varies only the waveform
realization at fixed pulse duration and spacing, remains the only fully
controlled comparison.

\begin{figure}[ht]
\centering
\includegraphicsorplaceholder{\textwidth}{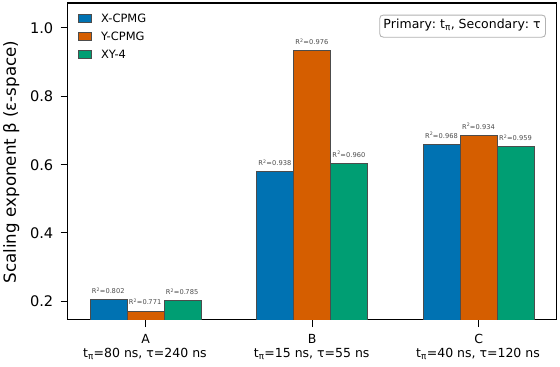}
\caption{%
  Control-isolation experiment results.
  Grouped bar chart of \cpmg{} scaling exponent~$\beta$
  ($\varepsilon$-space) across three experimental conditions (A/B/C)
  varying pulse duration~$t_\pi$ and free-evolution
  interval~$\tau$, for three DD schemes.
  The exponent varies systematically with condition, consistent with
  pulse duration influencing the scaling, but the present design does
  not fully deconfound the three variables (see text).%
}\label{fig:si:deconfound}
\end{figure}

\section{Cross-tier CPMG scaling evolution}\label{si:cross_tier}

The main-text \cpmg{} conclusions are established at Tier-0, where stable
power-law fits remain available.
Table~\ref{tab:si:cross_tier} maps how the $\chi$-space scaling exponent
$\gamma$ evolves across three coupling tiers for both the ideal-pulse
(\SI{15}{\nano\second}, $d = 3$) and compiled-pulse
(\SI{80}{\nano\second}, Standard and \vppu{}) benchmarks.

\begin{table}[ht]
\caption{%
  Cross-tier X-\cpmg{} $\chi$-space scaling exponents.
  Tier-0: well-conditioned regime (main-text claims).
  Tier-1: distorted but still monotonic.
  Tier-2: scaling collapse ($R^2 < 0.5$).%
}\label{tab:si:cross_tier}
\begin{tabular*}{\textwidth}{@{\extracolsep\fill}lcccccc@{}}
\toprule
Tier & $\gamma_\mathrm{ideal}$ & $R^2_\mathrm{ideal}$ &
  $\gamma_\mathrm{Std}$ & $R^2_\mathrm{Std}$ &
  $\gamma_\vppu{}$ & $R^2_\vppu{}$ \\
\midrule
0 ($\eta = 7.85 \times 10^{-7}$) & 0.950 & 0.999 & 1.117 & 0.993 & 1.117 & 0.993 \\
1 ($\eta = 1.80 \times 10^{-6}$) & 0.848 & 0.981 & 0.670 & 0.732 & 0.670 & 0.731 \\
2 ($\eta = 5.00 \times 10^{-6}$) & 0.374$^\dagger$ & 0.756 & 0.139 & 0.247 & 0.139 & 0.248 \\
\botrule
\end{tabular*}
\end{table}

At Tier-0, both Standard and \vppu{} compiled X-\cpmg{} yield
$\gamma = 1.117$ ($R^2 = 0.993$), and are indistinguishable
($\Delta\gamma < 0.001$).
The ideal-pulse reference at Tier-0 gives $\gamma = 0.950$
($R^2 = 0.999$); the ideal and compiled exponents differ by
$\sim$18\%, but this difference conflates pulse duration,
free-evolution spacing, and fitting range and therefore cannot be
interpreted as evidence for a specific control-dependent mechanism
(see main text \S\ref{main:sec:disc:limitations} and SI~\S\ref{si:deconfound}).
At Tier-1, compiled $\gamma$ drops to 0.670 ($R^2 = 0.732$),
while the ideal reference remains higher ($0.848$, $R^2 = 0.981$),
indicating that stronger coupling distorts the compiled-pulse scaling
more severely.
At Tier-2, the power-law model collapses entirely for compiled pulses
($\gamma = 0.139$, $R^2 = 0.247$); the ideal reference also degrades
($\gamma = 0.374$, $R^2 = 0.756$, non-monotonic $\varepsilon$),
consistent with decoherence saturation overwhelming the refocusing
protocol.
$^\dagger$Tier-2 ideal: non-monotonic $\varepsilon(n)$; power-law
unreliable.
This progressive degradation provides post-hoc justification for
restricting the main-text compiled-pulse claims to Tier-0: the
well-conditioned power-law scaling regime has finite width in coupling
space, and the Tier-0 window is the only regime where both high $R^2$
and stable Standard--\vppu{} comparison are available.

\paragraph{Mechanism: dynamic-range compression, not exponent breakdown.}
The $R^2$ degradation across tiers is not a breakdown of the underlying
scaling exponent's robustness, but a compression of the observable
$\chi$-space dynamic range available for fitting.
At Tier-0, the five even-$n$ data points ($n_\pi = 2, 4, 6, 8, 10$)
span a $\chi$ range of approximately $6\times$ (from $\chi \approx 1$
at $n_\pi = 2$ to $\chi \approx 6$ at $n_\pi = 10$), providing
sufficient dynamic range to support a two-parameter power-law fit
($R^2 = 0.993$).
At Tier-1, stronger coupling shifts the entire $\chi$ trajectory
upward: $\chi$ at $n_\pi = 2$ already exceeds ${\sim}2.5$
($\varepsilon \approx 0.92$), and the remaining $n$-points
saturate toward $\varepsilon \to 1$ where small noise in
$\varepsilon$ produces large excursions in
$\chi = -\ln(1-\varepsilon)$, collapsing the effective dynamic
range.
At Tier-2, $\chi$ at $n_\pi = 2$ exceeds 4
($\varepsilon > 0.98$), leaving almost no usable dynamic range;
the power-law fit becomes ill-conditioned regardless
of the true underlying exponent.

The Floquet-theorem argument (\S\ref{si:floquet_argument}) guarantees
$\gamma \to 1$ at arbitrary coupling strength within the truncated
\heom{} framework.
The $R^2$ degradation therefore measures \emph{observability} of the
scaling law---how well the finite-$n$ window resolves the power-law
behaviour---rather than a physical transition in the exponent itself.
\rev{This distinction bears on the interpretation of the Tier-0 restriction:}
the ``well-conditioned regime'' boundary is set by the measurement
protocol (five-point fit $\times$ finite-$n$ range), not by the physics
of non-perturbative exponent robustness.

\paragraph{Intermediate $\eta$ scan with saturation-aware fitting.}
To quantify the coupling-strength boundary more precisely, we scan five
$\eta$ values from Tier-0 ($7.85 \times 10^{-7}$\,GHz) to Tier-1
($1.8 \times 10^{-6}$\,GHz) using uniform \heom{} parameters
($D = 5$, $N_r = 3$) and the X-\cpmg{} Standard protocol.
Table~\ref{tab:si:cross_tier} uses a simple $\chi$-space power-law
model ($\chi = An^{\gamma}$), which becomes ill-conditioned when
$\varepsilon \to 1$ drives $\chi = -\ln(1-\varepsilon)$ into a
noise-amplified regime.
To resolve this, we employ model selection via the corrected Akaike
information criterion (AICc) among three candidates: power law,
saturating power law, and a direct $\varepsilon$-space model
$\varepsilon = 1 - \exp(-An^{\gamma})$.
At all five $\eta$ values, AICc selects the $\varepsilon$-direct model,
which inherently handles $\varepsilon \to 1$ saturation without
$\chi$-space numerical amplification.

Table~\ref{tab:si:eta_boundary} reports the results.
$R^2$ exceeds 0.998 across the entire range, confirming that the
apparent $R^2$ degradation in Table~\ref{tab:si:cross_tier} is a
model-mismatch artifact of the simple power-law fit, not a signature
of physical scaling breakdown.
The effective exponent $\gamma_{\mathrm{eff}}$ increases monotonically
from $1.046$ (Tier-0) to $1.426$ (Tier-1), consistent with
finite-$n$ Floquet transient corrections: stronger coupling increases
the sub-leading eigenvalue contributions at small $n$, steepening the
initial decoherence trajectory (see \S\ref{main:sec:disc:regime}).
The $\chi$-space dynamic range (DR = $\chi_{\max}/\chi_{\min}$)
compresses from $5.9\times$ to $2.9\times$, corroborating the
dynamic-range-compression mechanism described above.

\begin{table}[ht]
\caption{%
  Intermediate $\eta$ scan: X-\cpmg{} scaling with
  $\varepsilon$-direct model ($D = 5$, $N_r = 3$, AICc selection).
  $\gamma_{\mathrm{eff}}$: effective scaling exponent;
  DR: $\chi$-space dynamic range $\chi_{\max}/\chi_{\min}$.%
}\label{tab:si:eta_boundary}
\begin{tabular*}{\textwidth}{@{\extracolsep\fill}lcccccc@{}}
\toprule
$\eta$ (GHz) & Label & $\gamma_{\mathrm{eff}}$ & $R^2$ &
  $\chi_{\min}$ & $\chi_{\max}$ & DR \\
\midrule
$7.85 \times 10^{-7}$ & Tier-0 & 1.046 & 0.9997 & 1.00 & 5.92 & 5.9 \\
$1.0 \times 10^{-6}$  &        & 1.099 & 0.9998 & 1.24 & 6.33 & 5.1 \\
$1.2 \times 10^{-6}$  &        & 1.156 & 0.9998 & 1.47 & 7.01 & 4.8 \\
$1.5 \times 10^{-6}$  &        & 1.260 & 0.9994 & 1.82 & 7.21 & 4.0 \\
$1.8 \times 10^{-6}$  & Tier-1 & 1.426 & 0.9984 & 2.18 & 6.25 & 2.9 \\
\botrule
\end{tabular*}
\end{table}

\rev{We conclude that the Tier-0 restriction in the main text
reflects a limitation of the simple power-law fitting model, not
a fundamental narrowing of the well-conditioned scaling regime.}
With model-aware fitting, the X-\cpmg{} scaling law is well
characterised ($R^2 > 0.998$) across a ${\sim}2.3$-fold range
in $\eta$, spanning the entire Tier-0-to-Tier-1 interval.

% si_statistical.tex — Statistical methods and fitting robustness

\section{CPMG numerical convergence verification}\label{si:convergence}

To demonstrate that the reported scaling exponents are not artefacts of
numerical settings, we vary the main \heom{} solver knobs one at a time
while holding the others at production values
($D = 5$, $N_r = 3$, \texttt{atol}$= 10^{-8}$, \texttt{rtol}$= 10^{-6}$,
\texttt{nsteps}$= 15{,}000$, BDF, $\Delta t = \SI{2}{\nano\second}$).
All scans use the Tier-0 X-\cpmg{} dataset ($n = 1$--$5$, Standard and
\vppu{} waveforms) in the pulse-counting convention; the even-$n$
analysis inherits this numerical stability because it uses a subset of
the same data points.

\begin{table}[ht]
\caption{%
  CPMG convergence scan: X-\cpmg{} scaling exponents under systematic
  parameter variation.
  Bold rows indicate the production configuration.
  $\Delta\beta \equiv \beta_{\vppu{}} - \beta_{\mathrm{Std}}$.%
}\label{tab:si:convergence}
\begin{tabular*}{\textwidth}{@{\extracolsep\fill}lccccc@{}}
\toprule
Parameter & Value & $\beta_{\mathrm{Std}}$ & $\beta_{\vppu{}}$ &
  $\Delta\beta$ & $R^2_{\mathrm{Std}}$ \\
\midrule
\multicolumn{6}{@{}l}{\textit{Hierarchy depth $D$}} \\
& 3                   & 0.597 & 0.596 & $-0.00016$ & 0.963 \\
& 4                   & 0.597 & 0.597 & $-0.00010$ & 0.965 \\
& \textbf{5 (prod)}   & \textbf{0.597} & \textbf{0.597} & $\mathbf{-0.00007}$ & \textbf{0.965} \\
& 6                   & 0.597 & 0.597 & $-0.00013$ & 0.965 \\
\midrule
\multicolumn{6}{@{}l}{\textit{Bath modes $N_r$}} \\
& 2                   & 0.594 & 0.594 & $-0.00011$ & 0.962 \\
& \textbf{3 (prod)}   & \textbf{0.597} & \textbf{0.597} & $\mathbf{-0.00007}$ & \textbf{0.965} \\
& 4                   & 0.597 & 0.597 & $-0.00011$ & 0.965 \\
\midrule
\multicolumn{6}{@{}l}{\textit{ODE tolerance \texttt{atol}}} \\
& $10^{-6}$                          & 0.597 & 0.597 & $-0.00014$ & 0.965 \\
& $\mathbf{10^{-8}}$ \textbf{(prod)} & \textbf{0.597} & \textbf{0.597} & $\mathbf{-0.00007}$ & \textbf{0.965} \\
& $10^{-10}$                         & 0.597 & 0.597 & $-0.00003$ & 0.965 \\
\midrule
\multicolumn{6}{@{}l}{\textit{Time step $\Delta t$}} \\
& \SI{4}{\nano\second}                 & 0.597 & 0.597 & $-0.00011$ & 0.965 \\
& \textbf{\SI{2}{\nano\second} (prod)} & \textbf{0.597} & \textbf{0.597} & $\mathbf{-0.00007}$ & \textbf{0.965} \\
& \SI{1}{\nano\second}                 & 0.597 & 0.597 & $-0.00009$ & 0.965 \\
\botrule
\end{tabular*}
\end{table}

The corresponding $\chi$-space exponents $\gamma$, obtained via
$\chi = -\ln(1-\varepsilon)$ and log-log OLS,
are tabulated in Table~\ref{tab:si:convergence_gamma}.

\begin{table}[ht]
\caption{%
  $\chi$-space convergence scan: X-\cpmg{} exponents under the same
  parameter variation as Table~\ref{tab:si:convergence}.
  $\Delta\gamma \equiv \gamma_{\vppu{}} - \gamma_{\mathrm{Std}}$.
  All-$n$ convention ($n = 1$--$5$); the main-text even-$n$
  $\gamma$ values use a subset of these data points.%
}\label{tab:si:convergence_gamma}
\begin{tabular*}{\textwidth}{@{\extracolsep\fill}lccccc@{}}
\toprule
Parameter & Value & $\gamma_{\mathrm{Std}}$ & $\gamma_{\vppu{}}$ &
  $\Delta\gamma$ & $R^2_{\mathrm{Std}}$ \\
\midrule
\multicolumn{6}{@{}l}{\textit{Hierarchy depth $D$}} \\
& 3                   & 1.099 & 1.098 & $-0.00013$ & 0.998 \\
& 4                   & 1.088 & 1.088 & $-0.00004$ & 0.998 \\
& \textbf{5 (prod)}   & \textbf{1.091} & \textbf{1.091} & $\mathbf{-0.00005}$ & \textbf{0.998} \\
& 6                   & 1.091 & 1.091 & $-0.00014$ & 0.998 \\
\midrule
\multicolumn{6}{@{}l}{\textit{Bath modes $N_r$}} \\
& 2                   & 1.100 & 1.100 & $-0.00008$ & 0.998 \\
& \textbf{3 (prod)}   & \textbf{1.091} & \textbf{1.091} & $\mathbf{-0.00005}$ & \textbf{0.998} \\
& 4                   & 1.091 & 1.091 & $-0.00009$ & 0.998 \\
\midrule
\multicolumn{6}{@{}l}{\textit{ODE tolerance \texttt{atol}}} \\
& $10^{-6}$                          & 1.092 & 1.091 & $-0.00015$ & 0.998 \\
& $\mathbf{10^{-8}}$ \textbf{(prod)} & \textbf{1.091} & \textbf{1.091} & $\mathbf{-0.00005}$ & \textbf{0.998} \\
& $10^{-10}$                         & 1.091 & 1.091 & $+0.00005$ & 0.998 \\
\midrule
\multicolumn{6}{@{}l}{\textit{Time step $\Delta t$}} \\
& \SI{4}{\nano\second}                 & 1.091 & 1.091 & $-0.00011$ & 0.998 \\
& \textbf{\SI{2}{\nano\second} (prod)} & \textbf{1.091} & \textbf{1.091} & $\mathbf{-0.00005}$ & \textbf{0.998} \\
& \SI{1}{\nano\second}                 & 1.091 & 1.091 & $-0.00005$ & 0.998 \\
\botrule
\end{tabular*}
\end{table}

\paragraph{Convergence criterion.}
Across all tested parameter variations, the X-\cpmg{} scaling exponents
remain stable in both analysis spaces:
$\beta_\mathrm{Std}$ varies by $< 0.003$ ($\varepsilon$-space,
Table~\ref{tab:si:convergence}) and
$\gamma_\mathrm{Std}$ varies by $< 0.012$ ($\chi$-space,
Table~\ref{tab:si:convergence_gamma}).
Both $\Delta\beta$ and $\Delta\gamma$ are indistinguishable
from zero ($|\Delta\beta| < 0.0002$, $|\Delta\gamma| < 0.0002$ in all
configurations), confirming that the waveform-realization null result is
not an artefact of a particular hierarchy depth, bath decomposition,
solver tolerance, or temporal resolution.
The production configuration ($D = 5$, $N_r = 3$, \texttt{atol}$= 10^{-8}$)
sits within the converged plateau for all four dimensions in both spaces.
Since the even-$n$ analysis uses a subset of the same data points
($n = 2, 4$ from the $n = 1$--$5$ window), the main-text $\chi$-space
exponents $\gamma \approx 1.12$ inherit this stability directly.

\paragraph{Cross-solver pipeline check.}
The parameter-sweep convergence tests above vary numerical settings
within a single QuTiP~5.x \texttt{HEOMSolver} implementation and
therefore cannot detect systematic errors common to all runs of this
solver.
A complementary check is provided by the Lindblad \texttt{mesolve}
comparison (Table~\ref{tab:si:lindblad_tier0},
\S\ref{si:floquet_argument}), which substitutes only the dynamical
backend while retaining the identical platform pipeline ($d = 3$ qutrit
Hamiltonian, compiled pulse phases, waveform generation, $\ket{+}$
initial state, and coherence extraction).
Lindblad yields $\gamma_\mathrm{Lindblad} = 1.00$ for X-\cpmg{} and
XY-4 ($R^2 > 0.999$), with monotonic Y-\cpmg{}\rev{, matching the
Floquet-theorem prediction exactly.}
This confirms that the simulation pipeline does not introduce spurious
scaling artefacts and that the \heom{}-specific phenomena arise from
non-Markovian solver physics.
An independent \heom{} implementation (e.g.,
HierarchicalEOM.jl~\cite{huang2023heomjl}) would constitute a fifth
validation layer; this has not been performed and is noted as a
limitation (Discussion, \S\ref{main:sec:disc:limitations}).

\begin{figure}[ht]
\centering
\includegraphicsorplaceholder{\textwidth}{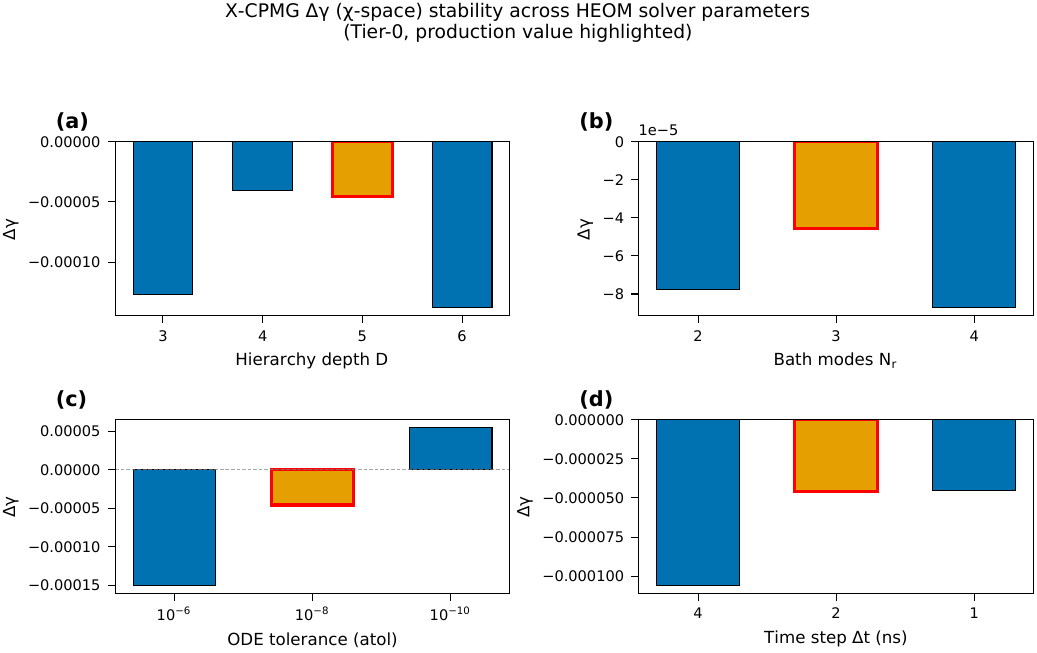}
\caption{%
  CPMG convergence verification.
  Each panel shows $\Delta\gamma$ ($\chi$-space, X-\cpmg{}, Tier-0) as
  a function of one solver parameter, with all others held at production
  values.
  Red-outlined bars mark the production configuration.
  The stability of $\Delta\gamma$ across all parameter sweeps confirms
  numerical robustness; the corresponding $\Delta\beta$ ($\varepsilon$-space)
  values in Table~\ref{tab:si:convergence} show identical stability.%
}\label{fig:si:convergence}
\end{figure}

\section{Fitting robustness analysis}\label{si:fitting_robustness}

The sensitivity of the fitted scaling exponents to methodological choices is
tested along three dimensions: fitting window, functional form, and
coordinate space.
The main text adopts the even-$n$ convention ($n_\pi = 2, 4, 6, 8, 10$)
with $\chi$-space exponents as the primary metric.
The tables below report the complementary pulse-counting ($n = 1$--$5$)
$\varepsilon$-space analysis with exponent $\beta$, \rev{which is a
robustness check and shows the sensitivity of }$\Delta\beta$\rev{ to
the inclusion of }\mbox{$n = 1$}.
The dual-space table (Table~\ref{tab:si:dual_space}) confirms that
$\Delta\beta$ and $\Delta\gamma$ track consistently; the even-$n$
$\chi$-space analysis adopted in the main text uses a subset of these
data points.

\paragraph{Odd--even modulation magnitude.}
To quantify the systematic effect removed by the even-$n$ convention,
we compute odd-$n$ residuals relative to the even-$n$ power-law trend
($\chi = Bn^{\gamma}$ fitted to $n_\pi = 2, 4, 6, 8, 10$ in log-log space).
Table~\ref{tab:si:odd_even} reports the fractional residuals for
X-\cpmg{} at Tier-0.

\begin{table}[ht]
\caption{Odd-$n$ residuals relative to even-$n$ power-law trend
  (X-\cpmg{}, Tier-0, Standard waveform).}
\label{tab:si:odd_even}
\begin{tabular*}{\textwidth}{@{\extracolsep\fill}lccc@{}}
\toprule
$n_\pi$ & $\chi_{\mathrm{actual}}$ & $\chi_{\mathrm{trend}}$ &
  $|\delta\chi/\chi_{\mathrm{trend}}|$ (\%) \\
\midrule
1 & 0.446 & 0.463 & 3.7 \\
3 & 1.503 & 1.579 & 4.9 \\
5 & 2.650 & 2.794 & 5.2 \\
7 & 3.867 & 4.069 & 5.0 \\
9 & 5.341 & 5.388 & 0.9 \\
\midrule
\multicolumn{3}{l}{Mean} & 3.9 \\
\botrule
\end{tabular*}
\end{table}

The dominant source of odd--even modulation is the net-rotation
parity: at odd $n$, the residual $\pi$~rotation about the
refocusing axis couples asymmetrically to the $d = 3$ transmon
transitions through the anharmonicity
($\alpha \approx \SI{-293}{\mega\hertz}$),
producing systematic population redistribution that alternates
sign with each additional $\pi$~pulse.
All five odd-$n$ points lie \emph{below} the even-$n$ trend,
confirming a coherent systematic shift rather than random scatter.

\paragraph{Fitting window sensitivity.}
Table~\ref{tab:si:fit_window} shows the effect of varying the pulse-number
fitting range on the extracted exponents.

\begin{table}[ht]
\caption{%
  Fitting window sensitivity (Tier-0, Standard vs \vppu{}).
  All fits use the pulse-counting convention including odd-$n$ points.
  Bold rows indicate the production window ($n = 1$--$5$).
  $R^2$ values are for the power-law fit $\varepsilon \propto n^{\beta}$.%
}\label{tab:si:fit_window}
\begin{tabular*}{\textwidth}{@{\extracolsep\fill}llccccc@{}}
\toprule
Window & Scheme & $\beta_{\mathrm{Std}}$ & $\beta_{\vppu{}}$ & $\Delta\beta$ &
  $R^2_{\mathrm{Std}}$ & $R^2_{\vppu{}}$ \\
\midrule
\multicolumn{7}{@{}l}{\textit{$n = 1$--$5$ (default, production window)}} \\
& \textbf{X-\cpmg{}} & \textbf{0.597} & \textbf{0.597} & $\mathbf{-0.00007}$ & \textbf{0.965} & \textbf{0.965} \\
& Y-\cpmg{}           & 0.893 & 0.617 & $-0.276^{\dagger}$ & 0.892 & 0.898 \\
& XY-4                & 0.748 & 0.586 & $-0.162^{\dagger}$ & 0.896 & 0.935 \\
\midrule
\multicolumn{7}{@{}l}{\textit{$n = 1$--$4$ (short)}} \\
& X-\cpmg{}           & 0.846 & 0.653 & $-0.193$ & 0.946 & 0.975 \\
& Y-\cpmg{}           & 1.044 & 0.739 & $-0.305$ & 0.937 & 0.966 \\
& XY-4                & 0.857 & 0.656 & $-0.202$ & 0.923 & 0.953 \\
\midrule
\multicolumn{7}{@{}l}{\textit{$n = 2$--$5$ (offset)}} \\
& X-\cpmg{}           & 0.426 & 0.410 & $-0.016$ & 0.987 & 0.986 \\
& Y-\cpmg{}           & 0.421 & 0.326 & $-0.095$ & 0.914 & 0.783 \\
& XY-4                & 0.359 & 0.352 & $-0.007$ & 0.928 & 0.960 \\
\midrule
\multicolumn{7}{@{}l}{\textit{$n = 1$--$7$ (extended, includes partial saturation)}} \\
& X-\cpmg{}           & 0.625 & 0.504 & $-0.121$ & 0.889 & 0.929 \\
& Y-\cpmg{}           & 0.671 & 0.452 & $-0.219$ & 0.787 & 0.768 \\
& XY-4                & 0.616 & 0.496 & $-0.120$ & 0.860 & 0.907 \\
\midrule
\multicolumn{7}{@{}l}{\textit{$n = 1$--$3$ (minimal, 3-point fit)}} \\
& X-\cpmg{}           & 0.957 & 0.716 & $-0.241$ & 0.955 & 0.981 \\
& Y-\cpmg{}           & 1.221 & 0.840 & $-0.382$ & 0.964 & 0.985 \\
& XY-4                & 1.036 & 0.764 & $-0.273$ & 0.970 & 0.983 \\
\botrule
\end{tabular*}
\medskip
\noindent{\footnotesize $^{\dagger}$Fitting artefact of odd-$n$
  contamination from net-rotation parity effects
  (see ``Odd--even modulation magnitude'' above).
  Under the even-$n$ convention adopted in the main text
  ($n_\pi = 2, 4, 6, 8, 10$), $\Delta\beta \approx 0$ for all
  three schemes (Table~\ref{tab:si:eps_exponents}).}
\end{table}

\paragraph{Functional form comparison.}
The default power-law model $\varepsilon(n) = A\,n^{\beta}$ is compared
with a stretched-exponential coherence model
$(1-\varepsilon)(n) = A\,\exp[-(n/n_0)^{\alpha}]$.
Table~\ref{tab:si:func_form} summarises the results.
For XY-4, the Standard--\vppu{} ordering is consistent between both forms:
$\beta_{\mathrm{Std}} > \beta_{\vppu{}}$ and $\alpha_{\mathrm{Std}} > \alpha_{\vppu{}}$.
For X-\cpmg{}, Standard and \vppu{} now yield essentially identical
power-law exponents ($\beta_{\mathrm{Std}} \approx \beta_{\vppu{}} =
0.597$), while the stretched-exponential $\alpha$ values remain
distinct ($\alpha_{\mathrm{Std}} = 0.725 < 0.936 = \alpha_{\vppu{}}$).
The $\alpha$ ordering inversion relative to $\beta$ reflects the
different physical quantities measured by each index: $\beta$ quantifies
the infidelity growth \emph{rate}, while $\alpha$ characterises the
\emph{shape} of the coherence decay (compressed vs.\ stretched
exponential).
A Standard waveform with $\alpha < 1$ undergoes a heavy-tailed coherence
decay, whereas $\alpha \approx 1$ for \vppu{} \rev{indicates near-exponential
decay; yet both waveforms accumulate infidelity at the same rate}
($\beta \approx 0.60$).
The two indices are therefore complementary rather than interchangeable.

\begin{table}[ht]
\caption{%
  Functional form comparison: power-law vs.\ stretched-exponential fits
  ($n = 1$--$5$, Tier-0).
  For X-\cpmg{}, the $\beta$ (power-law) ordering and the $\alpha$
  (stretched-exponential) ordering point in opposite directions because
  they measure different aspects of the decay.%
}\label{tab:si:func_form}
\begin{tabular*}{\textwidth}{@{\extracolsep\fill}llcccc@{}}
\toprule
Scheme & Waveform &
  $\beta_{\mathrm{PL}}$ & $R^2_{\mathrm{PL}}$ &
  $\alpha_{\mathrm{SE}}$ & $R^2_{\mathrm{SE}}$ \\
\midrule
X-\cpmg{} & Standard & 0.597 & 0.965 & 0.725 & 0.996 \\
           & \vppu{}  & 0.597 & 0.965 & 0.936 & 0.999 \\
\midrule
Y-\cpmg{} & Standard & 0.893 & 0.892 & 1.145 & 0.998 \\
           & \vppu{}  & 0.617 & 0.898 & 1.347 & 0.986 \\
\midrule
XY-4      & Standard & 0.748 & 0.896 & 0.749 & 0.994 \\
           & \vppu{}  & 0.586 & 0.935 & 0.656 & 0.999 \\
\botrule
\end{tabular*}
\end{table}

\begin{table}[ht]
\caption{%
  Even-$n$ Tier-0 \cpmg{} $\varepsilon$-space scaling exponents
  ($\varepsilon \propto n^{\beta}$) with
  95\% BCa bootstrap confidence intervals ($B = 10{,}000$).
  $n_\pi = 2, 4, 6, 8, 10$ ($N_\mathrm{d} = 5$ data points,
  $\nu = N_\mathrm{d} - p = 3$ residual degrees of freedom).
  $\Delta\beta \equiv \beta_{\vppu{}} - \beta_{\mathrm{Std}}$.
  Y-\cpmg{} values parenthesised: power-law model inapplicable
  (non-monotonic $\varepsilon$).
  This table is the $\varepsilon$-space companion to the
  $\chi$-space Table~\ref{main:tab:cpmg_gamma} in the main text;
  CI widths reflect the $N_\mathrm{d} = 5$ finite sample
  (Section~\ref{si:bootstrap}).%
}\label{tab:si:eps_exponents}
\footnotesize
\setlength{\tabcolsep}{3pt}
\begin{tabular}{@{}llcl@{}}
\toprule
Scheme & Realisation & $R^2$ & $\beta\;[95\%\;\mathrm{CI}]$ \\
\midrule
\multicolumn{4}{@{}l}{\textit{Well-conditioned scaling}} \\
X-\cpmg{} & Std  & 0.900 & $0.273\;[0.08,\;0.39]$ \\
           & \vppu{} & 0.900 & $0.273\;[0.08,\;0.39]$ \\
           & \multicolumn{3}{l}{$\Delta\beta = +9 \times 10^{-5}$} \\[2pt]
XY-4       & Std  & 0.912 & $0.251\;[0.05,\;0.39]$ \\
           & \vppu{} & 0.912 & $0.251\;[0.05,\;0.39]$ \\
           & \multicolumn{3}{l}{$\Delta\beta = -2 \times 10^{-5}$} \\
\midrule
\multicolumn{4}{@{}l}{\textit{Scaling-law breakdown (power law inapplicable)}} \\
Y-\cpmg{}  & Std  & 0.515 & $(0.144)\;[-0.13,\;0.30]$ \\
           & \vppu{} & 0.515 & $(0.145)\;[-0.12,\;0.30]$ \\
\botrule
\end{tabular}
\end{table}

\paragraph{$\varepsilon$-space versus $\chi$-space consistency.}
The pulse-counting ($n = 1$--$5$) $\beta$-space and $\chi$-space
analyses are summarised in Table~\ref{tab:si:dual_space}.
For X-\cpmg{}, both $\Delta\beta$ and $\Delta\gamma$ are
indistinguishable from zero, confirming that the waveform-realization
null result holds in both analysis spaces.
For Y-\cpmg{} and XY-4, the Standard~$>$~\vppu{} ordering is
preserved in both spaces ($\Delta\gamma = -0.311/-0.187$, tracking
the corresponding $\Delta\beta = -0.276/-0.162$);
however, these nonzero all-$n$ differences are fitting artefacts
of odd-$n$ net-rotation parity contamination
(see ``Odd--even modulation magnitude'' above),
not waveform-dependent physics.
Under the even-$n$ convention adopted in the main text,
the $\chi$-space exponent difference is
$\Delta\gamma \approx 0$ ($+0.000001/+0.0005$) for X-\cpmg{}/XY-4,
consistent with the pulse-counting result for X-\cpmg{}.
The waveform-realization null result is therefore robust for
X-\cpmg{} under both conventions; for Y-\cpmg{} and XY-4,
the all-$n$ differences vanish once the even-$n$ convention
eliminates systematic odd--even effects.

\begin{table}[ht]
\caption{%
  Dual-space consistency ($n = 1$--$5$, Tier-0).
  $\Delta\beta = \beta_{\vppu{}} - \beta_{\mathrm{Std}}$ in $\varepsilon$-space;
  $\Delta\gamma = \gamma_{\vppu{}} - \gamma_{\mathrm{Std}}$ in $\chi$-space.
  All three schemes show the same sign in both spaces.%
}\label{tab:si:dual_space}
\begin{tabular*}{\textwidth}{@{\extracolsep\fill}lcccc@{}}
\toprule
Scheme & $\Delta\beta$ & $\Delta\gamma$ & Same sign & $|\Delta\gamma|/|\Delta\beta|$ \\
\midrule
X-\cpmg{} & $-0.00007$ & $-0.00008$ & Yes & --- \\
Y-\cpmg{} & $-0.276$ & $-0.311$ & Yes & 1.13 \\
XY-4      & $-0.162$ & $-0.187$ & Yes & 1.16 \\
\botrule
\end{tabular*}
\end{table}

\section{Bootstrap methodology}\label{si:bootstrap}

All \cpmg{} scaling exponent uncertainties in this work are estimated via
non-parametric bootstrap resampling~\cite{efron1993introduction}.
The bootstrap analysis presented here uses the even-$n$ $\chi$-space
convention adopted in the main text ($n_\pi = 2, 4, 6, 8, 10$;
$\chi = -\ln(W_n/W_0)$; exponent~$\gamma$).

\paragraph{Resampling protocol.}
For each DD scheme and waveform realization:
\begin{enumerate}
  \item The original dataset consists of $n_{\max} = 5$ even-$n$
    pulse-number points ($n_\pi = 2, 4, 6, 8, 10$), each with a
    single \heom{}-computed decoherence function value $\chi(n)$.
  \item $B = 10{,}000$ bootstrap samples are generated by sampling
    $n_{\max}$ points \emph{with replacement} from the original dataset.
  \item For each bootstrap sample, OLS linear regression in
    $\log\chi$--$\log n$ space extracts the scaling exponent~$\gamma$.
  \item The difference $\Delta\gamma^{(b)} =
    \gamma_{\vppu{}}^{(b)} - \gamma_{\mathrm{Std}}^{(b)}$ is computed for
    each paired bootstrap draw.
\end{enumerate}

\paragraph{Confidence intervals.}
The 95\% confidence intervals are computed using the bias-corrected and
accelerated (BCa) method, which adjusts for both bias and skewness in the
bootstrap distribution.
BCa intervals are preferred over percentile intervals because the
small number of data points ($n_{\max} = 5$) can produce skewed bootstrap
distributions.

\paragraph{Degrees of freedom and model adequacy.}
Each even-$n$ power-law fit has $N_\mathrm{d} = 5$ data points and $p = 2$ free
parameters (intercept $\log B$ and exponent $\gamma$), yielding
$\nu = N_\mathrm{d} - p = 3$ residual degrees of freedom.
For X-\cpmg{}, the resulting $R^2 = 0.993$ with $\nu = 3$ confirms that
the power-law model captures the dominant trend; for XY-4,
$R^2 = 0.986$ similarly indicates adequate fit.
The fitting window (even-$n$ convention, $n_\pi = 2, 4, 6, 8, 10$) was
chosen \emph{a priori} to eliminate systematic odd--even effects
(Section~\ref{main:sec:meth:fidelity}), not optimised post hoc.
The complementary all-$n$ analysis ($N_\mathrm{d} = 10$, $\nu = 8$) in
Section~\ref{si:cpmg_params} provides a robustness check: the exponent
values shift by $< 5$\% and individual CIs narrow substantially (e.g.,
XY-4 ideal from $[0.28, 1.03]$ to $[0.98, 1.01]$), confirming that the
wide even-$n$ CIs reflect limited sample size rather than model
misspecification.

\paragraph{Small-sample coverage limitations.}
With $N_\mathrm{d} = 5$ data points, the number of distinct bootstrap compositions
is $N_\mathrm{d}^{N_\mathrm{d}} = 5^5 = 3{,}125$, which $B = 10{,}000$ draws samples
exhaustively.
The BCa method partially addresses small-sample coverage via bias and
acceleration corrections~\cite{efron1993introduction}, but exact nominal
coverage requires $N_\mathrm{d} \gg p$.
The practical effect is visible in CI width: X-\cpmg{} individual CIs
span 0.35 at $N_\mathrm{d} = 5$ but narrow to 0.055 at $N_\mathrm{d} = 10$ (ideal-pulse
analysis), confirming that width scales with sample size.
These limitations do not affect the paired-comparison conclusion,
because the paired $\Delta\gamma$ CIs (below) exploit cancellation of
systematic uncertainty.

\paragraph{Numerical results (Tier-0, $\chi$-space, even-$n$).}
\emph{Individual exponents.}\enspace
X-\cpmg{}: $\gamma_{\mathrm{Std}} = 1.117$ (95\% BCa CI:
$[1.04, 1.39]$, $R^2 = 0.993$);
$\gamma_{\vppu{}} = 1.117$ ($[1.04, 1.39]$, $R^2 = 0.993$).
XY-4: $\gamma_{\mathrm{Std}} = 0.950$ ($[0.28, 1.03]$, $R^2 = 0.986$);
$\gamma_{\vppu{}} = 0.951$ ($[0.28, 1.03]$, $R^2 = 0.986$).
Y-\cpmg{}: $\gamma_{\mathrm{Std}} = 0.342$
($[-1.01, 0.98]$, $R^2 = 0.213$);
the power-law model is inapplicable due to non-monotonic $\chi(n)$.
\par\noindent\emph{Paired differences ($\chi$-space).}\enspace
X-\cpmg{}: $\Delta\gamma = +5 \times 10^{-6}$
(paired 95\% BCa CI: $[-1.6 \times 10^{-4},\;+4.7 \times 10^{-4}]$,
$p = 0.86$).
XY-4: $\Delta\gamma = +5.3 \times 10^{-4}$
(paired 95\% BCa CI: $[-1.5 \times 10^{-4},\;+2.4 \times 10^{-3}]$,
$p = 0.10$).
Both paired CIs contain zero, confirming that the waveform-realization
difference is statistically undetectable
($|\Delta\gamma|/\gamma < 0.06$\%).
The narrow paired CIs relative to the wide individual CIs reflect
cancellation of systematic bath-induced uncertainty under identical
resampling indices: because both waveform realizations share the same
$n$-values and bath dynamics, correlated fluctuations in $\gamma$
cancel in the difference.
\par\noindent\emph{Ideal-to-compiled exponent difference
(confounded; see main text \S\ref{main:sec:disc:limitations}).}\enspace
$\Delta\gamma_{\mathrm{X}} = +0.167$
(ideal $0.950$ vs.\ compiled $1.117$);
$\Delta\gamma_{\mathrm{XY}} = -0.043$
(ideal $0.993$ vs.\ compiled $0.950$).

\paragraph{Statistical versus practical significance.}
In $\varepsilon$-space, X-\cpmg{} yields
$\Delta\beta = +0.00009$ with $p < 0.001$ (formally significant).
However, $|\Delta\beta|/\beta \approx 0.03$\%, a physically
inconsequential effect.
The \heom{} outputs are deterministic given identical inputs; the
bootstrap therefore detects real but numerically negligible differences
arising from the finite-precision arithmetic of the \vppu{} signal
processing chain.
\rev{The distinction between ``statistically detectable'' and
``experimentally meaningful'' matters here: the }$\Delta\gamma$\rev{ values
confirm waveform equivalence at the numerical noise floor, not a
physical effect that would be resolvable in any experimental measurement.}

\paragraph{\rev{Design choices.}}
The resampling unit is the pulse-number data point, not a synthetic noise
realization.
This reflects the experimental situation: each $n$-point represents a
distinct measurement configuration, and the bootstrap quantifies
uncertainty arising from the limited number of such configurations.
The fitting model (OLS power law in log-log space) is applied identically
across all comparisons (Standard, \vppu{}, ideal) to avoid introducing
systematic differences from model choice.

\end{appendices}
\fi

%%% References %%%
\bibliographystyle{unsrt}
\bibliography{references}

\end{document}